\documentclass[12pt,preprint]{aastex}

\newcommand{\chisq}{\ensuremath{\chi^2}}
\newcommand{\etal}{et~al.}
\newcommand{\om}{\ensuremath{\Omega_m}}
\newcommand{\ode}{\ensuremath{\Omega_{DE}}}

\newcommand{\si}{\ensuremath{\sigma_{\mathrm{int}}}}
\newcommand{\scriptm}{\ensuremath{\mathcal{M}}}
\newcommand{\scriptmo}{\ensuremath{\mathcal{M}_1}}
\newcommand{\scriptmt}{\ensuremath{\mathcal{M}_2}}
\newcommand{\scriptc}{\ensuremath{\mathcal{C}}}
\newcommand{\gmeg}{\ensuremath{g_M}}
\newcommand{\rmeg}{\ensuremath{r_M}}
\newcommand{\imeg}{\ensuremath{i_M}}
\newcommand{\zmeg}{\ensuremath{z_M}}
\newcommand{\snlsfilts}{\gmeg\rmeg\imeg\zmeg}
\newcommand{\up}{\ensuremath{u^{\prime}}}
\newcommand{\gp}{\ensuremath{g^{\prime}}}
\newcommand{\rp}{\ensuremath{r^{\prime}}}
\newcommand{\ip}{\ensuremath{i^{\prime}}}

\newcommand{\bd}{BD\ \ensuremath{17^{\circ}\ 4708}}

\newcommand{\allstatomwom}{\ensuremath{ 0.19^{+0.08}_{-0.10} }}
\newcommand{\allstatomww}{\ensuremath{ -0.90^{+0.16}_{-0.20} }}
\newcommand{\allstatwfixomw}{\ensuremath{ -1.031 \pm 0.058 }}

\newcommand{\allsysomwom}{\ensuremath{ 0.18 \pm 0.10 }}
\newcommand{\allsysomww}{\ensuremath{ -0.91^{+0.17}_{-0.24}}}
\newcommand{\allsyswfixomw}{\ensuremath{ -1.08^{+0.10}_{-0.11}}}

\begin{document}

\title{SUPERNOVA CONSTRAINTS AND SYSTEMATIC
 UNCERTAINTIES FROM THE FIRST 3 YEARS OF THE SUPERNOVA LEGACY 
 SURVEY$^1$}
\shorttitle{SN CONSTRAINTS AND SYSTEMATIC UNCERTAINTIES FROM SNLS3}
\shortauthors{Conley \etal}

\author{
 A.~Conley\altaffilmark{2,3}, J.~Guy\altaffilmark{4}, 
 M.~Sullivan\altaffilmark{5}, N.~Regnault\altaffilmark{4}, 
 P.~Astier\altaffilmark{4}, C.~Balland\altaffilmark{4,6},
 S.~Basa\altaffilmark{7}, R.~G.~Carlberg\altaffilmark{2},
 D.~Fouchez\altaffilmark{8}, D.~Hardin\altaffilmark{4},
 I.~M.~Hook\altaffilmark{5,9}, D.~A.~Howell\altaffilmark{10,11},
 R.~Pain\altaffilmark{4}, N.~Palanque-Delabrouille\altaffilmark{12},
 K.~M.~Perrett\altaffilmark{2,13}, C.~J.~Pritchet\altaffilmark{14},
 J.~Rich\altaffilmark{12}, V.~Ruhlmann-Kleider\altaffilmark{12},
 D.~Balam\altaffilmark{14}, S.~Baumont\altaffilmark{15},
 R.~S.~Ellis\altaffilmark{5,16}, S.~Fabbro\altaffilmark{14,17},
 H.~K.~Fakhouri\altaffilmark{18}, N.~Fourmanoit\altaffilmark{4},
 S.~Gonz\'alez-Gait\'an\altaffilmark{2}, M.~L.~Graham\altaffilmark{14},
 M.~J.~Hudson\altaffilmark{19}, E.~Hsiao\altaffilmark{18}, 
 T.~Kronborg\altaffilmark{4}, C.~Lidman\altaffilmark{20}, 
 A.~M.~Mourao\altaffilmark{17}, J.~D.~Neill\altaffilmark{21},
 S.~Perlmutter\altaffilmark{18,22}, P.~Ripoche\altaffilmark{18,4},
 N.~Suzuki\altaffilmark{18}, E.~S.~Walker\altaffilmark{5,23}
}

\altaffiltext{1}{
 Based on observations obtained with MegaPrime/MegaCam, a joint project
 of CFHT and CEA/DAPNIA, at the Canada-France-Hawaii Telescope (CFHT)
 which is operated by the National Research Council (NRC) of Canada,
 the Institut National des Sciences de l'Univers of the Centre National
 de la Recherche Scientifique (CNRS) of France, and the University of
 Hawaii. This work is based in part on data products produced at the
 Canadian Astronomy Data Centre as part of the Canada-France-Hawaii
 Telescope Legacy Survey, a collaborative project of NRC and CNRS.
}
\altaffiltext{2}{Department of Astronomy and Astrophysics, University
  of Toronto, 50 St. George Street, Toronto, ON M5S 3H4, Canada}
\altaffiltext{3}{Center for Astrophysics and Space Astronomy, University
  of Colorado, 593 UCB, Boulder, CO 80309-0593, USA}
\altaffiltext{4}{LPNHE, Universit\'e Pierre et Marie Curie Paris 6,
 Universit\'e Paris Diderot Paris 7, CNRS-IN2P3,
 4 place Jussieu, 75252 Paris Cedex 05, France}
\altaffiltext{5}{Department of Astrophysics, University of Oxford, 
  Keble Road, Oxford OX1 3RH, UK}
\altaffiltext{6}{Universit\'e Paris 1, Orsay, F-91405, France}
\altaffiltext{7}{LAM, CNRS, BP8, P\^ole de l'\'Etoile Site de 
  Ch\^ateau-Gombert,  38, rue Fr\'ed\'eric Joliot-Curie,
 13388 Marseille Cedex 13, France}
\altaffiltext{8}{CPPM, CNRS-IN2P3 and Universit\'e Aix-Marseille II,
 Case 907, 13288 Marseille Cedex 9, France}
\altaffiltext{9}{INAF - Osservatorio Astronomico di Roma, via
 Frascati 33, 00040 Monteporzio (RM), Italy}
\altaffiltext{10}{Las Cumbres Observatory Global Telescope Network,
 6740 Cortona Dr., Suite 102, Goleta, CA 93117, USA}
\altaffiltext{11}{Department of Physics, University of California, 
 Santa Barbra, Broida Hall, Mail Code 9530, Santa Barbara,
 CA 93106-9530, USA}
\altaffiltext{12}{CEA, Centre de Saclay, Irfu/SPP, F-91191
 Gif-sur-Yvette, France}
\altaffiltext{13}{Network Information Operations, DRDC Ottawa, 3701
 Carling Avenue, Ottawa, ON, K1A 0Z4, Canada}
\altaffiltext{14}{Department of Physics and Astronomy, University of
 Victoria, PO Box 3055 STN CSC, Victoria BC, V8T 1M8, Canada}
\altaffiltext{15}{LPSC, CNRS-IN2P3, 53 rue des Martyrs, 38026
 Grenoble Cedex, France}
\altaffiltext{16}{Department of Astrophysics, California Institute of
 Technology, MS 105-24, Pasadena, CA 91125, USA}
\altaffiltext{17}{CENTRA-Centro M.~de Astrofisica and Department of
 Physics, IST, Lisbon, Portugal}
\altaffiltext{18}{LBNL, 1 Cyclotron Rd, Berkeley, CA 91125}
\altaffiltext{19}{Department of Physics and Astronomy, University of
  Waterloo, 200 University Avenue West, Waterloo, Ontario N2L 3G1,
  Canada}
\altaffiltext{20}{Australian Astronomical Observatory, P.O.\ Box 296,
 Epping, NSW 1710, Australia}
\altaffiltext{21}{California Institute of Technology, 
  1200 E. California Blvd., Pasadena, CA 91125, USA}
\altaffiltext{22}{Department of Physics, University of California,
 Berkeley, 366 LeConte Hall MC 7300, Berkeley, CA 94720-7300, USA}
\altaffiltext{23}{Scuola Normale Superiore, Piazza dei Cavalieri 7,
 56126 Pisa, Italy}

\email{ alexander.conley@colorado.edu }

\begin{abstract}
We combine high redshift Type~Ia supernovae from the first 3 years of
the Supernova Legacy Survey (SNLS) with other supernova (SN) samples,
primarily at lower redshifts, to form a high-quality joint sample of
472 SNe (123 low-$z$, 93 SDSS, 242 SNLS, and 14 {\it Hubble Space
  Telescope}). SN data alone require cosmic acceleration at $>99.999
\%$ confidence, including systematic effects.  For the dark energy
equation of state parameter (assumed constant out to at least $z=1.4$)
in a flat universe, we find $w = -0.91^{+0.16}_{-0.20}\left(
\mathrm{stat} \right) ^{+0.07}_{-0.14} \left( \mathrm{sys} \right)$
from SNe only, consistent with a cosmological constant.  Our fits
include a correction for the recently discovered relationship between
host-galaxy mass and SN absolute brightness.  We pay particular
attention to systematic uncertainties, characterizing them using a
systematics covariance matrix that incorporates the redshift
dependence of these effects, as well as the shape-luminosity and
color-luminosity relationships. Unlike previous work, we include the
effects of systematic terms on the empirical light-curve models.  The
total systematic uncertainty is dominated by calibration terms.  We
describe how the systematic uncertainties can be reduced with soon to
be available improved nearby and intermediate-redshift samples,
particularly those calibrated onto USNO/SDSS-like systems.
\end{abstract}


\section{INTRODUCTION}
\nobreak 
\label{sec:introduction}
The Supernova Legacy Survey (SNLS) is a 5 year program to measure the
expansion history of the universe using Type~Ia supernovae (SNe~Ia).
The goal of this survey is to measure the time averaged equation of
state of dark energy $w$ to 0.05 (statistical uncertainties
only) in combination with other measurements, and to 0.10
including systematic effects.  The fundamental nature of dark energy,
which makes up 3/4 of the mass-energy budget of the universe, remains
almost completely mysterious. A solid measurement that $w \neq
-1$ (which would rule out the cosmological constant) would have
profound implications for cosmology and particle physics.  SNLS
completed data acquisition in 2008 June; this paper presents SN-only
cosmological results from the first 3 years of operation (SNLS3).

All analysis in this paper is in the context of standard cosmological
models -- i.e., we assume that the universe is homogenous on large
scales and that general relativity is correct.  SNe are used to
measure the cosmological parameters by comparing their apparent
brightnesses over a range of redshifts.  Hence, it is very useful to
include additional SN samples besides SNLS in the analysis,
particularly nearby SNe ($z < 0.1$). This paper has two primary goals.
The first is to place SNe from the literature on a common framework
and demonstrate the resulting constraints.  The second is to present
the systematic uncertainties on the SN measurements in detail.  This
is the second in a series of three SNLS cosmology papers
based on the first 3 years of data.  The
SNLS data sample used in this analysis is presented in
\citet[hereafter G10]{2010guylightcurves}, and the SN constraints are
combined with other measurements in \citet[hereafter
  S11]{2010sullivancosmo}, primarily the WMAP7 measurement of the
cosmic microwave background (CMB), and baryon acoustic oscillations measured
with galaxy redshift surveys.  In addition to
these three papers, the calibration of SNLS data is discussed in
\citet[hereafter R09]{2009regnaultcalib}, and the simulations used to
evaluate our survey selection effects in \citet{2010AJ....140..518P}.

Searches for additional parameters beyond light-curve shape and color
have been ongoing for two decades, and recently,
\citet{2010ApJ...715..743K} have found evidence that SN residuals from
the Hubble diagram are correlated with host-galaxy mass.  The physical
cause of this effect is as yet unknown.  \citet{2010MNRAS.406..782S}
analyze this issue using the SNLS3 sample and confirm this result at
higher significance.  They find that splitting the sample by
host-galaxy mass and allowing the peak absolute magnitudes of the two
samples to differ corrects for these effects.  We adopt this approach
in this paper; see \S\ref{subsec:twoscriptm} for details.  This effect
has also been confirmed in the Sloan Digital Sky Survey (SDSS) SN sample
\citep{2010arXiv1005.4687L}.

Here, we give a brief overview of the data sets used in this analysis.
SNLS combines photometry from the deep component of the
Canada-France-Hawaii Telescope (CFHT) Legacy survey with extensive
spectroscopic follow-up from the Keck, Very Large Telescope (VLT), and
Gemini telescopes to determine SN types and measure redshifts.  The
photometry is carried out with MegaCam, a 1 $\mathrm{deg}^2$ imager at
the prime focus of CFHT.  SNLS is a rolling search, like most other
modern high-redshift surveys, which means that the same telescope is
used to simultaneously find new SNe candidates and to follow those
already discovered, resulting in a large multiplex efficiency
advantage. Details of the procedures used to find and prioritize new
SNe can be found in \citet{2006AJ....131..960S}, and those used to
type candidates from their spectra in \citet{2005ApJ...634.1190H,
  2009A&A...507...85B}.

In addition to SNLS, we consider three additional data sets: low-$z$,
SDSS, and high-redshift SNe from the {\it Hubble Space Telescope}
({\it HST}).  The low-$z$ SNe, which we define here as coming from
surveys with the bulk of their SN below $z=0.1$, come from a
heterogeneous combination of non-rolling surveys which generally use
different telescopes to find and follow SN candidates.  The {\it HST}
SNe \citep{2007ApJ...659...98R} are at higher redshifts than any of
the other samples.  They contribute relatively little to measurements
of $\left<w\right>$, but are quite useful when trying to measure any
redshift evolution of $w$.  The SDSS SN survey \citep[][hereafter
  K09]{2008AJ....136.2306H, 2009ApJS..185...32K} occupies intermediate
redshifts ($0.1 < z < 0.4$) between SNLS and the low-$z$ SN, and is
also a rolling search.

As SN samples have grown in size, characterizing and
incorporating systematic uncertainties properly has grown in
importance.  There are many aspects of SNLS which are designed to
reduce the effects of systematic uncertainties compared with previous
SN projects, but they are still roughly comparable to the
statistical uncertainties.  A major lesson of this paper is that most
of the systematic effects which limit the current analysis are related
to the current low-redshift SN sample, particularly in terms of
cross-calibration requirements.  Absolute calibration is unimportant
for our purposes, but the relative calibration of different systems
and observations at different wavelengths is critical.  The current
low-$z$ sample is dominated by SNe largely calibrated to the Landolt
system \citep{1992AJ....104..340L}, which induces many complications
in our calibration (see R09 for more details).  As the low-$z$ SN sample
is replaced by better calibrated samples in the next few years,
it will be possible to cross-calibrate the various samples much more
accurately, which will substantially increase the legacy value of the
SNLS sample.  The dominant uncertainties will probably then relate to
the host-galaxy--SN brightness relation and SN modeling, particularly
the thorny issue of SN colors.

A critical step in any SN cosmological analysis is light-curve
fitting, the conversion of a time series of photometric (and possibly
spectroscopic) observations into a set of model parameters for each
SN which are used to estimate a relative distance.  Because
SN physics is sufficiently complicated, theoretical models have
so far offered relatively little guidance for this process.  As a
result, all current models used for light-curve fitting are empirical
in nature.  In this paper we consider updated versions of two models:
SiFTO \citep{2008ApJ...681..482C} and SALT2
\citep{2007A&A...466...11G}.  While these models share a common
overall philosophy, there are many significant differences between
them.  They are compared in detail in \S5 of G10, and we include the
differences in the light-curve parameters in our uncertainty budget.
Because these models are trained on SN data, they are affected by the
same sources of systematic effects as the cosmological analysis.  This
is the first analysis to include this effect in the analysis; previous
analyses have therefore underestimated their uncertainties
by holding the light-curve model fixed when modeling systematics.

There has been some variation in the literature as to how SN
systematic effects are treated.  Examples of the most common approach,
which we will refer to as the quadrature method, can be found in
\citet{1999ApJ...517..565P, 2006A&A...447...31A,2007ApJ...666..694W};
K09.  This approach has a number of disadvantages (and the advantage
that it is relatively easy to understand). The most important
disadvantage is that it is difficult for subsequent consumers of SN
relative distance moduli to incorporate systematic uncertainties into
their analyses.  In this paper we rectify these deficiencies by
modeling systematic effects using a covariance matrix.

In \S\ref{sec:datasets}, we describe the data sets included in this
analysis and the steps taken to bring them onto a common system with
the SNLS data.  We then describe the combined data set in
\S\ref{sec:combined}, presenting the statistical constraints on dark
energy from SNe~Ia alone, and the combined statistical and systematic
uncertainties in \S\ref{sec:application} along with a description of
our systematics methodology.  In section \S\ref{sec:systematics} we
describe individual systematic terms in detail, and in
\S\ref{sec:compare} we compare our analysis with previous ones.
Finally, in \S\ref{sec:improvement} we discuss ways in which the
effects discussed in this paper can be improved with the enhanced low-
and intermediate-redshift SN data sets which will be available in the
near future.

\section{DATA SETS}
\nobreak
\label{sec:datasets}
Ideally, all SNe would be observed with a single camera on a single
telescope.  While this may be possible with future programs such as
LSST or Pan-STARRS, at the moment it is not practical because no
individual telescope can currently obtain a large sample at both high
and low redshifts.  To obtain precision cosmological constraints,
particularly on $w$, we must incorporate additional sSN data
besides SNLS.  Rather than including all available SN data, we have
chosen to incorporate only external surveys which cover different
redshift ranges than SNLS.  Other surveys which cover substantially
the same redshift range as SNLS have fewer SN, larger photometric
uncertainties, less certain calibration, and have at most 2-band
coverage (compared with the 4 bands of SNLS).  Including them in our
analysis would reduce our statistical uncertainties only marginally,
and would introduce additional systematic effects which would have to
be analyzed in detail.  Nearby SNe are currently the most important
addition because the combination of low redshifts and the small
redshift range means that they provide a very good constraint on the
absolute magnitudes of SNe~Ia\footnote{More precisely, they constrain
  a combination of the absolute magnitudes and the Hubble constant,
  $H_0$; we refer to this combination as \scriptm .}, which is largely
independent of the density parameters and $w$.

There has been a recent encouraging trend toward providing photometry
in the natural system of the detectors used to obtain it -- that is,
rather than applying linear relations to transform instrumental
magnitudes into some standard system, the calibration is transferred
from the standard system to the instrumental system, which is much
more reliable.  Using natural system photometry involves slightly more
work in the analysis, but allows for a considerable improvement in
accuracy and precision, as long as the natural system response is well
measured.  We use the natural systems of the SNLS, {\it HST}, SDSS,
\citet{2009ApJ...700..331H}, and \citet{2010AJ....139..519C} samples
in our analysis, which together account for $\sim 90\%$ of our sample.

In this section, we present the data samples we include in our
analysis, and describe the steps we carry out to bring them to a
common system with the SNLS measurements.  The most critical items
relate to calibration, but in addition it is important to understand
and apply a correction for the selection biases of the external
samples (i.e., Malmquist bias).  The calibration used in this analysis
relies heavily on the CALSPEC program based on {\it HST} observations
of pure-hydrogen white dwarfs \citep{1996AJ....111.1743B}.  In
particular, we use \bd\ as our primary reference standard.  We
first describe our cuts (\S\ref{subsec:cuts}), introduce the data
samples (SNLS, nearby, {\it HST}, and SDSS,
\S\ref{subsec:snls}-\ref{subsec:sdssdata}), and then discuss how we
estimate selection effects for these samples (\S\ref{subsec:malm}) and
the peculiar velocity corrections we apply to nearby SNe
(\S\ref{subsec:pecvel}).  All of these have related systematic
uncertainties, which are discussed in \S\ref{sec:systematics}.

\subsection{Selection Requirements}
\label{subsec:cuts}
\nobreak 
The selection requirements (cuts) applied in our analysis have
already been described in \S4.5 of G10; these are designed to ensure adequate
phase and wavelength coverage to allow accurate parameter measurement.
A particularly important requirement is that each SN has data between
$-8$ and $+5$ rest-frame days of peak brightness to avoid biasing the
recovered light-curve parameters, which primarily affects the low-$z$
sample -- see G10 for further discussion.

As is the case for SNLS, we require spectroscopic confirmation of all
SNe.  Since each SN team has its own scheme for spectroscopic
classification, it is not clear how to best ensure uniform
  classification standards without re-examining all of the raw
spectra.  Based on direct comparison of spectra, the
  classification requirement applied to SNLS (SN~Ia$\star$
\citet{2009A&A...507...85B}, or CI=3 in the scheme of
\citet{2005ApJ...634.1190H}) lies somewhere between the ``gold'' and
``silver'' classification scheme of \citet{2007ApJ...659...98R} for
spectroscopically typed SN.  We have chosen to largely accept the
classifications of the original authors with a few caveats.  First, as
is the case for all SNLS SNe, we do not include candidates whose type
is based either purely on the photometric properties of the SN, or the
red color or elliptical morphology of its host galaxy; the latter
standard is particularly worrisome because some nearby ellipticals
show evidence of star formation, and this fraction may increase with
redshift.  Recently, \citet{2010Natur.465..326K} have reported the
discovery of an (unusual) core-collapse SN in an elliptical host,
which strengthens our caution.

Because we apply a peculiar velocity correction, we place our minimum
redshift cut at $z_{\mathrm{cut}} = 0.010$, somewhat lower than the usual value
of $0.015 - 0.020$.  There has been some recent controversy relating
to the minimum allowable redshift, related to the potential for a
discontinuous step in the local expansion rate (a so-called
Hubble bubble) detected by
\citet{2007ApJ...659..122J}. \citet{2007ApJ...659...98R} use $z_{\mathrm{cut}}
= 0.023$ in order to avoid this issue, which removes 40\% of the
nearby sample.  \citet{2007ApJ...664L..13C} argue that the
Hubble bubble is an artifact of the treatment of SNe colors combined
with selection effects in the nearby sample, and
\citet{2009ApJ...700.1097H} shows that adding more nearby SNe shifts
the position and sign of the putative bubble considerably.
K09 take a slightly more agnostic position (\S9.1 and 9.2), but
find significant variation in the cosmological parameters with the
minimum redshift cut, and therefore adopt $z_{\mathrm{cut}} = 0.02$ while
including the variation as a major source of systematic uncertainty.
We also see similar variation with $z_{\mathrm{cut}}$ in our sample, but find
it to be consistent with shot noise, and see no evidence
that such conservatism is warranted (\S\ref{subsec:bubble}).  SNe with
$z < z_{\mathrm{cut}}$ are still used to train the light-curve fitters, since
we do not use distance information in this process.

We exclude known peculiar SNe by hand, such as SNe 2000cx and 2002cx,
rather than using automated quality of fit (\chisq) cuts.  In our
experience, the uncertainties for low-$z$ photometry are sufficiently
inaccurate that such cuts are often misleading.  Furthermore, many
low-$z$ SNe have the occasional outlying photometric observation which
has little to no effect on the derived parameters but which drives the
\chisq\ of the light-curve fit to large values.  A \chisq\ based cut
will also affect different SN samples very asymmetrically because the
signal-to-noise of the photometry varies strongly between samples, and
hence may introduce bias.  Since SNe~Ia are not perfect standard
candles, we should include some additional scatter in their corrected
peak magnitudes for cosmological purposes.  This is often called
``intrinsic scatter'' (\si) even though it probably represents the gaps
in our understanding of SN physics and photometry rather than anything
truly intrinsic.  It is impossible for any fully automated set of cuts
to catch all the pathologies of such a large and diverse collection of
SNe, so we have also inspected all of the SNe by eye to remove
problems not caught by our cuts.  In any case, we remove the known
spectroscopically peculiar SNe 2000cx \citep{2001PASP..113.1178L},
2001ay (Krisciunas et al., 2011, in preparation), 2002cx
\citep{2003PASP..115..453L}, 2005hk \citep{2007PASP..119..360P},
2005gj \citep{2006ApJ...650..510A}, 03D3bb
\citep{2006Natur.443..308H}, 05D1by (which is spectroscopically
similar to SN 2001ay), and SN 2006X because it shows evidence for a
light-echo \citep{2008ApJ...677.1060W}.  Furthermore, we require all
SNe to have estimated photometric uncertainties.

We apply Chauvenet's criterion \citep{1997ieas.book.....T} to reject
outliers, removing SNe for which we could expect less than half of an
event in our full sample (assuming a Gaussian distribution of
intrinsic luminosities).  This corresponds to a cut at about $3.2
\sigma$, which removes six~SNe: one low-$z$ (SN 2006cj), one from SDSS
(SDSS5635), three from SNLS (03D4au, 04D4gz, 05D2ei), and one from the
{\it HST} sample (McEnroe), many more than the 0.5 objects
one would expect if the distribution were truly Gaussian.  Of
the SNLS SNe, two of these have slightly less secure spectroscopic
identifications (SNIa$\star$) and are faint relative to the best
fit, so may either be non-SN~Ia or spectroscopically peculiar
SN~Ia that our spectra were not high enough signal-to-noise to
identify.  Note that lensing effects are too small to explain
these outliers, as the expected number of $3.2\sigma$ or
greater outliers due to lensing is much less than 1 for our
sample.

\begin{figure}
\plottwo{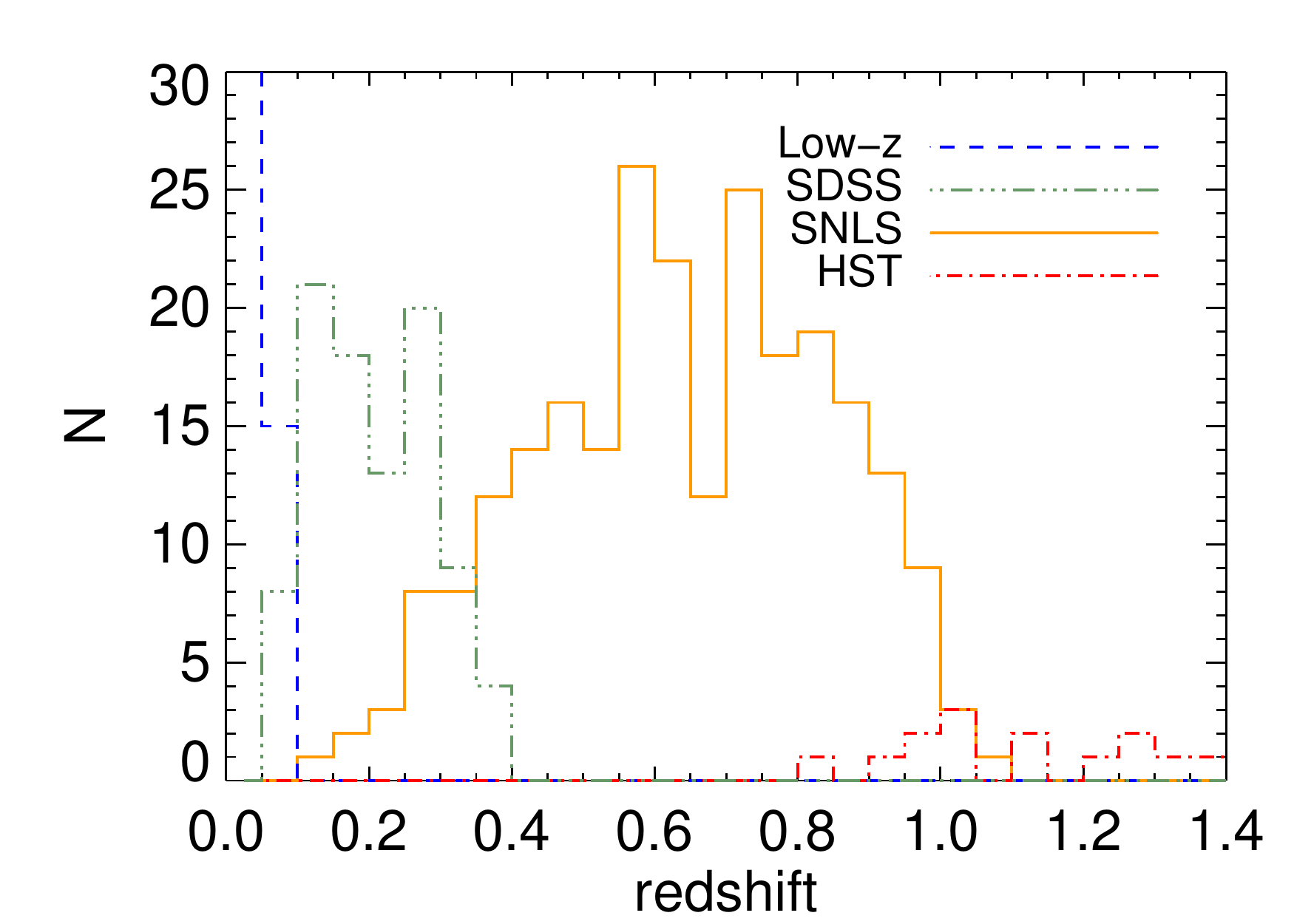}{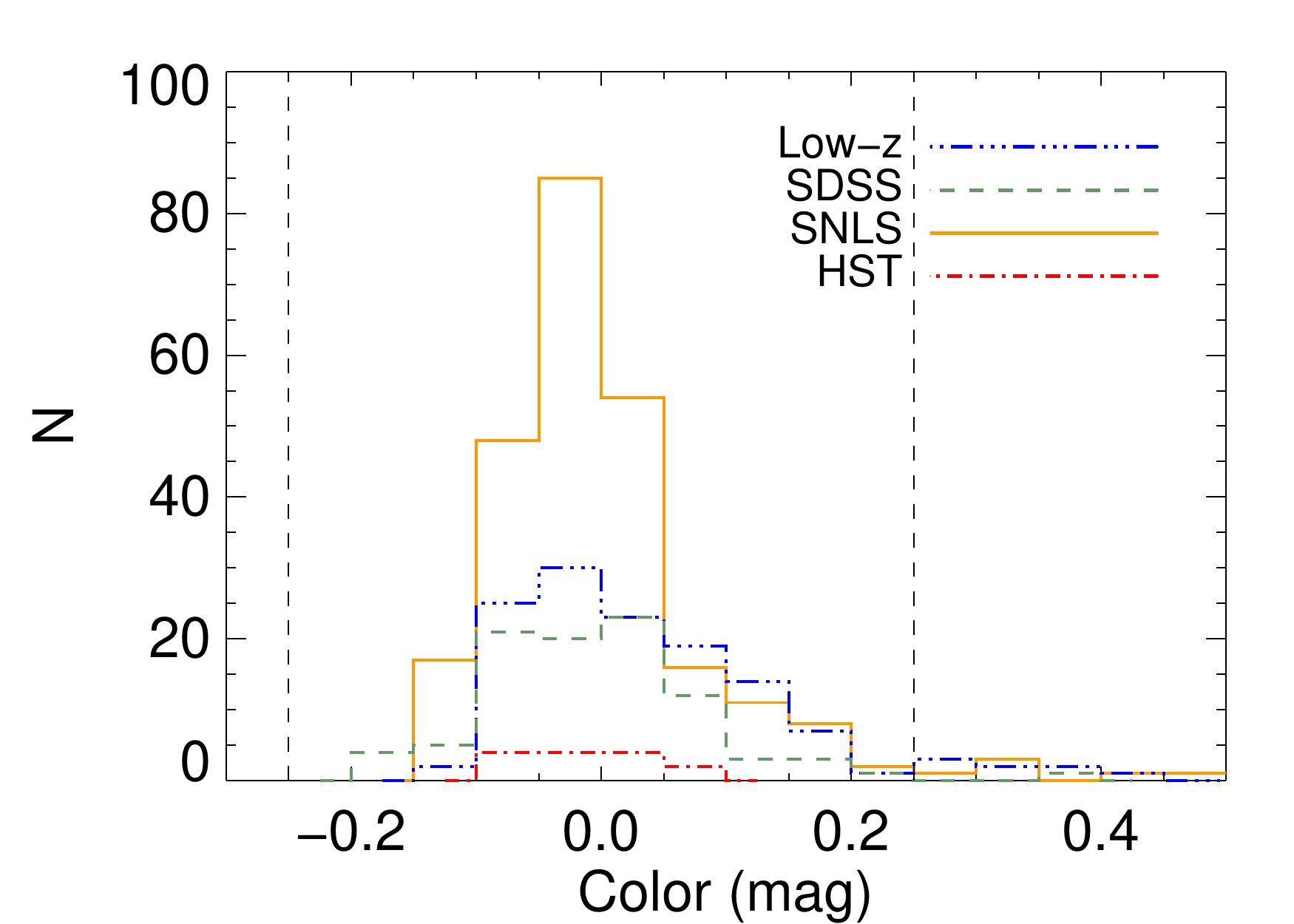}
\caption[Redshift and color histograms of the input SN samples]
{Redshift and color histograms of the samples used in this
analysis.  The left panel shows the redshift histogram after all 
cuts are applied. The first bin for the low-$z$ SNe contains 108 SNe.
The right panel shows the color histogram before the color
cut is applied (but removing known peculiar SNe as well as those
with insufficient coverage), and the cut is indicated by the vertical 
dashed lines. \label{fig:zclrhisto}}
\end{figure}

We require the Galactic reddening along the line of sight to satisfy
$E\left(B - V\right)_{MW} < 0.2$ mag because of concerns that the
assumed Galactic value of $R_V = 3.1$ might not be appropriate for
highly extinguished objects.  Next, we require $0.7 < s < 1.3$, where
$s$ (stretch) is the light-curve width parameter for SiFTO; neither
SiFTO nor SALT2 produce reliable fits for SNe outside that stretch
range.  Finally, we require $-0.25 < \left(B - V\right)_{\mathrm{Bmax}} < 0.25$
because of concerns that very blue SNe are not represented in our
training sample, and that the colors of very red SNe may represent a
combination of different effects, both affecting the peak magnitudes
of the SNe; see \citet{2008ApJ...681..482C} for further discussion of
the latter.  Histograms for the redshift and color distributions of
the input samples are shown in figure~\ref{fig:zclrhisto}, and the
effects of the cuts are given in table~\ref{tbl:cuteffects}.  The
requirement of good coverage near the peak luminosity has the largest
effect, mostly on the low-$z$ sample.  We also considered the effects of
the sharp color cut because of concerns that a high-redshift SN with
poorly measured colors might migrate across the cut boundary, but
found that the effect was negligible.

\begin{deluxetable}{lrr|rrrrrrr}
\tablewidth{0pt}
\tabletypesize{\small}
\tablecaption{Effects of Cuts on Different Samples}
\tablehead{
 \colhead{Dataset} & 
 \colhead{Initial\tablenotemark{a}} & 
 \colhead{Final\tablenotemark{b}} &
 \colhead{Coverage\tablenotemark{c}} &
 \colhead{Fail\tablenotemark{d}} & \colhead{$z$ Cut} &
 \colhead{$s$} & \colhead{Color} & \colhead{Outliers}
}
\startdata
 low-$z$ & 323 & 123 & 99 &  9 & 54 & 32 &  31 & 1 \\
 SDSS  & 101  &  93 & 6  & 0  & 0  & 4  &  1 & 1 \\
 SNLS  & 279 & 242 & 16  & 10 & 0  & 6  &  9 & 3 \\
 {\it HST} & 35 & 14 & 5 & 12 & 11 & 3 & 0 & 1 \\
\enddata
\tablecomments{The number of SN removed by each 
  selection criterion. Many SNe fail multiple cuts. \label{tbl:cuteffects} }
\tablenotetext{a}{The initial number of SN, after the removal of known peculiar
 SNe, SN with clear photometric inconsistencies, and
 SN which have better photometry from other samples.}
\tablenotetext{b}{The number of SN satisfying all selection criterion.}
\tablenotetext{c}{SNe which did not have data close enough to peak
 brightness.}
\tablenotetext{d}{SNe for which the fits could not provide reliable
 light-curve parameters, either because of an insufficient number
 of epochs of data or insufficient wavelength coverage to measure a
 color.  }
\end{deluxetable}

\subsection{SNLS SNe}
\label{subsec:snls}
\nobreak 
Systematics control is fundamental to the design of SNLS.
\begin{itemize}
\item Because SNe are both discovered and photometrically followed
  with only one telescope, we are able to concentrate our efforts on
  thoroughly understanding that system (R09).  We can also avoid the
  difficulties associated with combining observations from many
  telescopes onto a single system.
\item The survey is in four passbands (\gmeg \rmeg \imeg \zmeg ),
  allowing us to measure colors for all SNe in our survey.  We measure
  different rest-frame passbands at different redshifts; at low
  redshifts we are mostly sensitive to $B-V$, and at high redshifts to
  $U-B$.  These measurements must be used in a consistent fashion at
  all redshifts to obtain accurate results.  We use
  intermediate-redshift SNLS SNe as a consistency check of this
  process, since they have high-quality measurements of both $U-B$ and
  $B-V$ (G10).
\item There are four survey fields (D1-D4) distributed in RA to allow
 year-round coverage.  The results from the four fields can be
 compared with each other (S11). 
\item The survey is quite deep, which limits the effects of
 Malmquist bias \citet{2010AJ....140..518P} at $z<0.6$, the sweet spot
 for measuring a constant $w$.
\item We obtain spectra for all of the SNe used in our analysis, which
  allows us to limit non-Ia contamination, to search for peculiar SNe
  such as 03D3bb \citep{2006Natur.443..308H}, and the comparison of
  the spectral energy distributions (SEDs) of nearby and distant SNe
  as a test for evolution
  \citep{2008A&A...477..717B,2009A&A...507...85B,2010walkerspectra}.
\item We have obtained much higher signal-to-noise spectra of a subset
 of our SNe which can be used to study the near-UV properties of SNe~Ia
 in detail, and allows more detailed spectroscopic comparisons
 \citep{2008ApJ...674...51E,2009ApJ...693L..76S}.
\item Because SNLS is a rolling search, it is possible to go back after
 an SN is discovered and study early-time, pre-discovery photometry.  
 Therefore almost all of our SNe have very good early-time coverage, 
 which can be used to test evolutionary models which predict changes 
 in the early-time behavior \citep{2006AJ....132.1707C}.
\item Because the duration of our survey is much longer than the month
  timescale of SNe~Ia, we can construct very deep SN-free image stacks
  and get accurate colors for the host galaxies.  These can be turned
  into estimates of the mass and star formation history and used to
  study the relation between SN properties and host galaxy environment
  to search for host-dependent systematic effects
  \citep{2006ApJ...648..868S,2010MNRAS.406..782S}.
\item Our photometry allows us to construct improved empirical models
  of SNe~Ia lightcurves and spectra
  \citep{2007A&A...466...11G,2008ApJ...681..482C}, which is particularly
  important in the near-UV.
\end{itemize}
SNLS is designed to obtain sufficient quality data to allow us to
investigate systematic effects within our own data, as well as
enabling a great deal of non-cosmological SN science too extensive
to list here.

We take the 3 year SNLS sample from G10 with only one modification: we
have updated the light-curve models to account for the 2010 February update
to the CALSPEC calibration library.  This update has almost no effect
on the SNLS, SDSS, and low-$z$ SNe, but does have some on the {\it HST}
SNe, which are observed in the near-IR.

\subsection{low-$z$ SNe}
\label{subsec:nearbydata}
\nobreak
The current nearby SN set is dominated by five main samples:
Cal\'an/Tololo \citep[29 SNe]{1996AJ....112.2408H}, CfAI \citep[22
  SNe]{1999AJ....117..707R}, CfAII \citep[44
  SNe]{2006AJ....131..527J}, CfAIII \citep[185
  SNe]{2009ApJ...700..331H}, and CSP \citep[35
  SNe]{2010AJ....139..519C}.  In addition, we make some use of 37 SNe
with modern photometry from a mixture of other papers.  As the numbers
above show, the CfAIII sample provides almost half of the nearby
sample, and it was tempting to include only these SNe to simplify
  our analysis. However, the other samples contribute to the training
of the light-curve models disproportionately because some have denser
phase coverage or sample slightly different wavelengths, and since it
is necessary to characterize their properties and systematic
uncertainties to include them in the training, there is no point in
excluding them from the cosmological analysis.  The effects of the
cuts of \S\ref{subsec:cuts} are broken down by each nearby sample
  in table~\ref{tbl:lowzsamplenumbers}.  We do not use rest-frame
observations in the $U$-band of the nearby sample in our analysis for
the reasons discussed in \S\ref{subsec:lowzu}.

\begin{deluxetable}{lrrr}
\tablecaption{Contributions to the Low-$z$ SN Sample}
\tablehead{
 \colhead{Source} & \colhead{Initial Number\tablenotemark{a}} & 
 \colhead{All Uses} & \colhead{Cosmology Fit}
}
\startdata
 Cal\'an/Tololo & 29  & 19 & 17 \\
 CfAI  & 22  & 10 & 7  \\
 CfAII & 43  & 18 & 15 \\
 CfAIII & 172 & 60 & 58 \\
 CSP  & 20  & 20 & 14 \\
 Other                       & 37  & 23 & 12 \\
\hline
 Total                       & 323 & 150 & 123 \\
\enddata
\tablenotetext{a}{After the removal of duplicates, known peculiar
 SNe, and SNe with problematical photometry.}
\tablecomments{The relative contributions of various
 sources to the low-$z$ sample used in this paper.
 \label{tbl:lowzsamplenumbers}}
\end{deluxetable}

A handful of the nearby SNe have photometry from multiple sources.
This is useful for estimating the uncertainty in the zero points of
different samples (\S\ref{subsubsec:lowzzp}), but also forces us to
make some decisions about which data to include.  Generally, nearby
SNe have dense enough light-curve coverage that adding additional
points does not improve the uncertainties significantly.  Since
\si\ dominates the uncertainty budget for these SNe\footnote{SN~Ia are
  not perfect standard candles even after correction for the empirical
  width- and color-luminosity relations.  $\sigma_{\mathrm{int}}$
  represents the remaining scatter in distance moduli, and is
  discussed further in \S3.1 and 3.4.}, including data from multiple
sources is not beneficial unless the data samples complement each other
in wavelength or light-curve phase.  Unless this is the case, or one
of the data sets is clearly superior, we generally prefer to only use
data from one of the five large surveys, where we have a better
understanding of the systematic uncertainties.

For the Cal\'an/Tololo, CfAI and CfAII samples, the data were
transformed by the authors from the natural instrumental system into the
the \citet{1992AJ....104..340L} system using linear transformations
derived from stars in a limited color range.  Since SN and stellar
spectra are quite different, these linear transformations are
incorrect when applied to SNe, and will introduce an error in the
reported magnitudes which is correlated across the SN sample.  In
order to use these samples in a precision analysis, we must
determine the effective passbands of the Landolt system,
described in Appendix~\ref{apndx:filterslowz}.

For the CfAIII and CSP samples, natural system photometry is
available, which we use in our analysis.  To use the natural
systems, we need the magnitudes of our fundamental flux standard in their
natural system.  For the CfAIII sample we use the linear
transformations given in table~2 of \cite{2009ApJ...700..331H}, after
verifying that the arbitrary additive constants are identically zero.
Fortunately, our fundamental flux standard (\bd ) lies in the color
range well measured by these transforms so they should be reliable,
which would not necessarily be the case for very blue stars such as
Vega or the white dwarfs used in the CALSPEC program.  For the CSP
sample, the magnitudes of \bd\ are already given in
the natural system.

\subsection{{\it HST} SNe}
\label{subsec:hstdata}
\nobreak SNe above $z = 1$ are difficult to observe from the ground.
Therefore, the most successful searches in this redshift range have
been carried out with {\it HST}. SNe at such high redshifts are useful
when studying any possible time variation of $w$, so we include the 2
SNe from \citet{2003ApJ...589..693B}, 16 from
\citet{2004ApJ...607..665R}, and the 22 from
\citet{2007ApJ...659...98R} in our sample; the latter also re-reduces
the photometry from the previous papers, taking into account the
nonlinearity of the NICMOS camera.  The {\it HST} sample extends from
$z=0.2 - 1.55$.  For $z<0.7$ SNe, rest-frame $B$ is measured by the
F606W filter, which has a sufficiently broad response that the
$U$-band is effectively included as well.  This is unlike any of the
other SNe in our sample, and therefore may introduce additional
systematic effects in our light-curve fitting.  Because these SNe have
virtually no impact on the cosmological parameters when compared with
the hundreds of SNe from other samples in this redshift range, we
exclude the 10 {\it HST} SNe with $z<0.7$.  Requiring spectroscopic
type confirmation eliminates five SNe, but ensures a consistent
analysis.  We also exclude SNe with $z > 1.4$ because we cannot
estimate the Malmquist bias above this redshift (\S\ref{subsec:malm}),
which eliminates one more SN.

\begin{figure}
\plotone{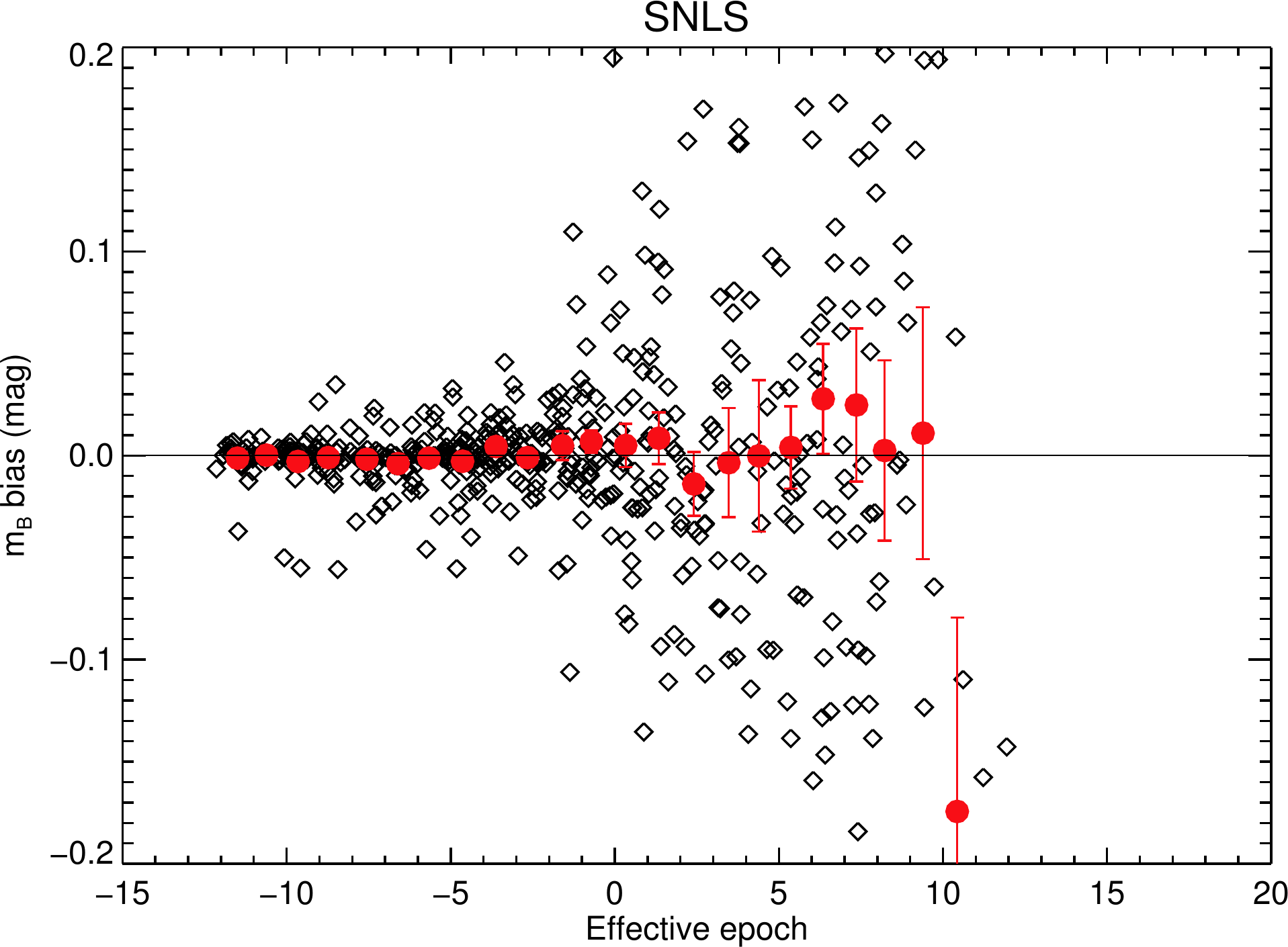}
\caption[Bias in $m_B$ as a function of first epoch relative to max
  for {\it HST}-like data]{Bias in $m_B$ as a function of first
  epoch relative to maximum after day $-15$ for {\it HST}-like data.
  The diamonds are the individual SNe used in this test, the red
  circles with errors are the values averaged in 1 day bins.
\label{fig:firstepoch_hst} }
\end{figure}

One cut which should be revisited for the {\it HST} sample is the
requirement of having photometry close to the epoch of peak flux.  In
\S4.5 of G10, we quantitatively studied the effects on the recovered fit
parameters of not having photometry before a certain epoch, and
concluded that an observation near peak between rest-frame epochs $-8$
and $+5$ was necessary to avoid bias. Unlike any of the other samples
we consider, in many cases the {\it HST} sample has data earlier than
rest-frame day $-8$ but not near peak, because it is the only
(psuedo-)classical high-$z$ search we have included (i.e., it is not a
rolling search, and therefore there may be a large gap between the
reference and search image).  We did not consider this possibility in
our original simulations.  On the other hand, the signal-to-noise of
the {\it HST} photometry is much lower than any of the other samples
and the cadence less frequent, which may increase the bias.
Therefore, we have revisited this study, this time including a single
early-time data point between $-20$ and $-15$ with a representative
signal-to-noise in F850LP (the filter that the {\it HST} searches were
carried out in).  Since it is imperative to use real data in this test
to reflect the diversity of light-curve shapes, to simulate the lower
signal-to-noise of the {\it HST} data we use SNLS SNe above $z = 0.65$
and further degrade their uncertainties to match the {\it HST} data.
As for the other samples, only $m_B$\footnote{$m_B$ is the peak
  magnitude in the rest-frame $B$ band and is one of the critical
  parameters we use to estimate relative distances; see
  \S~\ref{sec:combined}.}  shows significant bias, which is shown in
figure~\ref{fig:firstepoch_hst}.  Comparing with the results shown in
G10, the extra early epoch significantly reduces the bias.  Therefore,
for {\it HST} SNe we require either an early observation between $-20$
and $-15$ and at least one between $-8$ and $+9$, or, with no early
time point, an observation between $-8$ and $+5$ (all values rest
frame epochs). Altogether, we are left with a sample of 14 {\it HST}
SNe from $z=0.7 - 1.4$.

The {\it HST} data set considered here consists of observations
obtained with ACS in wide-field mode and camera 2 of the NICMOS
instrument.  The calibration of the NICMOS data is defined by the
solar analog P330e (A~.G.~Riess, 2009, private communication) having
the magnitudes 11.91 and 11.45 in the F110W and F160W filters,
respectively.  We do not have reliable Landolt system magnitudes for
P330e, so it is not suitable as a primary standard.  Fortunately
it has a CALSPEC SED on the same system as our primary flux standard,
so we can use spectrophotometry to align these observations with our
calibration.  Contrary to convention in the near-IR, this is {\it not}
consistent with Vega having zero magnitude and zero color in these
passbands, but instead implies a magnitude of about $-0.057$ in both.
It is not clear if other analyses have taken this into account (K09,
for example, do not, while \citet{2010ApJ...716..712A} do), but
this does not imply that the \citet{2007ApJ...659...98R} analysis is
in error.  We have also updated the ACS bandpasses and zero points to
match \citet{2007acs..rept....6B} and again for the 2009 January and 2010 
February CALSPEC updates.  Together, these imply $2-3\%$ zero point changes
for ACS\footnote{The ACS calibration involves setting the magnitude of
  the CALSPEC SED of Vega to zero using synthetic photometry, so
  updates on the SED or bandpasses change the zero point.}.  Before
these corrections are applied, the {\it HST} SNe have unusually red
colors and are significant outliers in the Hubble diagram.

\subsection{SDSS SNe}
\label{subsec:sdssdata}
\nobreak 
The SDSS Supernova Survey has recently released light curves for 146
intermediate redshift ($z < 0.4$) SNe observed in their first year of
operation \citep{2008AJ....136.2306H}. Perhaps the most important
aspect of this sample is that the calibration is better understood
than will likely ever be possible with the bulk of the current nearby
SN sample.  Furthermore, because the SDSS filter system is similar to
the MegaCam system, many inter-calibration issues with SNLS can be
reduced significantly.  We have not yet fully carried out this
program, but expect to do so in a future collaboration with the SDSS
and CSP SN teams.  This paper presents the results of a preliminary
combination of these data sets with our own which should be at least
as good as any previous combinations of different data sets, but which
does not take full advantage of the calibration improvements possible
with these three samples.

A visual inspection of the SDSS light curves reveals severe systemic
problems in the $u$ band, and lesser ones in $z$.  We therefore follow
the advice of K09 and only include $gri$ in
our fits.  We follow \citet{2008AJ....136.2306H} in only including
points with data quality flag $< 1024$.  Of the 146 SNe present in
the full sample, our data quality cuts combined with removing
obviously peculiar SNe remove 45 SNe.

The SDSS SN relative calibration is based on the catalog of
\citet{2007AJ....134..973I}, and tied to an absolute calibration (AB)
via observations of three solar-analog stars with high-quality CALSPEC
SEDs.  Unfortunately, these stars were too bright to observe directly
with the SDSS imager.  Instead, they were calibrated by transferring
observations from the SDSS monitor telescope
\citep{2006AN....327..821T} using linear magnitude and color
transformations derived from standard stars.  These stars were chosen
not because they are solar analogs, but because they have colors
similar to those of the standard stars and hence the linear
transformations are applicable.  Because these data are ultimately
compared with data calibrated onto the Landolt system, ideally these
three stars would also have Landolt magnitudes.  Unfortunately, this
is not the case.  Since the SEDs, effective bandpasses, and SDSS
magnitudes are known, synthetic photometry can be used to compute the
offset between SDSS and AB \citep{1996AJ....111.1748F} magnitudes.  We
can then place the derived light-curve parameters on the Landolt
system (i.e., $m_B$) using our fundamental flux standard (\bd ), for
which we have Landolt magnitudes and whose SED shares the same
calibration as the solar analogs.  The last step is sensitive to the
$V$ magnitude of \bd , unlike the rest of our analysis -- avoiding
this dependence is why the current SNLS calibration is tied to the
Landolt system, despite the large differences between the MegaCam and
Landolt passbands.  Offsets to the AB system are given in
\citet{2008AJ....136.2306H}.  As was the case for the {\it HST} data,
we update these offsets for the Feb 2010 CALSPEC update, which changed
the SEDs of the solar analogs slightly and thus the SDSS to AB
conversion factor by 1\% ; when combined with the changes in the SED
of \bd , the net effect of the CALSPEC update is to change the flux
calibration of the SDSS data by about 0.3\% in $g$ and less in $ri$; the
corresponding AB offsets $m_{AB}-m_{SDSS}$ are 0.021, 0.005, and 0.018
mag, respectively.  Here and elsewhere we use the mean SDSS bandpasses
of \citet{2010AJ....139.1628D}.

\subsection{$U$-band Observations of Nearby SNe}
\label{subsec:lowzu}
\nobreak 
K09 devote considerable attention to discrepancies between
low-$z$ rest-frame $U$-band data and higher redshift surveys (\S10.1.3
and 10.2.4), while finding that SNLS and SDSS SNe at $z=0.3$ were
consistent.  We will not repeat all of this discussion here, but we
independently confirm their result on a larger nearby sample, finding
strong evidence for issues with observer-frame $U$ data on several
fronts.  First, the observer-frame $U$ data shows considerably more
scatter in light-curve fits than can be accounted for by the
observational uncertainties, which is not the case for higher-redshift
samples in rest-frame $U$.  In fact, despite their larger photometric
uncertainties, the rms of the rest-frame $U$-band data for the
higher-z samples is much lower than that for the nearby ones.  Second, in
color-color space (i.e., rest-frame $U-B$ versus $B-V$ at the epoch of
maximum $B$ brightness) low-$z$ SNe with observer-frame $U$ observations
show much more scatter and a large systematic offset ($\sim 0.1$ mag)
compared with the other samples.  Third, if observer-frame $U$-band
photometry is included in the cosmological fits, then there is
significant ($\sim 2 \sigma$) tension in the residuals from the best
fit cosmology between the low-$z$ SNe and the others, which all agree
quite well with each other, while without the $U$-band data, there is
good agreement (\S\ref{subsec:tension}).

In addition to $\sim 90$ low-$z$ SNe observed in the $U$-band, we also
have 14 nearby CSP SNe with rest-frame \up-band data.  These SNe have
virtually identical selection effects as the CfAII and CfAIII samples,
since all three samples are primarily composed of KAIT-discovered SNe.
Tellingly, the CSP sample displays none of the above pathologies --
the scatter around the light-curve template is small, the SNe are
completely consistent with the higher redshift samples in color-color
space, and this introduces no tension in the residuals from the
best-fit cosmology.  We can quantify this by looking for offsets in
the rest-frame $U-B$ versus $B-V$ relation.  The low-$z$ $U$ band sample
is offset from the SNLS+SDSS samples by $0.047 \pm 0.005$ mag 
toward redder values of $B-V$ for a fixed $U-B$ --
nearly $10\sigma$ -- while the CSP \up\ sample is offset by $0.001 \pm
0.006$ mag; not only are the low-$z$ $U$-band data inconsistent with the SNLS
and SDSS data, they are also quite inconsistent with the low-$z$
\up\ data.  Interpreted in terms of a change in $U-B$, which is
natural because the $B-V$ versus$V-R$ relation shows no problems,
this corresponds to a blueward shift of $\sim 0.1$ mag in $U-B$
for the low-$z$ SN relative to the high-$z$ ones, which is consistent
in sign and magnitude with that seen by K09.

We are uncertain where the problem lies.  $U$-band data are notoriously
difficult to calibrate due to the wide range of detector and filter
responses, the fact that the blue end of the bandpass is affected by
the atmospheric cutoff, which can be highly variable, and the fact
that many $U$-band filters suffer from red leaks, but none of these
seem adequate to fully explain the effect.  An interesting possibility
may be that the fundamental flux calibration is deficient in the UV,
which would only affect the nearby sample.  However, this would not
explain the large observed scatter.  Furthermore, the consistency of
the CSP sample argues that if there is any problem it is not with the
spectral calibration itself, but rather with the $U$-band magnitudes
of the flux standards.  Note that the offset between a
Vega-based flux calibration and one based on \bd , as used in this
analysis, is not nearly large enough to explain these effects.  A
third possibility is that the response of the effective Landolt
$U$-band is very different than that given by
\citet{1990PASP..102.1181B}, which is usually used as a starting point
in SN analyses.  This is suggested by \citet{2006AJ....131.1184M}, who
find a substantially different $U$-band response curve using an {\it HST}
spectroscopic database; however, the passband suggested there
does not alleviate these problems.  Finally, it is possible that this
discrepancy is evidence for evolution in SN properties with redshift,
but if so this evolution must be extremely sudden, turning on abruptly
around $z=0.25$, and then there is no additional evolution out
to at least $z=1$, which seems unlikely.  Such evolution must also
somehow not affect the CSP sample.  \citet{2008ApJ...674...51E} find
no evidence for evolution in a spectroscopic study of UV spectra, but
this is hindered by the lack of low-$z$ rest-UV spectra.  This issue
deserves further study, and with the repairs to {\it HST} it will now
be possible to improve this test.

K09 argue that it is likely that this is due to some unidentified
issue with the low-$z$ observer frame $U$-band, and includes this effect as
their major source of systematic uncertainty.  We concur with the
former, but take the more aggressive position of excluding all
observer-frame $U$-band observations from our analysis, while
retaining rest-frame UV data for higher redshift SN.  Interestingly,
two recent spectroscopic studies comparing low- and high-redshift SNe
show some evidence for spectral evolution in the near-UV
\citep{2010arXiv1010.2211C,2010arXiv1010.2749F}, although
there is some concern about selection effects.  However, in both cases
the evolution is in the opposite sense to that seen here -- that is,
the low-$z$ SNe are redder in $U-B$ in these studies, while they appear
bluer in our comparison, which only exacerbates the problem.

\subsection{Selection Effects for External Samples}
\label{subsec:malm}
\nobreak 
SN samples are expected to suffer from Malmquist bias
\citep{1936Malmquist}, which we explicitly correct for in our
analysis.  In addition to the usual peak flux bias, there should be
additional effects associated with the stretch and color of the SNe:
for a given peak luminosity, higher stretch, more slowly declining
SNe~Ia will stay above the detection threshold longer and hence be
easier to find.  Because our survey has different depths in different
passbands, at the high redshift end bluer SNe will have more of their
flux in bands we are more sensitive to, and hence will also be easier
to detect.  The effects of Malmquist bias differ from survey to
survey; some of the low-$z$ surveys, which are galaxy-targeted, suffer
from little Malmquist bias when searching for SN candidates, while
others are traditional flux-limited surveys and are strongly affected.
The simulations used to calculate these corrections for SNLS are
described in \citet{2010AJ....140..518P}, and are based on inserting
2.4 million fake SNe into the real images and re-running the
extraction pipeline.  We are not in a position to apply a similar
analysis to the external samples, although we do compute and correct
for Malmquist-type effects in these samples as described below.  The
mean Malmquist bias for the SNLS sample is about 0.03 mag at $z=1$ for
$m_B$ after correction for light-curve shape and color.

For the {\it HST} and SDSS samples, we make use of the detection and
spectroscopic selection models of K09 (\S6), which are expressed in terms of
the redshift and derived light-curve parameters.  These models assume
certain distributions of stretch, color, and \si .  In order to ensure
consistency with our SNLS estimates, we generate our own simulated
samples using the same distributions as the SNLS simulations
and apply the K09 selection models to the results in order to estimate
the Malmquist bias.

Our cosmological analysis corrects for both SN light-curve shape and
color, which significantly reduces the effects of Malmquist
bias. However, the relationship between color, shape, and peak
magnitude is not perfect, so it is useful to study residual effects.
For example, for an SN at $z=0.45$ where $\rmeg \mapsto B$ and $\imeg
\mapsto V$, because the SNLS \imeg\ images are deeper, an SN which
fluctuates red relative to the mean color relation may be harder to
detect without affecting the peak rest-frame $B$ magnitude, $m_B$.  In
some previous studies
\citep[e.g.,][]{1999ApJ...517..565P,2003ApJ...598..102K} it sufficed
to treat Malmquist bias as a simple offset between the nearby and
distant SN samples. Because the measurement of $w$ is more sensitive
to the shape of the luminosity-distance relationship within our
sample, this approach no longer suffices, and we map out the redshift
dependence for all but the nearby sample.  Applying selection effects
as priors when fitting the SN parameters is unwieldy in the SALT2 and
SiFTO frameworks, but is an equally valid approach often used with the
MLCS light-curve fitting package \citep[as
  in][K09]{2007ApJ...666..694W}.  Instead, we apply the mean
corrections to the relative distance moduli as a function of redshift
to each sample.  Our tool for calculating these corrections are Monte
Carlo simulations of artificial SNe tuned to match the observed
properties of the SN samples.

\subsubsection{Malmquist Bias in the Nearby Sample}
\label{subsubsec:nearbymalm} 
\nobreak 
A nice feature of the nearby sample is that we do not need to know the
redshift dependence of the Malmquist bias, at least for standard
cosmological models, because of the very simple form of the 
luminosity-distance-redshift relation.  Because of the simple nature of the
luminosity-redshift relation, if a nearby sample has a sharp cutoff in
magnitude, then the mean Malmquist bias is 0.032 mag, independent of
the cutoff.  In reality, however, any cutoff is unlikely to be
perfectly sharp (which tends to decrease the amount of bias), some
nearby samples are poorly described by magnitude limits (those based
on searching known galaxies), and sample cuts tend to reduce the amount
of bias.

We have two tools for analyzing the Malmquist bias in the nearby
samples.  First, the bias can roughly be measured by
examining at what point on the light-curve the SNe were discovered --
if most of the SNe were discovered around peak brightness, then the
sample is more biased than one in which SNe are mostly discovered
before peak.  The bias for the Cal\'an/Tololo sample was considered in
\citet{1999ApJ...517..565P} using this approach, who found a bias of
0.04 mag.  Since this is larger than the value obtained for a sharp flux
cutoff ($0.032$ mag), we adopt the latter.  The CfAI sample was
partially discovered by searches which targeted specific host
galaxies, and therefore should suffer from less Malmquist bias than
the Cal\'an/Tololo survey.  An examination of the light-curves
suggests that this data set is not entirely free from bias, since many
of the SNe were discovered near peak.  Applying the same technique
as was used for the Cal\'an/Tololo sample, we estimate a Malmquist
bias of 0.02 mag for this sample.  The CfAII sample was
almost completely based on galaxy-targeted searches, and hence should
suffer from even less Malmquist bias.  Indeed, most of the SNe from
this sample used in our analysis were discovered prior to peak
luminosity.  Therefore, we adopt a value of 0 mag here.  The CSP is
selected from a combination of the same galaxy-targeted survey and the
SDSS survey.  For the latter, at the redshifts of these SNe, Malmquist
bias should be completely negligible, so we also adopt a value of 0
mag.

The size of the CfAIII sample allows us to adopt a more sophisticated
treatment.  This sample was also largely selected from galaxy-targeted
surveys, and therefore the detection efficiency should be fairly flat
with redshift.  This is confirmed by the fact that the earliest epoch
of photometry relative to maximum is mostly uncorrelated with
redshift.  However, the spectroscopic selection criterion did make use
of a magnitude cut of $\sim 18.5$ mag, which introduces Malmquist
bias.  This is in turn confirmed by the fact that the mean color of
the sample changes with redshift, with few of the faint/red SNe in the
upper half of the redshift range, which would not be the case in the
absence of selection effects.

We model the spectroscopic selection function of the CfAIII
sample by comparing the predicted number of SNe for an unbiased survey
to that actually obtained as a function of redshift.  We generate 32
million fake SNe using the same color and stretch distributions as the
SNLS Malmquist simulations distributed evenly per comoving volume
element\footnote{We include a small correction for the evolution in
  the SN rate with $z$ over this range from the $A+B$ model fits of
  Perrett et al.\ 2011, in preparation, although this has negligible
  effect.}  and then attempt to fit the parameters of a selection
function to match the observed redshift distribution.  We use a
logistics function to model the selection probability $P$ in terms of
the peak magnitude: $ P = 1 / \left(1 + \exp \left[ \gamma \left( m -
  m_0 \right) \right] \right)$ .  Most low-$z$ SN searches are carried
out either unfiltered or in very broad bands, so for $m$ we consider
both $V$ and $R$, finding identical parameters.  We obtain $\gamma =
2.76 \pm 0.31, m_0 = 16.29 \pm 0.20$ mag.  The amount of bias depends
weakly on $\gamma$, but is independent of $m_0$.  These imply a mean
Malmquist bias for this sample of 0.027 mag.

\subsubsection{Malmquist Bias in the {\it HST} Sample}
\nobreak 
\cite{2004ApJ...613..200S} argue that the {\it HST} searches are
sufficiently deep to suffer from little or no Malmquist bias out to
the maximum redshift SN discovered.  Because of the small size of the
{\it HST} sample, we compute the Malmquist bias for each SN
individually by constructing 10,000 constrained realizations of our
SN model consistent with the observed properties of that SN
given the observational uncertainties, and then apply the models of
K09. This only models the search efficiency,
and not the spectroscopic selection.  \citet{2007ApJ...659...98R}
argues that there are no spectroscopic losses below $z = 1.4$, so we
cut the {\it HST} sample at this redshift because we cannot model the
Malmquist bias for the single more distant SN.  We find that the
Malmquist bias for the remaining SNe is small, less than 0.01 mag in
all cases.

However, this conflicts with the fact that the majority of the $z>1$
{\it HST} SNe were discovered at or near maximum light -- if there were
no Malmquist bias, we would expect many more of the SNe to be
discovered before maximum, as is seen with the lower-redshift {\it
  HST} SNe.  We do not understand the reason for this discrepancy,
but it has almost no effect on our fits to (constant) $w$.

\subsubsection{Malmquist Bias in the SDSS Sample}
\begin{figure}
\plotone{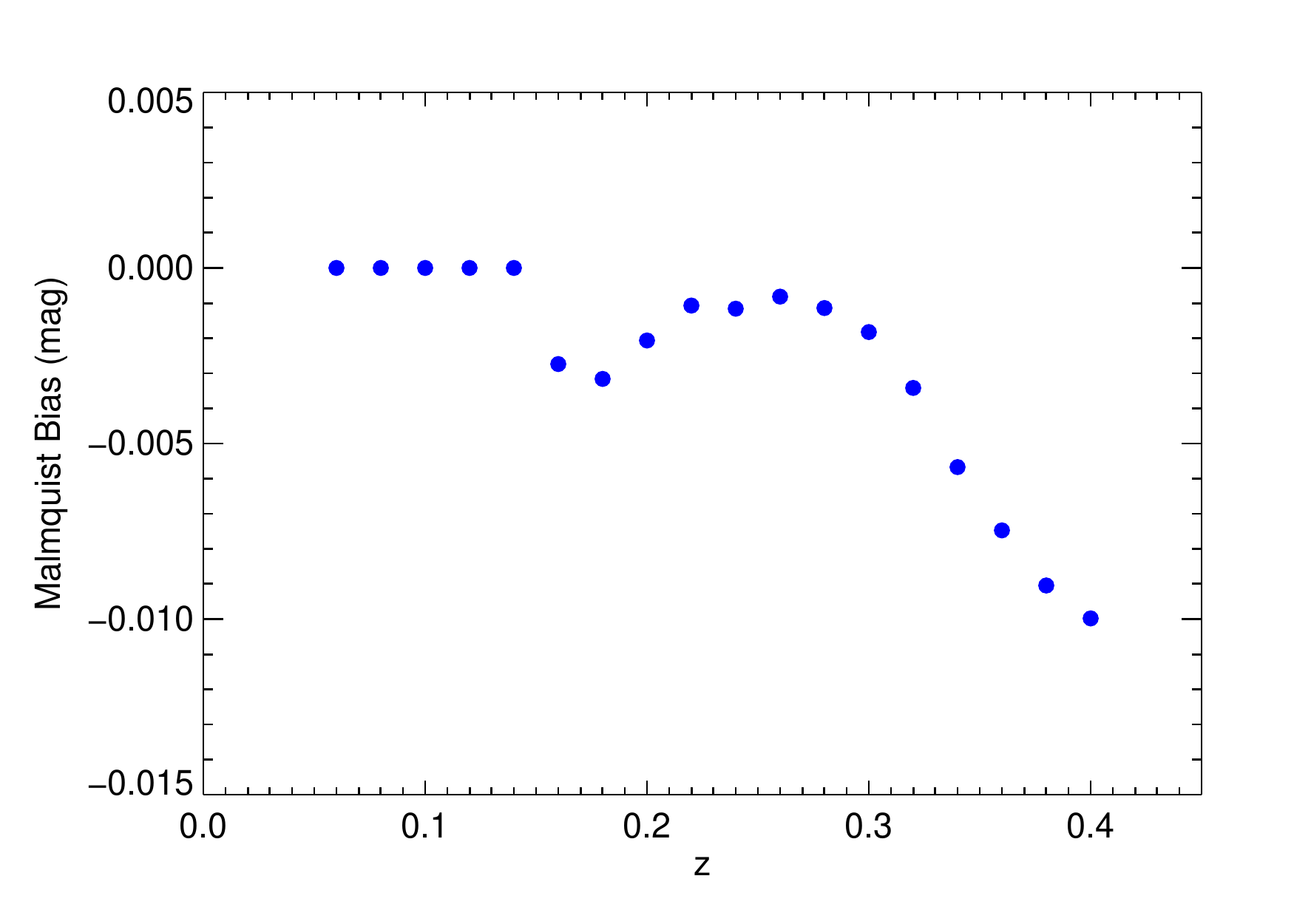}
\caption[SDSS Malmquist bias]{Mean Malmquist bias as a function
of redshift for the SDSS sample.  The sharp feature at $z=0.15$ is
an artifact of the discontinuous spectroscopic efficiency model of
K09 and has little effect on the cosmological
constraints. \label{fig:sdssmalm} }
\end{figure}

\nobreak 
Our method for modeling the SDSS Malmquist bias is described above.
The resulting mean bias per redshift bin is shown in
figure~\ref{fig:sdssmalm}.  The mean amount of bias is relatively
small, perhaps because for SDSS it was easy to obtain spectra of even
the faintest detected candidates with 8m class telescopes.

\subsection{Peculiar Velocities}
\nobreak
\label{subsec:pecvel}
Most analyses have attempted to include the effects of SN peculiar
velocities (relative to the Hubble flow) by assuming a large intrinsic
velocity dispersion as an additional source of redshift uncertainty.
These are only important for nearby SNe.  This approach neglects any
correlations in the flows.  We instead follow the discussion presented
in \citet{2007neillpecvel} and explicitly correct for peculiar
velocities on an SN-by-SN basis. The peculiar velocity model used in
this analysis \citep{2004MNRAS.352...61H} is based on the galaxy
density field in the nearby universe ($z<0.06$) and accounts for
infall into nearby superclusters (e.g.\ Virgo, Hydra-Centaurus and
Perseus-Pisces) as well as a large-scale bulk flow. The accuracy of
the corrections estimated to be $\pm 150$ km $\mathrm{s}^{-1}$ for an
individual SN, but this is a random uncertainty whose importance can
be reduced by observing multiple SNe.  Applying this model adjusts the
CMB frame redshift of nearby SNe, but also applies a (much smaller)
correction to their peak magnitudes because the standard 
luminosity-distance relation is not quite correct in the presence of peculiar
velocities, although this is only important at low-$z$ -- see
\citet{2006PhRvD..73l3526H} for details.

\section{COSMOLOGICAL RESULTS AND STATISTICAL UNCERTAINTIES WITH THE 
COMBINED SN SAMPLE}
\nobreak
\label{sec:combined}
After the selection criterion are applied, the combined SN sample
for our primary analysis consists of 472 SNe~Ia: 123 nearby, 93 from
SDSS, 242 from SNLS, and 14 from {\it HST}.  These numbers do not
include those nearby SNe that are not in the Hubble flow but which are
used for training the light-curve fitters.  The parameters of each
subsample are given in table~\ref{tbl:surveysummary}.

\begin{deluxetable}{llll}
\tablecaption{Summary of SN Samples}
\tablehead{
 \colhead{Sample} & \colhead{Redshift Range\tablenotemark{a}} & 
 \colhead{$N_{\mathrm{SN}}$} &
 \colhead{$\left<N_{\mathrm{points}}\right>$\tablenotemark{b}}
}
\startdata
low-$z$ & 0.01 -- 0.10 & 123 & 32 \\
SDSS  & 0.06 -- 0.4  & 93  & 27 \\
SNLS  & 0.08 -- 1.05 & 242 & 37 \\
HST   & 0.7  -- 1.4  & 14  & 10 \\
\enddata
\tablenotetext{a}{Redshift range from this sample included in our
 analysis.}
\tablenotetext{b}{Average number of photometric epochs in the range
 $-25$ to $+30$ days, the range used in our light-curve fits.}
\label{tbl:surveysummary}
\end{deluxetable}

\subsection{Model for Relative Distance Moduli}
\nobreak
The \chisq\ of the data relative to our model is similar to that of
A06. In the absence of any covariances between SNe,
\begin{equation}
 \chisq = \sum_{\mbox{SNe}} \frac{ \left( m_{B} - m_{\mathrm{mod}} \right)^2 }
                                 { \sigma^2 },
 \label{eqn:basechisq}
\end{equation}
where $m_B$ is the rest-frame peak $B$ band magnitude of an
SN, $m_{\mathrm{mod}}$ is the predicted magnitude of the SN given the
cosmological model and two other quantities (stretch and color) which
describe the light-curve of the particular SN, and $\sigma$ includes
both the uncertainties in $m_B$ and $m_{\mathrm{mod}}$.  The model
magnitude is given (for SiFTO) by
\begin{equation}
 m_{\mathrm{mod}} = 5 \log_{10} \mathcal{D}_L \left(z_{\mathrm{hel}}, 
   z_{\mathrm{cmb}}, w, \om, \ode \right)
   - \alpha \left(s - 1\right) +
   \beta \scriptc + \scriptm .
 \label{eqn:modeqn}
\end{equation}
where $\mathcal{D}_L$ is the Hubble-constant free luminosity distance,
$z_{\mathrm{cmb}}$ and $z_{\mathrm{hel}}$ are the CMB frame and
heliocentric redshifts of the SN, $s$ is the stretch (a measure of
the shape of the SN light-curve), and \scriptc\ is color measure for
the SN.  For SALT2, the expression is similar, with
$\left(s-1\right) \mapsto x_1$, where $x_1$ is a different measure of
the light-curve shape. $\alpha$ and $\beta$ are nuisance parameters
which characterize the stretch-luminosity and color-luminosity
relationships, reflecting the well-known broader-brighter and
bluer-brighter relationships, respectively.  \scriptm\ is another
nuisance parameter representing some combination of the absolute
magnitude of a fiducial SN~Ia and the Hubble constant; even if one of
these is known from some other measurement, the other still has to be
marginalized over.  

$m_B$, $s$, and \scriptc\ will naturally all be correlated for an
individual SN because they are determined from the same data.
Furthermore, their values are correlated between different SNe in the
presence of systematic effects and statistical uncertainties. For
example, a systematic uncertainty in the \rmeg\ zero point will
directly affect $m_B$ for all SNLS SNe between $z=0.35$ and $0.55$
(since rest-frame $B \mapsto \rmeg$ in this redshift range), and will
have indirect effects even for an SN without $r_M$ measurements because
it affects our empirical SN models by changing the templates and also
changes the measured color-luminosity relationship ($\beta$).
Introducing a vector of model residuals over the SN sample $\Delta
\vec{\mathbf{m}} = \vec{\mathbf{m}}_B -
\vec{\mathbf{m}}_{\mathrm{mod}}$ then a better expression for
\chisq\ is
\begin{equation}
 \chisq = \Delta \vec{\mathbf{m}}^T \cdot \mathbf{C}^{-1} \cdot 
 \Delta \vec{\mathbf{m}} . \label{eqn:chieqn}
\end{equation}
It is useful, for reasons explained in \S\ref{subsec:method}, to
  further factor the total covariance matrix as:
\begin{equation}
 \mathbf{C} =
  \mathbf{D}_{\mathrm{stat}} + \mathbf{C}_{\mathrm{stat}} +
  \mathbf{C}_{\mathrm{sys}} . \label{eqn:cdef}
\end{equation}
The first term is the purely diagonal part of the statistical uncertainty 
given by
\begin{equation}
 \mathbf{D}_{\mathrm{stat},\ i i}  
  = \sigma^2_{m_B , i} + \alpha^2 \sigma^2_{s,i}
  + \beta^2 \sigma^2_{\mathcal{C},i} + \sigma^2_{\mbox{int}}
  + \left( \frac{ 5 \left(1 + z_i\right)}
     {z_i \left(1+ z_i/2 \right) \log 10} \right)^2 \sigma^2_{z,i}
  + \sigma_\mathrm{lensing}
  + \sigma_\mathrm{host\, correction} + C_{m_B s \mathcal{C}, i} .
  \label{eqn:vstatd}
\end{equation}
$D_{\mathrm{stat},\ i i}$, or something similar, is the only term
that has been included in most analyses.  Here $\sigma_{m_B , i},
\sigma_{s,i}, \sigma_{\mathcal{C},i}$ are the errors on the fitted
light-curve parameters of the $i^{th}$ SN, and \si\ represents the
intrinsic scatter of SNe~Ia.  $C_{m_B s \mathcal{C}, i}$ represents
the covariance terms between $m_B$, $s$, and $\mathcal{C}$ for each
SN, and is a function of $\alpha$ and $\beta$.  We will discuss how to
construct the off-diagonal parts of the covariance matrix
($\mathbf{C}_{\mathrm{stat}}$ and $\mathbf{C}_{\mathrm{sys}}$) in
\S\ref{subsec:method}.

We use the empty-universe approximation for the relation between
redshift uncertainty and the associated magnitude uncertainty (with
$z_i$ being the redshift in the rest frame of the CMB).  This is
appropriate because $\sigma_z$ is only important for the lowest
redshift SNe in our sample.  Future surveys which make use of more
uncertain photometric redshifts may have to use the full form for the
redshift uncertainty, which depends on the cosmological parameters.
The lensing term represents the statistical uncertainty caused by
gravitational lensing (\S\ref{subsec:statresults}), and the host
correction term relates to the statistical uncertainty in the
host-galaxy correction (\S\ref{subsec:twoscriptm}).

\subsection{Correcting for Host-galaxy Properties}
\label{subsec:twoscriptm}
\nobreak
\citet{2010MNRAS.406..782S} explore various approaches for including
host-galaxy information in the cosmological fits to correct for the
dependence of SN residual from the Hubble relationship; see that paper
for the details of how we measure host-galaxy parameters using
broad band photometry and the Z-PEG package
\citep{2002A&A...386..446L}.  Without a clear physical understanding
of the cause, or even which descriptive host parameter best accounts
for the variation (metallicity, host mass, or star formation rate),
they find that simply splitting the sample on one of these parameters
and allowing the absolute magnitude (\scriptm\ in our analysis) to be
different between the two samples is sufficient to describe the
observations.  It seems likely that the true relationship is both more
complex and more continuous, but the current data do not require a
more sophisticated model.  Therefore, we adopt this approach here.
This is the first analysis to incorporate these corrections.

Since all that is available for the vast majority of the SN hosts in
our combined sample is broad-band photometry, the host galaxy
parameters can have considerable measurement correlations, and there
is little evidence to favor one over the others.  Therefore, we split
our sample based on host-galaxy stellar mass at $10^{10} M_{\odot}$,
since it is the most directly constrained parameter.  The effects of
changing this selection are included as a systematic as described in
\S\ref{subsec:environment}.  For systematic tests, instead of
allowing \scriptm\ to vary based on host type, $\alpha$ and $\beta$
might be allowed to differ between subsamples, as explored in
\citet{2010MNRAS.406..782S,2010arXiv1005.4687L}. Because the SNe
light-curve properties are physically correlated with host type (i.e.,
higher-stretch SNe are preferentially found in star-forming hosts),
these approaches are almost entirely degenerate with the
\scriptm\ model for our purposes, although they have different
physical implications.

Not including this correction has a substantial effect on the measured
cosmological parameters.  The difference in \scriptm\ between the two
samples is $\sim 0.075$ mag \citep{2010MNRAS.406..782S}.  In addition
to the systematic uncertainties discussed in
\S\ref{subsec:environment}, we include the statistical uncertainties
in the measured host masses by multiplying this magnitude difference
by the probability that a given SN host is assigned to the wrong
subsample due to measurement uncertainties.  The overall mass scale
does not affect our correction since it will affect all host masses
identically.

\subsection{Combining SALT2 and SiFTO}
\nobreak We fit all SNe using both SiFTO and SALT2.  Since our
analysis gives us no clear reason to choose either fitter, we combine
results from both in our final analysis, although we show the results
from each. The details are discussed in \S5 of G10.  We also include the
difference between the two fitters in our systematic uncertainty
budget (\S\ref{subsec:lcfitters}).  We give the parameters from each
fitter an equal weight in the combination.  It would be incorrect to
treat the parameters from each as independent since they are fit to
the same data; this would unjustifiably reduce our uncertainties by
$\sim 1/\sqrt{2}$.  One approach would be to treat the relative
distance moduli from each fitter as perfectly correlated.  However,
this has the side effect of making it impossible to ask questions like
what the uncertainty on the combined peak magnitude is, and is
complicated to extend to the systematics analysis.  We have therefore
adopted a simpler approach, which is to average the covariance
matrices for each statistical and systematic term derived for
SiFTO and SALT2.  This satisfies our requirement that the final
uncertainty should be larger than the minimum of the SALT2 and
SiFTO uncertainties for each SN for every individual term.

\subsection{Statistical Results}
\nobreak
\label{subsec:statresults}
As mentioned previously, we add some additional, intrinsic scatter
(\si ) to the peak magnitudes of SNe~Ia to match observations.  This
value will include both true intrinsic scatter, related to our
imperfect understanding of SN physics, and any mis-estimates of the
photometric uncertainties, uncorrected selection biases, etc.  We have
no way to distinguish between these contributions.  Therefore, we
allow different values of \si\ for each sample (e.g., SNLS, low-$z$,
SDSS), as is supported by the data.  The \si\ values are derived by
requiring the \chisq\ of the best fitting $\Omega_m, w$ cosmological
fit to a flat universe to be one per degree of freedom for each
sample, including systematic effects, and are given in
table~\ref{tbl:sigmaint}.  The cost is that we weaken our ability to
detect any deviations from our cosmological parameterization that are
much smaller than the intrinsic scatter -- but such deviations would
be difficult to detect in any case.  The \si\ are held fixed at these
values for all other fits (e.g., $\Omega_m$, $\Omega_{\Lambda}$).  If
the data turn out to be inconsistent with a flat universe, fixing
these values unduly penalize some data sets -- see S11 for fits that
allow for spatial curvature.

\begin{table}
\centering
\caption{\si\ and rms values for various 
  samples \label{tbl:sigmaint}}
\begin{tabular}{l|ll}
\tableline\tableline
Sample & \si & rms \\
\tableline
  low-$z$      & 0.113 & 0.153 \\
  SDSS       & 0.099 & 0.143 \\
  SNLS       & 0.068 & 0.156 \\
  {\it HST}  & 0.082 & 0.242 \\
\end{tabular}
\tablecomments{\centering The uncertainty in each value is about 0.005 mag,
 and the RMS is around the best fit cosmology.}
\end{table}

All SNe are corrected for Galactic extinction using the maps of
\cite{1998ApJ...500..525S}, including the estimated 10\% random
uncertainty for each SN (there is also a correlated systematic
uncertainty discussed in \S\ref{subsec:mwextcorr}).  We correct for
peculiar velocities in the nearby sample (\S\ref{subsec:pecvel}),
Malmquist bias effects (\S\ref{subsec:malm}) for all samples, and
assign a random peculiar velocity uncertainty of 150 km
$\mathrm{s}^{-1}$, as is appropriate after the peculiar velocity
correction.  Our statistical uncertainties also include the random
uncertainty in the SN model, as described in Appendix A.3 of G10; this
means that the statistical covariance matrix between SNe is not
diagonal: $\mathbf{C}_{\mathrm{stat}} \neq \mathbf{0}$.  We include the
measurement uncertainties in the host-galaxy masses as described in
\S\ref{subsec:twoscriptm}, and random, uncorrelated scatter due to
lensing following the prescription of \citet{2010MNRAS.405..535J}:
$\sigma_{\mathrm{lens}} = 0.055\, z$.

\begin{figure}
\plotone{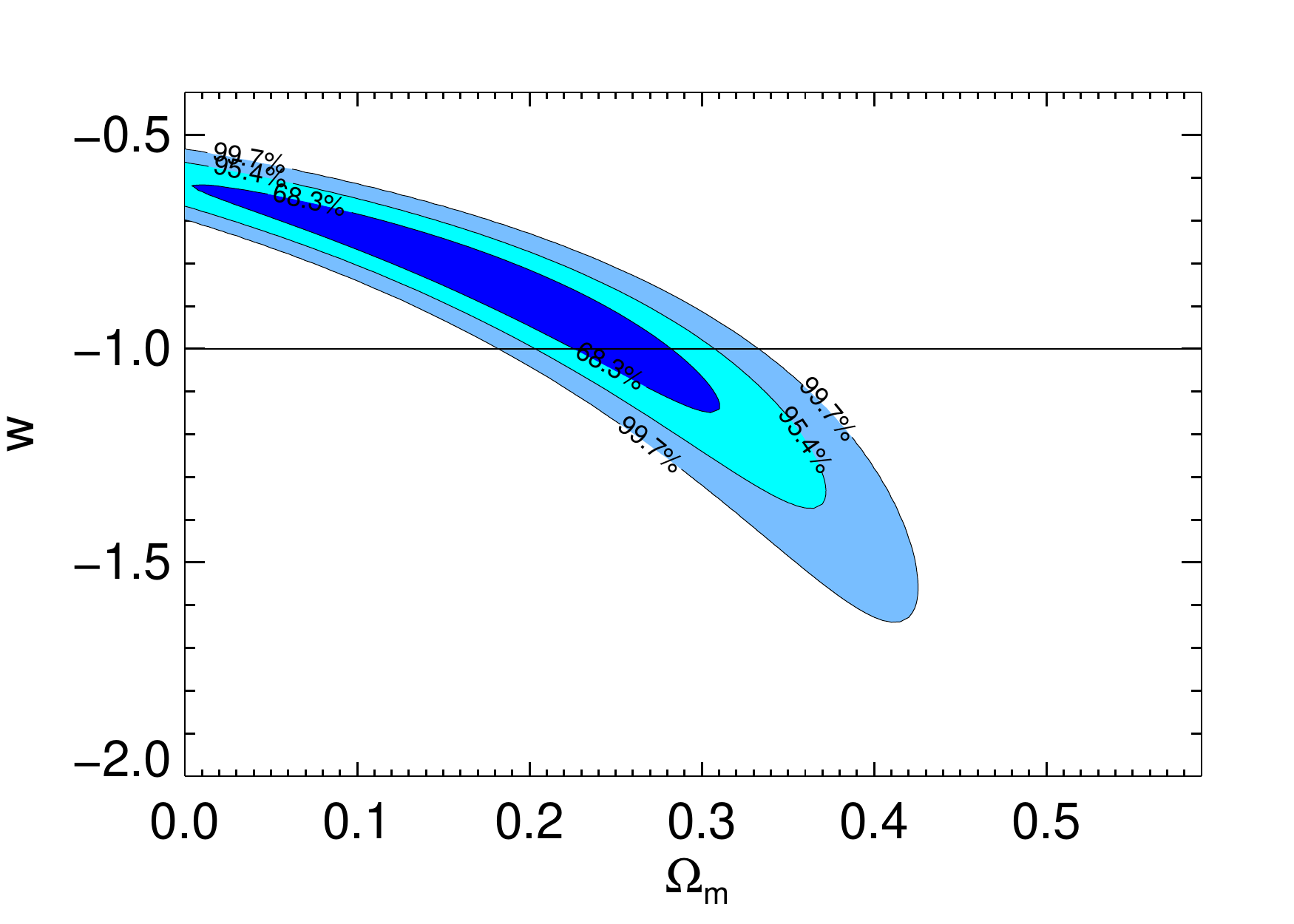}
\caption[Statistical constraints from SN]{Statistical SN only constraints on
$\Omega_m$, $w$ assuming a flat universe and constant dark energy
equation of state.  \label{fig:omwstat} }
\end{figure}

\begin{figure}
\plotone{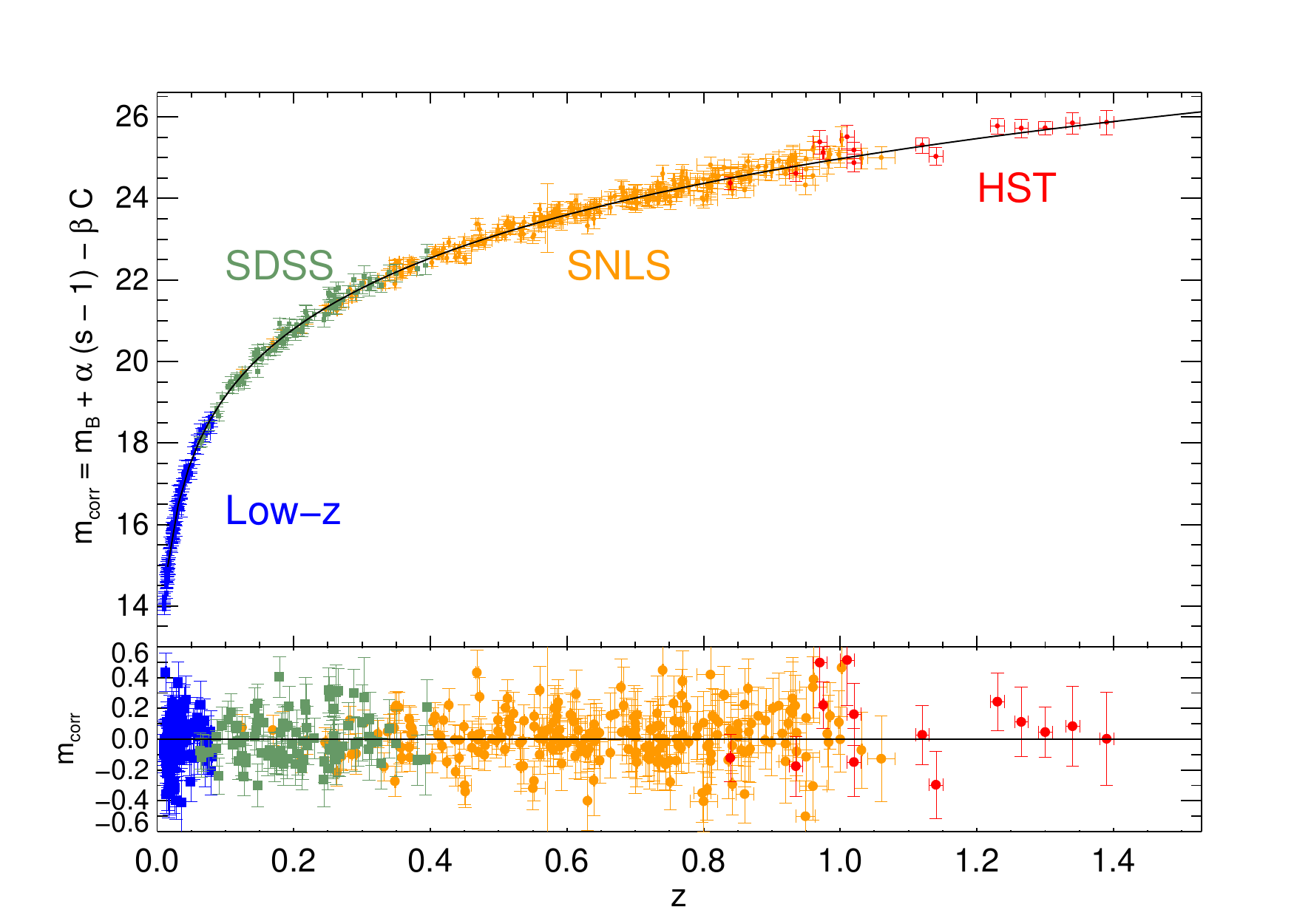}
\caption[Hubble diagrams]{Hubble diagram of the combined
 sample.  The residuals from the best fit are shown in the
 bottom panel. \label{fig:hd} }
\end{figure}

We perform two types of fits: one in which we compute probabilities
over a grid and then report the mean value of the marginalized
parameters, and a \chisq\ minimization routine that reports the best
fit.  We should not expect the two to agree as they have different
meaning, but as such it is useful to provide both -- see
Appendix~\ref{apdx:margmin} for further details.  The light-curve
parameters for the combined sample are given in
table~\ref{tbl:bigtable}.  The luminosity distance integral does not
converge for $\Omega_m < 0$, so both fits effectively have the (very
reasonable) prior of $\Omega_m \geq 0$.  Our statistical constraints
on $\Omega_m, w$ in a flat universe for a constant dark energy
equation of state are shown in figure~\ref{fig:omwstat} using the
marginalization approach, and are summarized in
table~\ref{tbl:cosmoresults}.  We find $w=-0.90^{+0.16}_{-0.20}$,
consistent with a cosmological constant ($w=-1$).  The
Hubble diagram is shown in figure~\ref{fig:hd}.  The error introduced
by using a simplified treatment of the nuisance parameters $\alpha$
and $\beta$ (such as holding them fixed at their best fit values) is
described in \S\ref{subsec:wrongway}.

\begin{deluxetable}{lrrrrcrrl}
\rotate
\tabletypesize{\small}
\tablewidth{0pc}
\tablecaption{Light-curve Parameters of the Combined SN Sample}
\tablehead{
 \colhead{Name} & \colhead{$z_{\mathrm{cmb}}$\tablenotemark{a}} & 
 \colhead{$m_B$\tablenotemark{b}} & \colhead{$s$} & \colhead{$\mathcal{C}$} &
 \colhead{$\log_{10}\, M_{\mathrm{host}}$ \tablenotemark{c}} &
 \colhead{ $\mathrm{MJD_{max}}$ } & \colhead{Filters} & 
 \colhead{Reference} }
\startdata
 sn2004s & 0.010(0.000) & 14.183(0.042) & 0.973(0.026) &  0.035(0.025) & 12.07 & 53039.56(0.60) & $BVR$ & 1 \\
 sn1999ac & 0.010(0.000) & 14.130(0.030) & 0.987(0.009) &  0.056(0.018) &  9.92 & 51249.82(0.07) & $BVR$ & 2 \\
 sn1997do & 0.011(0.000) & 14.317(0.036) & 0.983(0.023) &  0.056(0.025) & 12.07 & 50765.81(0.14) & $BVR$ & 2 \\
 sn2006bh & 0.011(0.001) & 14.347(0.021) & 0.814(0.008) & $-0.045$(0.019) & 10.91 & 53833.01(0.06) & $BgVr$ & 3 \\
 sn2002dp & 0.011(0.000) & 14.597(0.030) & 0.973(0.029) &  0.113(0.024) & 10.47 & 52450.45(0.20) & $BVR$ & 4 \\
\enddata
\label{tbl:bigtable}
\tablecomments{Combined SiFTO and SALT2 light-curve parameters for
 full data sample, with uncertainties in parentheses.  Note that the 
 individual variables are correlated, as are the values for 
 different SNe due to systematic and light-curve
 fitter training uncertainties.}
\tablenotetext{a}{Uncertainty does not include residual peculiar
 velocity uncertainty.}
\tablenotetext{b}{Includes the lensing term and the effects of the
 statistical uncertainty in $\log_{10}\, M_{\mathrm{host}}$.}
\tablenotetext{c}{Host stellar mass in solar masses.}
\tablerefs{(1) \citet{2007AJ....133...58K}, (2) \citet{2006AJ....131..527J}, 
 (3) \citet{2010AJ....139..519C}, (4) \citet{2009ApJ...700..331H}, 
 (5) \citet{2004MNRAS.349.1344A}, (6) \citet{2000ApJ...539..658K}, 
 (7) \citet{1996AJ....112.2408H}, (8) \citet{2004AJ....128.3034K}, 
 (9) \citet{2001AJ....122.1616K}, (10) \citet{1999AJ....117..707R}, 
 (11) \citet{2007MNRAS.377.1531P}, (12) \citet{2004AJ....127.1664K}, 
 (13) \citet{2005ApJ...632..450L}, (14) \citet{2006AJ....131.1639K}, 
 (15) \citet{2002AJ....124.2905S}, (16) \citet{2008ApJ...686..749K}, 
 (17) \citet{2008AJ....136.2306H}, (18) G10, 
 (19) \citet{2007ApJ...659...98R}}
\end{deluxetable}

The \si\ and rms values for each sample are summarized in
table~\ref{tbl:sigmaint}.  The rms residuals are similar for the low-$z$,
SNLS, and SDSS samples, which is impressive considering the range in
flux densities from $z=0.01$ to $z=1$.  That of the {\it HST} sample
is larger, but this is unsurprising due to the larger photometric
uncertainties for such distant and difficult to observe SNe.  
K09 carry out the same analysis (tables~11 and 15) and find a similar
rms for the SDSS sample as we do (0.15 versus 0.16 mag), but a smaller
\si\ (0.08 versus 0.10 mag).  However, they find significantly larger values
of both measures for the other samples.  Note that the SNLS data in this paper
benefits from improved calibration and photometry, and both the low-$z$
and SNLS samples are not the same as those in K09.

\begin{figure}
\plotone{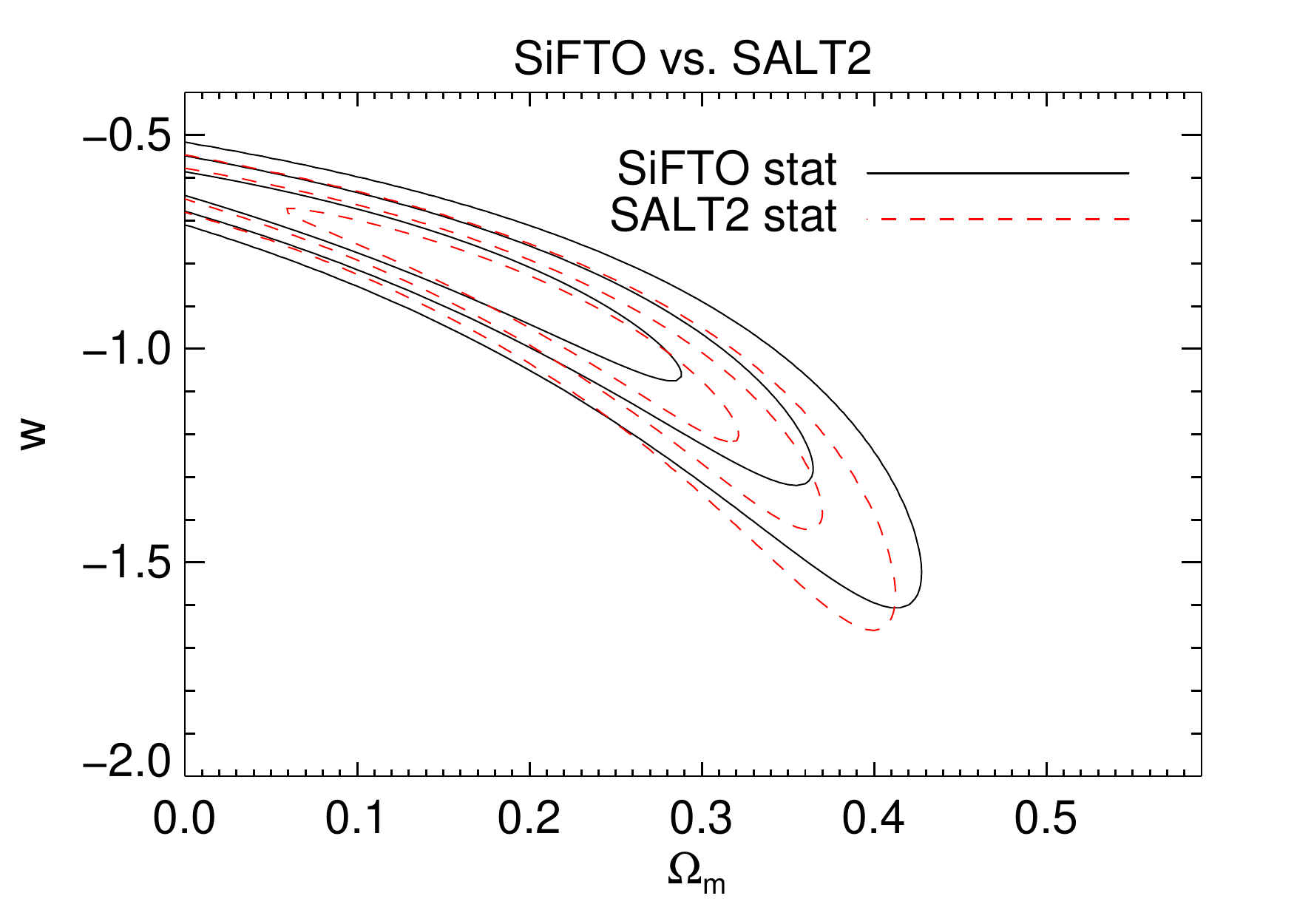}
\caption[Comparison of SiFTO and SALT2 statistical constraints]
{Comparison of SN-only statistical constraints for SiFTO and SALT2
 $\Omega_m$, $w$ assuming a flat universe and constant dark energy
 equation of state. \label{fig:salt2vssifto} }
\end{figure}

Comparing the individual light-curve fitters, for SiFTO we find $\om =
0.173^{+0.095}_{-0.098}$ and $w=-0.85^{+0.14}_{-0.20}$ and for SALT2
$\om = 0.214^{+0.072}_{-0.097}$ and $w=-0.95^{+0.17}_{-0.19}$ (all
uncertainties statistical only).  The contours are directly compared in
figure~\ref{fig:salt2vssifto}. We include the difference between the
light-curve parameters from the two fitters in our systematic uncertainty
budget as described in \S\ref{subsec:lcfitters}.  

\begin{table*}
\centering
\caption{Results from SN-only fits \label{tbl:cosmoresults}}
\renewcommand{\arraystretch}{1.5}
\begin{tabular}{lllll}
\tableline\tableline
 Uncertainties & \om & $w$ & $\alpha$ & $\beta$ \\
\tableline
 \multicolumn{5}{c}{Marginalization fits} \\
 \tableline
 Stat Only  & \allstatomwom & \allstatomww &
 $1.45^{+0.12}_{-0.10}$ & $3.16^{+0.10}_{-0.09}$ \\[0.1cm]
 Stat plus Sys & \allsysomwom & \allsysomww &
 $1.43^{+0.12}_{-0.10}$ & $3.26^{+0.12}_{-0.10}$ \\[0.2cm]
\tableline
\tableline
 \multicolumn{5}{c}{\chisq\ minimization fits} \\
 \tableline
 Stat Only & $0.19^{+0.09}_{-0.12}$ & $-0.86^{+0.17}_{-0.19}$ &
  $1.397^{+0.085}_{-0.083}$ & $3.152^{+0.095}_{-0.093}$ \\[0.1cm]
 Stat plus Sys & $0.17^{+0.10}_{-0.15}$ & $-0.86^{+0.22}_{-0.23}$ &
  $1.371^{+0.086}_{-0.084}$ & $3.18 \pm 0.10$ \\[0.1cm]
\end{tabular}
\renewcommand{\arraystretch}{1}
\end{table*}

The value for $\beta$ is larger than that found by A06 (who found
$\beta = 1.57 \pm 0.15$), although it remains inconsistent with a value
of 4.1 expected if SN colors are primarily caused by MW-like dust in
the host galaxies.  There are several reasons for the increase.
First, our current light-curve fitting frameworks handle the intrinsic
variation between SNe~Ia much better than in A06, particularly the
observed scatter in the relations between different rest-frame
wavelengths -- see Appendix A.2 of G10 for details.  This causes a significant
increase in the measurement uncertainties for SN colors, and deweights
high-$z$ SNe which are primarily measured in the near-UV.  These SNe,
coupled with a simpler near-UV model, were driving the fits in A06 to
low values of $\beta$.  Modeling the uncertainty in the intrinsic
uncertainty in the SN color relations is a complicated subject which
can also present itself as apparent evolution in $\beta$; see
\S\ref{subsec:evolution} for more details.  Second, the method used
to marginalize over $\alpha$ and $\beta$ in A06 was intrinsically
biased toward low values, as shown in \citet{2008ApJ...686..749K}.

\section{COSMOLOGICAL RESULTS INCLUDING SYSTEMATIC UNCERTAINTIES}
\label{sec:application}
\nobreak 
The SNLS search and analysis proceeds through two independent
pipelines, one in France and one in Canada, which can then be checked
against each other.  This has resulted in many improvements to both
pipelines.  In this section we explain the methodology we use for
analyzing systematic uncertainties and how they impact our
measurements (\S\ref{subsec:method}), present the combined results
including both statistical and systematic effects
(\S\ref{subsec:statplussysresults}), explore tension between
subsamples (\S\ref{subsec:tension}), and discuss the consequences of
simplified treatments (\S\ref{subsec:wrongway}).  The detailed
descriptions of individual systematic terms are given in
\S\ref{sec:systematics}.

A difficulty in estimating systematic uncertainties is how to handle
effects for which there is no clear physical model.  The obvious
example for SNe~Ia is the possibility of undetected and uncorrected
evolution in the SN population.  While considerable effort and
ingenuity have been devoted to making specific theoretical predictions
for what form evolution might take and what signatures it might
produce, currently these predictions frequently disagree with each
other even as to the sign of possible effects.  In the absence of a
physical model, it will {\it always} be possible to imagine an
evolutionary scenario which will pass all available tests, yet which
will bias the cosmology by an arbitrary amount.  We could artificially
add some uncertainty in $w$ to try to take this into account, but this
is highly unsatisfying because the adopted value would be essentially
arbitrary.  In this paper we will concentrate on systematic effects
for which we have some sort of model, which includes some simple
models of evolution (\S\ref{subsec:evolution}).

\subsection{Systematics Methodology}
\nobreak
\label{subsec:method}
The difference between statistical and systematic uncertainties is not always
entirely clear -- what one author labels a systematic another may
label a statistical uncertainty.  Here we (mostly) adopt an effective
definition which is that terms whose effects on our final uncertainty budget
could be simply reduced by increasing the SN sample size are
statistical uncertainties.  Thus, for example, the uncertainty in the
redshift of a particular SNLS SN is a statistical uncertainty, while the
uncertainty in our zero points is a systematic uncertainty, since it affects
multiple SNe in a correlated fashion and its importance cannot be
reduced by simply observing more SNe.

There is no standard method for handling SN systematic effects, but the most
common approach works as follows: first, a fit without any systematic
effects is performed.  Then, the effect under consideration (e.g., a
zero point offset) is applied to the SN sample and the cosmological
parameters are re-measured.  The difference in each parameter is taken
as the systematic uncertainty in that parameter. Finally, the values for all
of the known effects are then added in quadrature to form the final
systematic uncertainty; we shall therefore refer to this shift-and-add
approach as the quadrature method.  Not all SN papers follow this
approach, but with a few exceptions noted below, the alternatives have
generally been even less sophisticated.

The main advantage of this method is that it is simple to implement
once one has a list of systematic effects and estimates of their size
(in reality, of course, this is by far the most difficult part of the
analysis), but it has several disadvantages.  One is that individual
effects can be shown graphically, but the combined effects of multiple
terms and their interrelations are difficult to visualize, relying on
many-panel plots or graphical distortions that may not capture the
full effects (see \citet[][K09]{2003ApJ...598..102K,
  2006ApJ...644....1C} for examples).  A more
serious disadvantage is that it is difficult for others to include
these systematics in their analyses. Even were the individual modified
distance moduli made available, to apply this information to a
new cosmological parameter space, or to include additional datasets,
an entirely new cosmological fit has to be carried out for each
systematic term, which can be prohibitively expensive. The result has
been that SN papers themselves include systematic uncertainties, but
most subsequent analyses ignore them; for example, the
WMAP7 analysis of \citet{2010arXiv1001.4538K} provides constraints
that do not include SN systematics because they had no simple way to
do so.  Because systematic effects induce correlations between
different SNe, in principle the best fit point should be modified by
their presence, so ignoring these effects not only underestimates the
uncertainties, but can lead to biased results.  Additional criticism
of the quadrature approach can be found in
\citet{2003physics...6138B}.

One way to overcome both problems is to marginalize over all of the
systematic terms during the fit.  This marginalization can be carried
out explicitly (using the uncertainty estimates for each term of the
following sections as priors), or by adding a systematics covariance
matrix to the statistical covariance matrix as in
equation~\ref{eqn:cdef}.  We have used both methods in this analysis,
but only the results of the latter are presented here.  This
separation of the covariance matrix has been used before in the SN
literature, although in abbreviated form.  In
\citet{1999ApJ...517..565P,2003ApJ...598..102K} a similar approach was
used for the zero point uncertainties and their effects on $m_B$, with
all other systematic effects handled via the quadrature method.

The statistical and systematic covariance matrices,
$\mathbf{C}_{\mathrm{stat}}$ and $\mathbf{C}_{\mathrm{sys}}$ of
equation~\ref{eqn:cdef}, are generally not diagonal, which have also
been neglected in most SN analyses.  We compute them using
standard techniques \citep[e.g.,][]{2005PhRvC..72e5502A}:
\begin{equation}
 \mathbf{C}_{\mathrm{sys},\, i j} = \sum_{k=1}^K
   \left( \frac{ \partial m_{\mathrm{mod}\, i} }{ \partial S_k } \right)
   \left( \frac{ \partial m_{\mathrm{mod}\, j} }{ \partial S_k } \right)
   \left( \Delta S_k \right)^2
\end{equation}
where the sum is over the $K$ systematics $S_k$, $\Delta S_k$ is the
size of each term (for example, the uncertainty in the zero point), and
$m_{\mathrm{mod}}$ is defined as in equation~\ref{eqn:modeqn}.  A
similar expression is used for $\mathbf{C}_{\mathrm{stat}}$.  An
example of such a term in $\mathbf{C}_{\mathrm{stat}}$ is the
uncertainty in the SN model.  Since these uncertainties could
be reduced with more SNe, they are a statistical uncertainty rather
than a systematic one, but are correlated between different SNe and
therefore cannot be included in $\mathbf{D}_{\mathrm{stat}}$.  Note
that all of the components of $\mathbf{C}$ are functions of the
nuisance parameters $\alpha$ and $\beta$, but not \scriptm\ because
the relative distance moduli are independent of its value.  Since
$\mathbf{D}_{\mathrm{stat}}$ is diagonal and the off-diagonal pieces
of $\mathbf{C}$ have the right form, we can take advantage of the
Sherman-Morrison-Woodbury formula \citep{1996maco.book.....G} to
compute $\mathbf{C}^{-1}$ in $N^2$ steps.

This approach must be modified slightly for effects which cannot be
smoothly parameterized.  An example is contamination by non-SNe~Ia,
which has the effect of removing an SN from the sample
rather than changing $m_B, s$ or $\mathcal{C}$.  In such cases, we
resort to studying SNe in redshift bins, where there are enough
objects in each bin that the effect is sufficiently continuous for our
purposes, and then reproject the systematic onto the individual SNe
(e.g., non-Ia contamination \S\ref{subsec:nonia}) by determining what
systematic uncertainty, assumed perfectly correlated between all SNe in
that bin, would produce the same uncertainty in the mean bin
parameters as the specified effect and then assigning it to those SNe.

\citet{2008ApJ...686..749K} have recently presented an alternative to
the standard scheme for handling systematic effects which shares some
characteristics with that adopted here.  They treat all systematic
effects as offsets between the peak magnitudes of nearby and distant
SNe on a sample-by-sample basis.  This could be replicated in our
scheme by forcing $\mathbf{C}_{\mathrm{sys}}$ to consist of blocks of identical
values when ordered by SN sample, and in fact this transformation
was used in \citet{2010ApJ...716..712A}.  This is a good approximation when
each SN sample covers a very small range in redshift, since it does
not allow for any effects within a sample, but this is not true of
current high-$z$ SN samples.  Both papers also hold $\alpha$ and $\beta$ fixed
while calculating $\mathbf{C}_{\mathrm{sys}}$, which both biases their values,
and more importantly underestimates the uncertainty on $w$, as discussed 
in \S\ref{subsec:wrongway}.

\subsection{Light-curve Training}
\nobreak 
All current light-curve analysis frameworks are trained on SN data
which share systematics in common with the data used to derive the
cosmological constraints.  While many previous analyses have included
some estimates of the uncertainty in the light-curve models, none have
properly considered the interplay of systematic effects with the
training process, and hence have underestimated their systematic
uncertainties.  A unique feature of the current analysis is that both of the
frameworks used for SNLS3 (SiFTO and SALT2) are trained on
high-redshift SN data.  This is only practical because the training
process for both does not assume any relationship between redshift and
distance (and, in fact, makes no use of distance information),
and so are completely independent of the cosmological parameters.  The
benefit of including the high-redshift data in the training process is
that it allows us to probe further into the blue, where the SNLS
calibration of rest-frame near-UV data is much more secure than the
corresponding low-$z$ data.  Because SNLS systematics affect the
light-curve model, they will affect the derived parameters of even
nearby SNe, and vice-versa.  Excluding high-$z$ data from the training
sample would only increase the overall uncertainty budget.

We proceed in two steps for each systematic effect.  First, we
calculate the effect on the light-curve model.  We then apply the
modified model to all of the data, deriving a new set of light-curve
parameters.  A similar procedure could easily be applied to other
light-curve techniques (e.g., MLCS2k2 or $\Delta m_{15}$).

\subsection{Method for Presenting Systematic Effects}
\nobreak There are 134 individual systematic terms considered in this
paper.  Attempting to compare the importance of each term is
difficult, and we have not found any entirely satisfactory method.  In
the parameter space we are most interested in for this paper
($\Omega_m, w$ for a flat universe with a constant dark energy
equation of state) the size of the uncertainties on the cosmological
parameters depends strongly on the values of those parameters.
Therefore, introducing a correlated uncertainty which shifts the
results can actually reduce the uncertainties.  This effect is
mitigated when we combine SN data with external constraints such as
those from the Wilkinson Microwave Anisotropy Probe ({\it WMAP}) because
they are nicely orthogonal to the SN ones, and therefore prevent the
constraints from shifting too much.  Therefore, some effects which
seem very large when only SN are considered are minor when combined
with {\it WMAP} and BAO results (see S11).  Also, some systematic effects
appear minor simply because the SNe they affect have little weight in
the fits considered in this paper, which may not be the case for a
different cosmological model.

\begin{figure}
\plotone{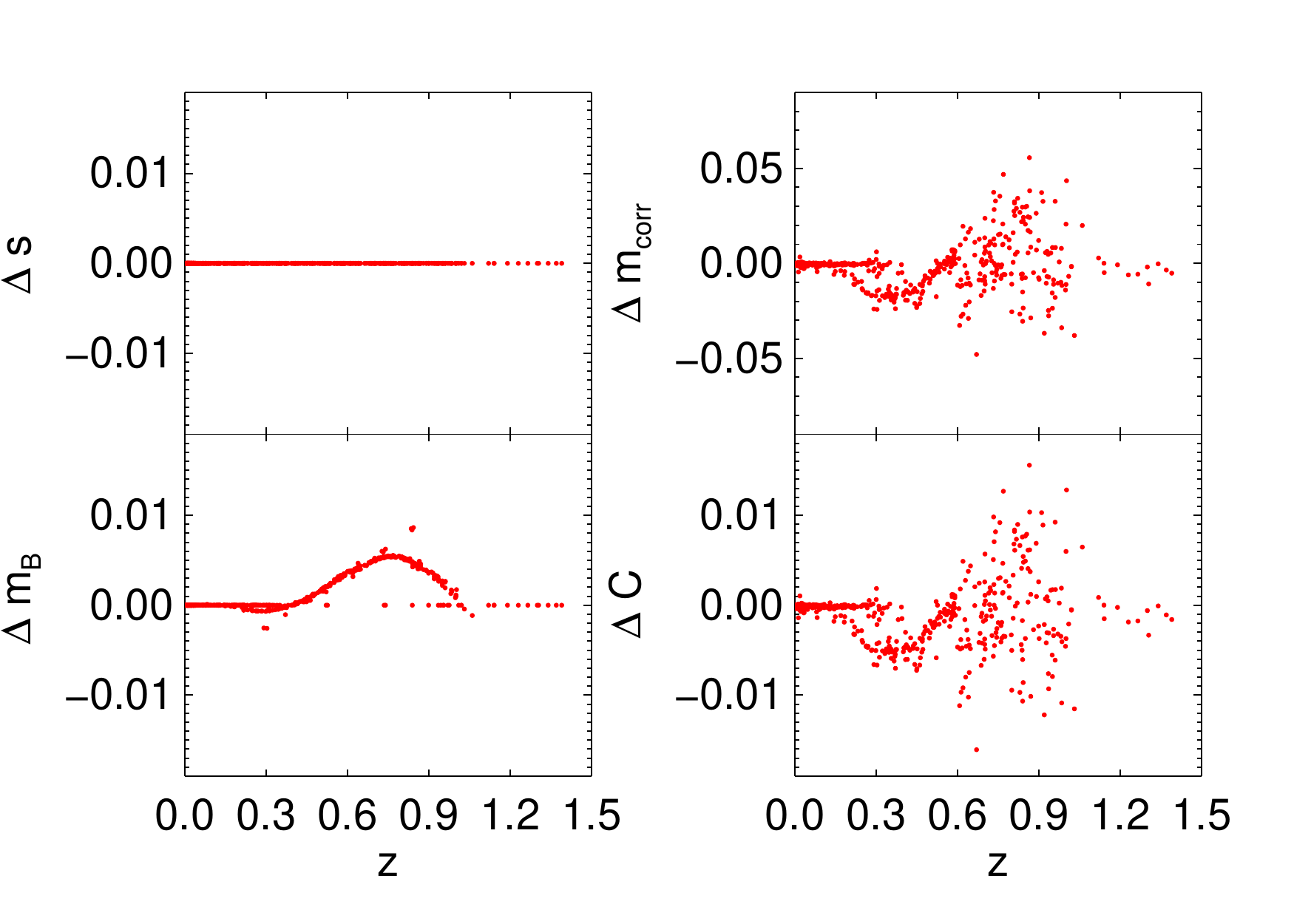}
\caption{The effects of changing the $i_M$ zero point by 6.1 mmag on
  various light-curve parameters and the corrected peak magnitude
  $m_{\mathrm{corr}} = m_B + \alpha \left(s-1\right) -\beta
  \mathcal{C}$, as a function of redshift.  Note that because this
  affects our SN models (SALT2 and SiFTO), as the training sample
  includes SNLS data, this alters the corrected peak magnitudes of all
  SNe, not just those in the SNLS sample.  Furthermore, because
    of the changes in the light-curve model, the offset in the derived
    color can actually be larger than the shift in the zero point.
 \label{fig:sysexample} }
\end{figure}

With those caveats in mind, we present the effects of our systematics
in three ways: first, in terms of the effects on $\Omega_m, w$ for SNe
only, and second in terms of the effects on $w$ with $\Omega_m$ fixed.
The latter approximates the effects of including the BAO and {\it
  WMAP} constraints, which mostly improve the measurement by
constraining \om.  In both cases, we compare the uncertainties on the
cosmological parameters from a fit that only includes statistical
uncertainties with one that includes statistical uncertainties plus
only that systematic term. We also provide the relative size of the
contour that encloses 68.3\% of the probability, compared with the
statistical only contours.  This is similar in spirit to the Dark
Energy Task Force figure of merit \citep{2006astro.ph..9591A}, and is
perhaps the simplest way of expressing the importance of each term.
Because of the curvature of the SN-only constraints, the area of the
inner contour can actually increase while the marginalized
uncertainties decrease. We caution against the practice of comparing
the shifts in the best fit as a useful method of measuring systematic
effects, as it can be misleading.  An updated version of the effects
on the cosmological parameters is given in S11 combined with external
constraints such as baryon acoustic oscillations and CMB measurements.
An example of the effects of one systematic (the \imeg\ SNLS
zero point, discussed in \S\ref{subsubsec:snlszp}) on the light-curve
parameters as a function of redshift is shown in
figure~\ref{fig:sysexample}.

\subsection{Combined Statistical and Systematic Results}
\label{subsec:statplussysresults}
\begin{figure}
\plotone{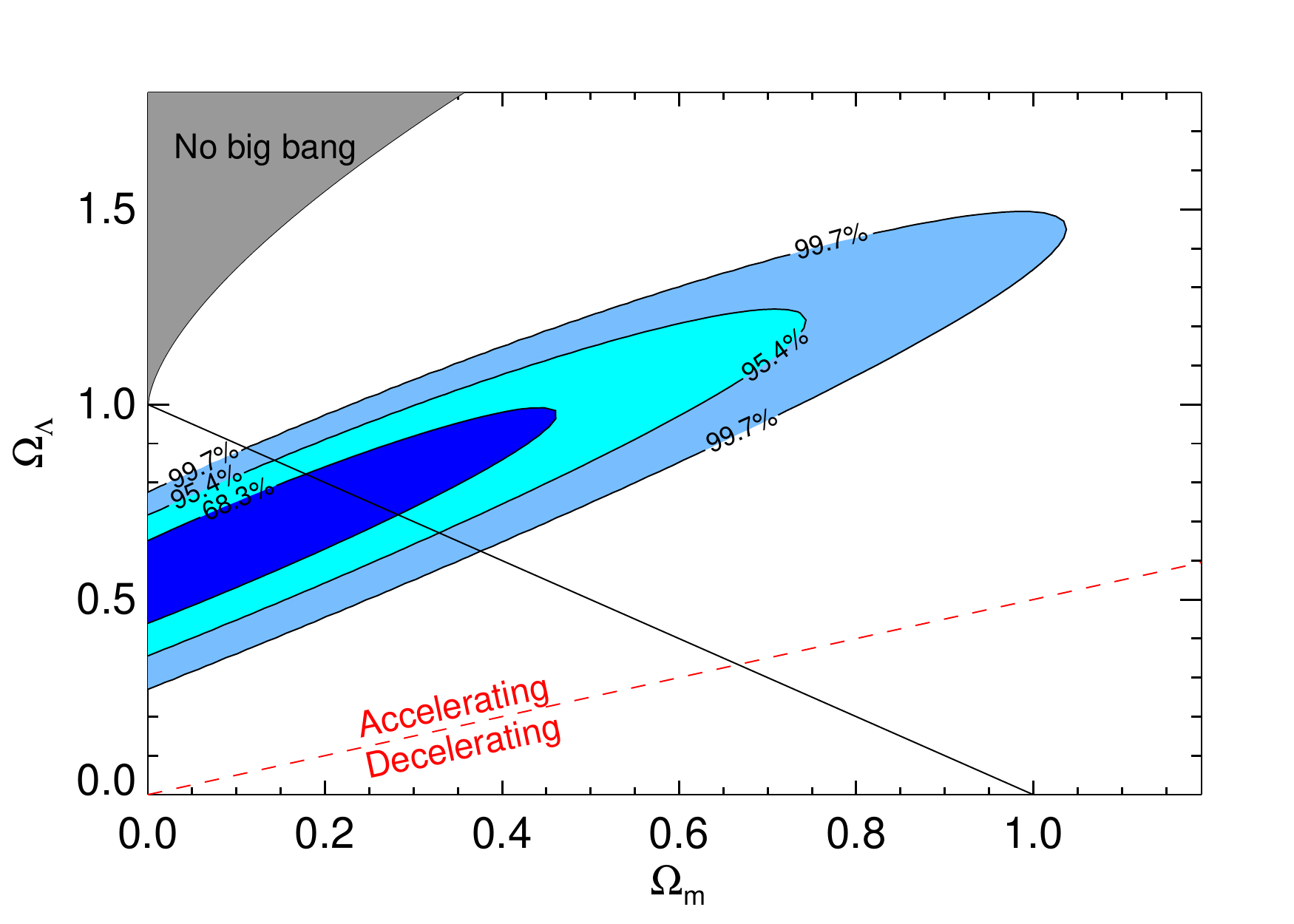}
\caption[$\Omega_m, \Omega_{\Lambda}$ contours]
 {$\Omega_m, \Omega_{\Lambda}$ (i.e., $w=-1$, but allowing for non-zero
 spatial curvature) contours including all identified systematic uncertainties.
 \label{fig:omol} }
\end{figure}

\begin{figure}
\plotone{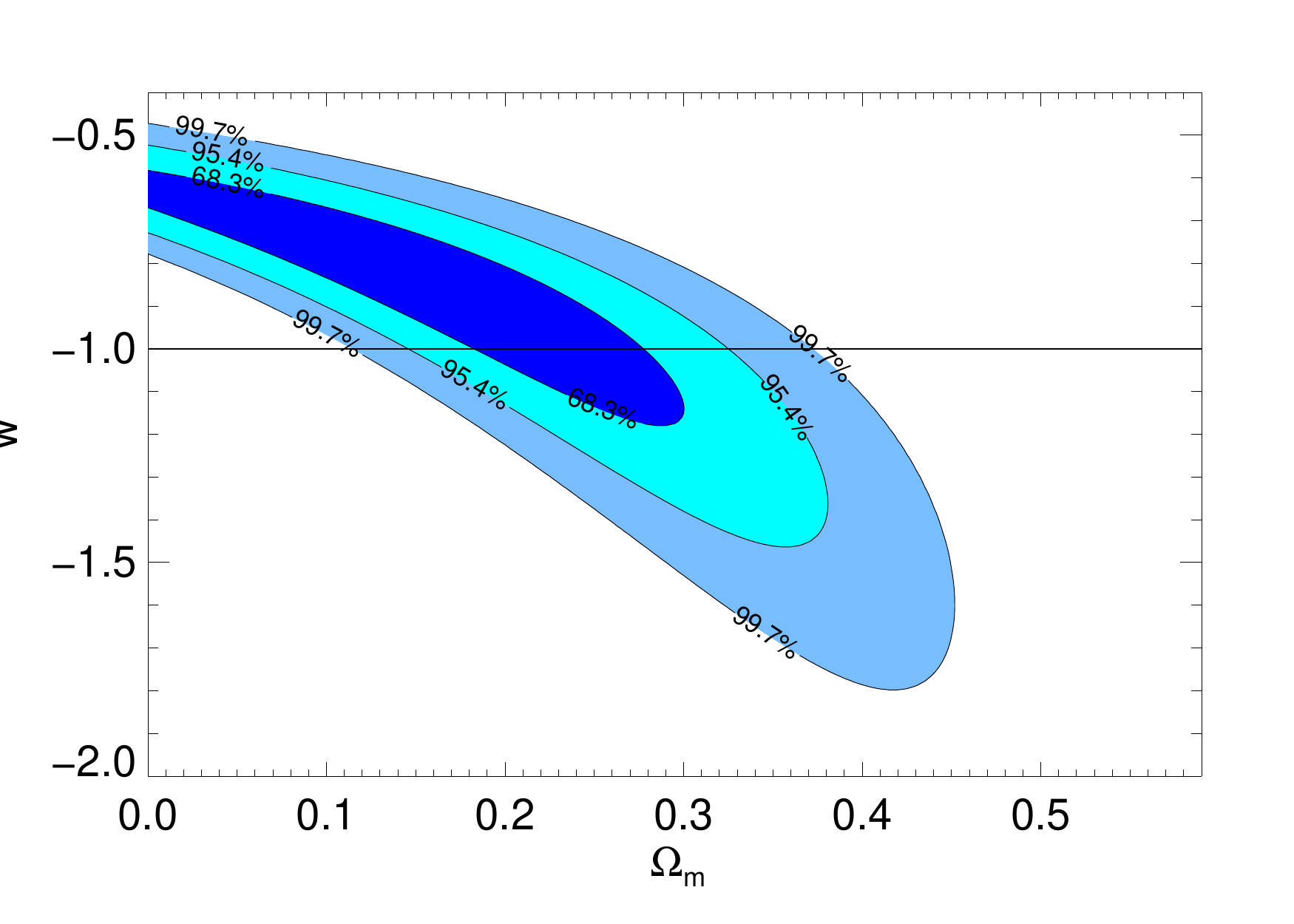}
\caption{Constraints on $\Omega_m, w$ in a flat universe 
 including all identified systematic uncertainties.  \label{fig:snsys} }
\end{figure}

\nobreak
Including all identified systematic effects, the results for an
$\Omega_m$, $\Omega_{\Lambda}$ fit are shown in figure~\ref{fig:omol}.
The SN data alone require acceleration at high significance.  The
results for a flat universe with a constant dark energy equation of
state are summarized in table~\ref{tbl:cosmoresults}, and the contours
are shown in figure~\ref{fig:snsys}. We find $w =
-0.91^{+0.16}_{-0.20}\left( \mathrm{stat} \right) ^{+0.07}_{-0.14}
\left( \mathrm{sys} \right)$, again consistent with a
cosmological constant.  An overview of the importance of each class of
systematic effect is given in table~\ref{tbl:overallsystematics} --
calibration effects are by far the dominant type of identified
systematic uncertainty.  Overall, the systematic uncertainties degrade
the area of the uncertainty ellipse by a bit less than a factor of 2
relative to the statistical-only constraints.  Excluding the SDSS and
CSP SNe (calibrated to a USNO system) increases the area of the
uncertainty ellipse by about 10\%; the improvement from including this
data should be increased once the full benefits of cross calibrating
these two samples with SNLS are realized.

\begin{table*}
\centering
\caption{Identified systematic uncertainties \label{tbl:overallsystematics}}
\renewcommand{\arraystretch}{1.5}
\begin{tabular}{l|ccc|c|l}
\tableline\tableline
 Description & \om & $w$ & Rel.\ Area \tablenotemark{a} &
 $w$ for \om=0.27 & Section \\
\tableline
 Stat only           & \allstatomwom & \allstatomww & 1 &
  \allstatwfixomw & \\
 All systematics     & \allsysomwom & \allsysomww & 1.85 &
  \allsyswfixomw & \S\ref{subsec:statplussysresults} \\[0.1cm]
\tableline
 Calibration & $0.191^{+0.095}_{-0.104}$ & $-0.92^{+0.17}_{-0.23}$ & 1.79
     & $-1.06 \pm 0.10$ & \S\ref{subsec:calibration} \\
 SN model             & $0.195^{+0.086}_{-0.101}$ & $-0.90^{+0.16}_{-0.20}$ & 
        1.02 & $-1.027 \pm 0.059$ & \S\ref{subsec:lcfitters} \\
 Peculiar velocities  & $0.197^{+0.084}_{-0.100}$ & $-0.91^{+0.16}_{-0.20}$ & 
        1.03 & $-1.034 \pm 0.059$ & \S\ref{subsec:bubble} \\
 Malmquist bias & $0.198^{+0.084}_{-0.100}$ & $-0.91^{+0.16}_{-0.20}$ & 1.07
   & $-1.037 \pm 0.060$ & \S\ref{subsec:Malmquistsys} \\
 non-Ia contamination & \allstatomwom & \allstatomww & 1 &
  \allstatwfixomw & \S\ref{subsec:nonia} \\
 MW extinction correction & $0.196^{+0.084}_{-0.100}$ & $-0.90^{+0.16}_{-0.20}$ & 
  1.05 & $-1.032 \pm 0.060$ &\S\ref{subsec:mwextcorr} \\
 SN evolution & $0.185^{+0.088}_{-0.099}$ & $-0.88^{+0.15}_{-0.20}$ & 1.02 &
   $-1.028 \pm 0.059$ & \S\ref{subsec:evolution} \\
 Host relation & $0.198^{+0.085}_{-0.102}$ & $-0.91^{+0.16}_{-0.21}$ &
   1.08 & $-1.034 \pm 0.061$ & \S\ref{subsec:environment} \\ 
\tableline
\tablenotetext{a}{Area relative to statistical only fit of the contour
  enclosing 68.3\% of the total probability.}
\end{tabular}
\tablecomments{Results including statistical and identified systematic
  uncertainties broken down into categories.  In each case the
  constraints are given including the statistical uncertainties and
  only the stated systematic contribution.  The importance of each
  class of systematic uncertainties can be judged by the relative area
  compared with the statistical-only fit.}
\renewcommand{\arraystretch}{1}
\end{table*}

\begin{figure}
\plotone{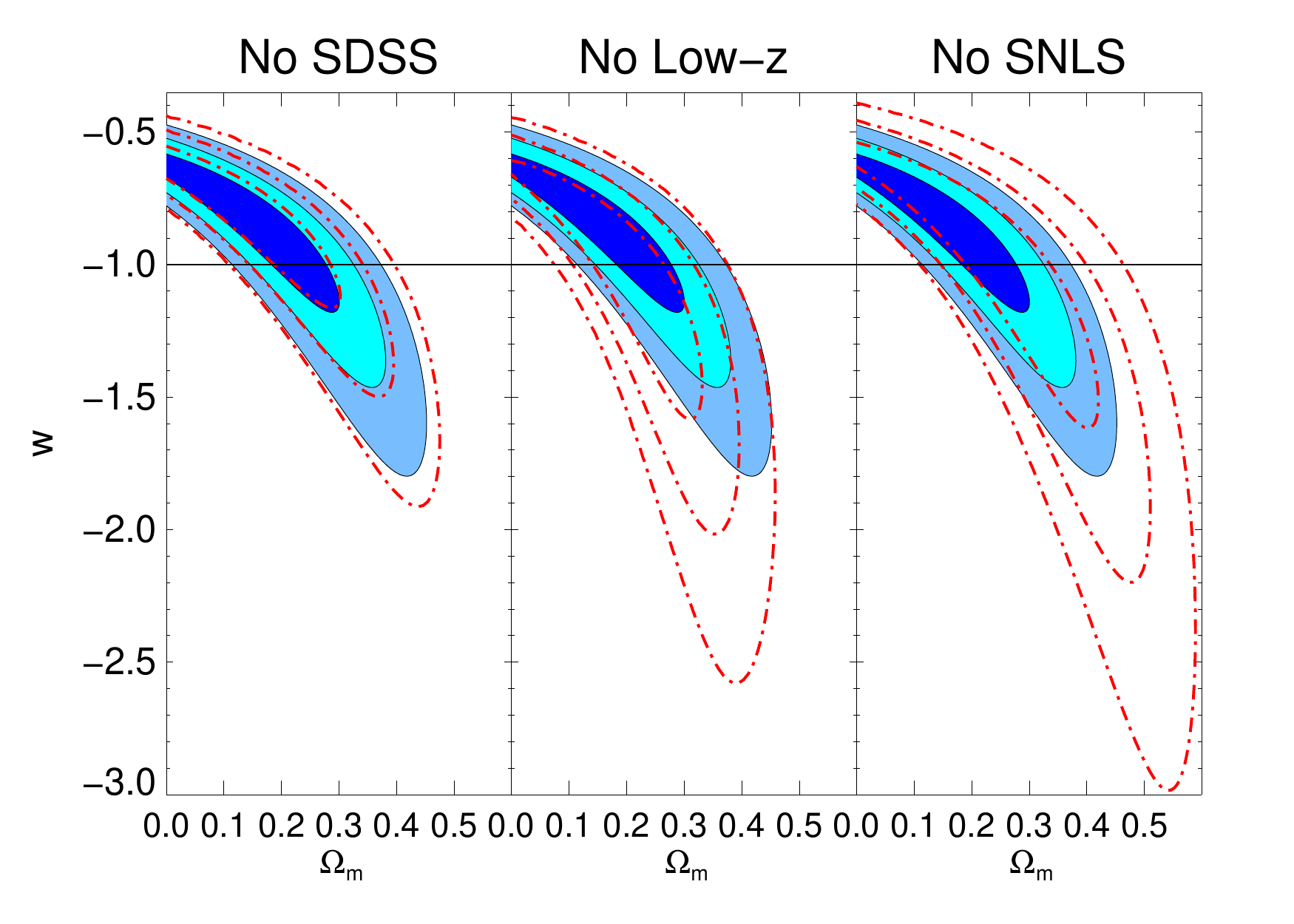}
\caption{Comparison of the constraints on $\Omega_m, w$ in a flat
  universe excluding various samples.  First, in the left panel,
  we exclude the SDSS sample (so the included samples are SNLS, {\it
  HST}, and low-$z$), in the middle the low-$z$ sample, and on the
  right the SNLS data. The fits include all identified systematic
  uncertainties.  The filled contours are the constraints with all
  samples, and the dashed contours exclude the labeled
  sample. \label{fig:nolowz} }
\end{figure}

It is interesting to compare the constraining power of the low-$z$,
SDSS, and SNLS samples.  This was also explored in K09, but with a
smaller low-$z$ sample (33 versus 123 SNe).  The resulting constraints
without each sample are shown in figure~\ref{fig:nolowz}; as can be
seen, the first year of SDSS data is not a good replacement for the
nearby sample, although the full 3 year sample may alter this
situation.  Excluding the SNLS sample has a significant negative
impact on the cosmological constraints.

\subsection{Tension Between Data Sets}
\label{subsec:tension}
\nobreak As a test of whether our estimates for systematic effects are
reasonable, we compute the mean offsets in the residuals from the
cosmological fit between different samples.  We compute the weighted
mean residual for each sample, including the statistical and
systematic covariance matrices and assuming the best fit values of
$\alpha$ and $\beta$.  The results are summarized in
table~\ref{tbl:tension}, and show no significant evidence for any
disagreement between samples (the apparent increase with $z$ is not
statistically significant). Note that our estimates for each
systematic term were constructed before this test was carried out.  We
have also compared the rms around the best fit for different sources
of the low-$z$ sample.  Generally, these are consistent, with the
exception of the CfAII sample (which has an rms of 0.20 mag, compared
with 0.153 mag for the other samples).  This is mostly due to a single
SN which just barely passes our outlier rejection, SN 1999dg.  Without
this SN, the rms of the CfAII sample is 0.163 mag.  The CfAI, CfAIII,
and CSP samples show slightly below average rms (about 0.14 mag), but
this is not statistically significant.  Note that the relative weights
of the samples in this comparison should not be taken too seriously
because of the somewhat arbitrary way in which systematics were
attributed to particular samples, although test for
tension is meaningful.  Specifically, the systematic uncertainties
in the cross-calibration between the low-$z$ and other
samples are always assigned to the latter (SDSS, SNLS, etc.) which
decreases their apparent weight; it would be just as valid to assign
the cross-calibration uncertainties to only the low-$z$ sample, which
would make the other samples appear to have much more weight.
The actual weights should be judged in terms of the consequences
of removing each sample, as in figure~\ref{fig:nolowz}.

\begin{table}
\centering
\caption{Tension between different SN samples\label{tbl:tension}}
\begin{tabular}{l|lll}
\tableline\tableline
Sample & Mean offset (mag) & Uncertainty & N \\
\tableline
  low-$z$      & -0.027 & 0.024 & 123 \\
  SDSS       & 0.020 & 0.027 & 93 \\
  SNLS       & 0.023 & 0.023 & 242 \\
  {\it HST}  & 0.043 & 0.072 & 14 \\[0.1cm]
\hline
 Cal\'an/Tololo & -0.027 & 0.046 & 17 \\
 CfAI  & 0.064 & 0.062 & 7  \\
 CfAII & 0.051 & 0.049 & 15 \\
 CfAIII & -0.047 & 0.034 & 58 \\
 CSP  & 0.052 & 0.057 & 14 \\
 Other & 0.052 & 0.057 & 12 \\
\hline
\end{tabular}
\end{table}

\subsection{The Consequences of Simplified Treatments}
\label{subsec:wrongway}
\begin{figure}
\plotone{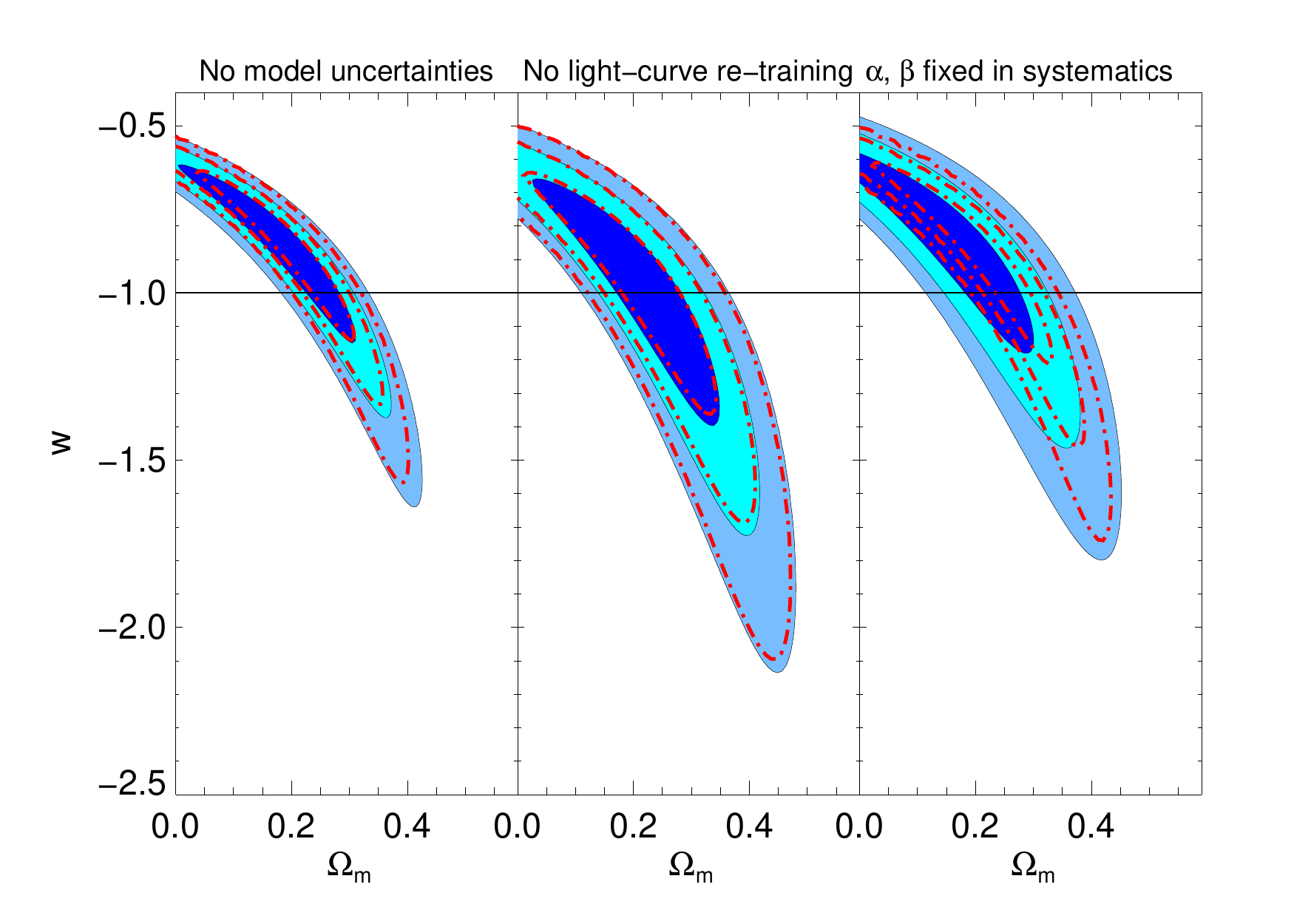}
\caption{Effects of simplified systematic and statistical
  treatments of SN data.  In the left panel, the statistical uncertainties
  with (filled contours) and without (dashed contours) the model
  statistical uncertainties are shown. In the center panel, the results
  of not including the effects of systematics on the light-curve model
  are shown for the combined statistical and systematic errors using
  the same labeling scheme.  Finally, the right panel shows the
  consequences of holding $\alpha$ and $\beta$ fixed when computing
  the systematics covariance matrix.  The center panel is for SiFTO
  only, since it is easier to separate the retraining effects, but the
  other two panels show combined SALT2 and SiFTO results.
  The dashed contours are smaller than the filled contours,
  indicating that each simplification will underestimate the errors
  by some amount.
  \label{fig:wrongway} }
\end{figure}

\nobreak 
In this section we describe the consequences of various simplifying
assumptions in the analysis.  Figure~\ref{fig:wrongway} shows the
effects of not including uncertainties related to the light-curve
models, specifically omitting the model statistical uncertainty and
the effects of the systematic effects on the model training.  The
consequences are not as severe as the effects of fixing the nuisance
parameters, but still will obviously underestimate the uncertainties.
The general effects are to underestimate the total uncertainty budget,
although the size of the effect depends on the simplification.

$\alpha$ and $\beta$ are correlated with the cosmological parameters,
with correlation coefficients of about 0.2 for the $\om, w$ fit;
dropping the assumption of flatness or investigating time varying $w$
generally increases these correlations.  Therefore, the treatment
sometimes found in the literature of fixing $\alpha$ and $\beta$ at
their best fit values and not fitting for them explicitly both
underestimates the uncertainties and results in biased parameter
estimates.  A related simplification is allowing $\alpha$ and $\beta$
to vary for the statistical uncertainty ($\mathbf{D}_{\mathrm{stat}}$
of equation~\ref{eqn:vstatd}), but holding it fixed when computing the
systematics, as in \citet{2008ApJ...686..749K,2010ApJ...716..712A}.
This simplification also biases $\alpha$ and $\beta$, and
therefore the cosmological parameters, as shown in the right hand
panel of figure~\ref{fig:wrongway}, and underestimates the
uncertainties; for this sample, it amounts to underestimating the
size of the inner uncertainty contour by $\sim 40\%$, although the
effects on the marginalized uncertainties are modest.

\section{INDIVIDUAL SYSTEMATIC TERMS}
\label{sec:systematics}
\begin{figure}
\plotone{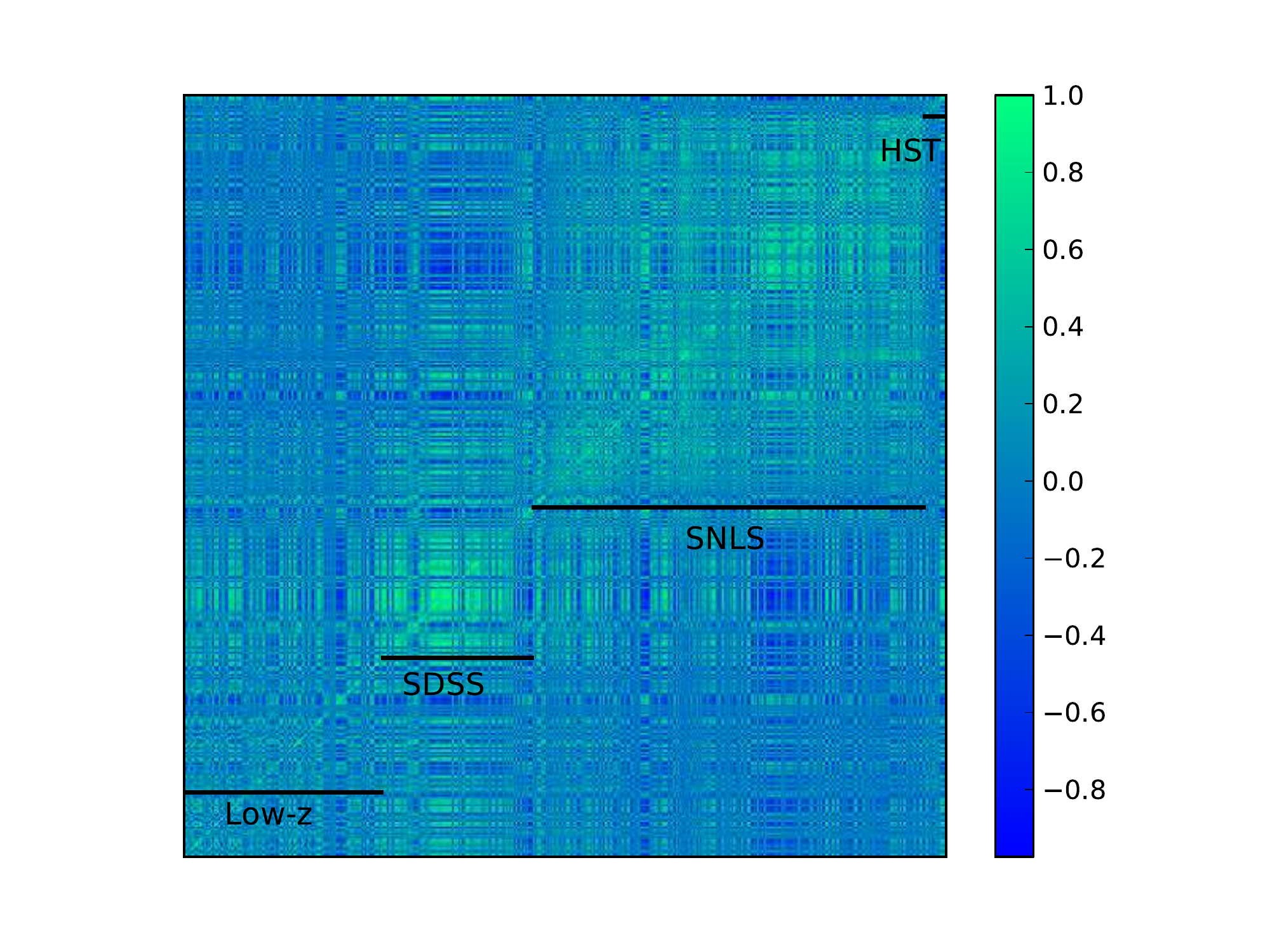}
\caption[Combined correlation matrix]{Combined correlation matrix
  ($\mathbf{C}_{\mathrm{stat}} + \mathbf{C}_{\mathrm{sys}}$ of
  \S\ref{subsec:method}) for the best-fit values of $\alpha, \beta$
  sorted by sample, and by reshift within each sample.  The purely
  diagonal statistical uncertainty term of equation~\ref{eqn:vstatd}
  is not included.  The redshift and sample dependent structure of the
  covariance information is quite complicated, and is not well
  represented by simple $(1+z)^n$ type models.
\label{fig:corrmat}}
\end{figure}

\nobreak 
In this section we discuss the individual systematic terms in detail,
also presenting their effect on the parameter estimates.  Different
categories of systematic effects are summarized in
table~\ref{tbl:overallsystematics}.  The systematic correlation matrix
is shown in figure~\ref{fig:corrmat} to convey some of the sample- and
redshift-dependent structure of the systematic effects.  When
summarizing individual terms, for brevity we only provide the effects
on a fit for $w$ assuming a flat universe and $\om = 0.27$ and the
area relative to statistical only fit of the contour enclosing 68.3\%
of the total probability.

\subsection{Calibration}
\label{subsec:calibration}
\nobreak 
The calibration of the SNLS data onto a standard photometric system is
described in detail in R09.  Achieving the target accuracy of
$\approx 1\%$ has proven quite difficult, and remains the major source
of systematic uncertainty in our measurement.  In addition to the SNLS
calibration, we also consider the calibration of the external
samples included in our analysis.  It is useful to separate the
effects into a number of categories.  Calibration can be considered to
consist of two steps: first, the observations are standardized
onto some photometric system, and second, they are converted from
the standard system into (relative) fluxes in order to compare SNe at
different redshifts.  Items related to the first step are hereafter
referred to as zero-point uncertainties, and those related to the
second as flux calibration uncertainties, which includes both bandpass
uncertainties and uncertainties in the magnitudes and SED of our
fundamental flux reference, \bd .  In many other treatments these two
terms are lumped together into a single ``calibration error'' term.

The details of how most of these uncertainties are derived are given
in \S12 of R09.  Here we present them in a somewhat different (but
equivalent) form, breaking them down into various contributions in
order to highlight which particular items cause the largest effect on
the cosmological parameters.  In addition to the effects discussed in
R09, we also present calibration uncertainties for the external
samples and uncertainties in the bandpasses.  The various
contributions are summarized in table~\ref{tbl:calib}.  Calibration is
the largest component of our systematics uncertainty budget.  This is
not due to the fact that we train our light-curve fitters on our own
data, as can be inferred from the center panel of
figure~\ref{fig:wrongway}.  The most important sub-terms are the
colors and SED of \bd\ and the SNLS zero points.

\begin{table}
\centering
\caption{Calibration Systematics \label{tbl:calib}}
\renewcommand{\arraystretch}{1.5}
\begin{tabular}{l|cc|l}
\tableline\tableline
 Description & $w$ for \om=0.27 & Rel area & Section \\
\tableline
 Stat only           & \allstatwfixomw & 1 & \\
 All calibration     & $-1.06 \pm 0.10$ & 1.79 &
                     \S\ref{subsec:calibration} \\[0.1cm]
\tableline
 Colors of \bd      & $-1.075 \pm 0.075$ & 1.31 & \S\ref{subsubsec:bdcolors}\\
 SED of \bd         & $-1.026 \pm 0.073$ & 1.23 & \S\ref{subsubsec:bdsed} \\
 SNLS Zero Points    & $-1.030 \pm 0.069$ & 1.21 & 
     \S\ref{subsubsec:snlszp} \\
 low-$z$ Zero Points   & $-1.044 \pm 0.065$ & 1.13 & 
     \S\ref{subsubsec:lowzzp} \\
 SDSS Zero Points    & $-1.028 \pm 0.060$ & 1.02 & \S\ref{subsubsec:sdsszp} \\
 MegaCam Bandpasses & $-1.017 \pm 0.066$ & 1.20 &
     \S\ref{subsubsec:snlsfilts} \\
 low-$z$ Bandpasses   & $-1.027 \pm 0.059$ & 1.04 & 
    \S\ref{subsubsec:otherfilts} \\
 SDSS Bandpasses    & $-1.026 \pm 0.059$ & 1.02 & 
    \S\ref{subsubsec:otherfilts} \\
 {\it HST} Zero Points & $-1.027 \pm 0.058$ & 1.03 & 
    \S\ref{subsubsec:hstcalib} \\
 NICMOS Nonlinearity & $-1.029 \pm 0.059$ & 1.05 &
   \S\ref{subsubsec:hstcalib} \\
\tableline
\end{tabular}
\renewcommand{\arraystretch}{1}
\tablecomments{\centering Individual calibration systematics.}
\end{table}

\subsubsection{SNLS Zero Points}
\label{subsubsec:snlszp}
\nobreak Since SNLS observes in four filters and there are four fields
which each have their own zero point, there are 16 zero-point terms to
consider.  However, most of the important terms are strongly
correlated between different fields, and in fact in some cases between
different filters.  These are described in the first section of
table~12 of R09 in combination with the uncorrelated random
uncertainties of table~8 of that paper.  The net effect is an
uncertainty of around $2.5, 2.4, 5.9,$ and $3.2$ mmag in \snlsfilts ,
with correlations.  These uncertainties are fairly small, as benefits
the considerable effort involved in the SNLS calibration; this is an
advantage of large, single instrument surveys, which can devote
considerably more effort to internal calibration than is practical for
small surveys or those which make use of many instruments.  In
addition to these uncertainties, there is a small additional
uncertainty of about 2 mmag in each filter arising from the
differences between the point-spread function and aperture photometry
used on the SNe and stars, as discussed in \S3.2 of G10.

The net effect on the measured parameters is given in table~\ref{tbl:zp}.
Because of the correlations, the breakdown by individual filter should
only be considered approximate, but the largest effect stems from the
$i_M$ filter, which maps to rest-frame $B$ near the redshift range
where SNe are most sensitive to $w$.

\begin{table}
\centering
\caption{Zero Point Uncertainty Systematics \label{tbl:zp}}
\renewcommand{\arraystretch}{1.5}
\begin{tabular}{ll|ll}
\tableline\tableline
 Description & Uncertainty (mag) & $w$ for \om=0.27 & Rel Area \\
\tableline
 Stat only           & \nodata & \allstatwfixomw & 1 \\
\tableline
 SNLS zero points     & \nodata & $-1.031 \pm 0.070$ & 1.21 \\
 $g_M$ only  & 0.0028 & $-1.030 \pm 0.059$ & 1.00 \\ 
 $r_M$ only  & 0.0028 & $-1.012 \pm 0.060$ & 1.12 \\
 $i_M$ only  & 0.0061 & $-1.034 \pm 0.064$ & 1.10 \\
 $z_M$ only  & 0.0035 & $-1.049 \pm 0.060$ & 1.02 \\
\tableline
 low-$z$ zero points  & \nodata & $-1.044 \pm 0.065$ & 1.13 \\
 CfAIII only        & \nodata & $-1.047 \pm 0.061$ & 1.08 \\
 CSP only           & \nodata & $-1.033 \pm 0.059$ & 1.01 \\
 $B$               & 0.015 & $-1.027 \pm 0.060$ & 1.03 \\
 $V$               & 0.015 & $-1.024 \pm 0.060$ & 1.01 \\
 $R$               & 0.015 & $-1.030 \pm 0.059$ & 1.00 \\
\tableline
 SDSS zps         & \nodata & $-1.028 \pm 0.060$ & 1.02 \\
 $g$ only         & 0.003 & $-1.029 \pm 0.058$ & 1.00 \\
 $r$ only         & 0.004 & $-1.029 \pm 0.059$ & 1.01 \\
 $i$ only         & 0.007 & $-1.030 \pm 0.059$ & 1.00 \\
 Scale from $V$   & 0.0062 & $-1.030 \pm 0.059$ & 1.00 \\
\tableline
 {\it HST} zero points & \nodata & $-1.027 \pm 0.058$ & 1.03 \\
\tableline        

\end{tabular}
\renewcommand{\arraystretch}{1}
\tablecomments{The SNLS zero points do not include the transformational
 uncertainties for \bd , which are given in table~\ref{tbl:bd}.  The
 combined uncertainties are given in table D.2 of R09. SDSS
 values do not include the effects of the filter mean wavelength
 uncertainties on the conversion to AB magnitudes, given in
 table~\ref{tbl:filts}.  The {\it HST} zero-point uncertainties do include
 the effects of passband uncertainties.  The zero-point uncertainties
 for the CSP sample are not broken down into individual terms because
 it constitutes a small fraction of the overall sample.}
\end{table}

\subsubsection{Low-$z$ Data Set Zero Points}
\label{subsubsec:lowzzp}
\nobreak
The zero-point uncertainties of the nearby SN samples are generally not
as well understood as those of SNLS.  These samples are generally less
homogenous than the SNLS sample, and are frequently not reported in
the natural system of the telescope/detector.  On the other hand,
since the data come from many sources, both in terms of the
telescope/detector systems and the techniques applied to reduce the
data, some of the calibration errors may partially average out across
multiple samples, resulting in a smaller uncertainty. We are not in a
position to model the zero-point uncertainties of the low-$z$ data at the
same level of sophistication as the SNLS zero points as we do not have
access to the raw data, and the inhomogeneity would make this quite
difficult in any case.

As discussed previously, the nearby SN sample is dominated by
five large datasets.  The observations for each of these sets
are generally from 1-3 instruments at a single site, with a few
exceptions.  In addition to the raw measurement uncertainties in the
calibrations, we expect uncertainties from at least two other sources.
First, three of the five samples present photometry in the Landolt
system rather than the natural system by applying linear
transformations to the natural system magnitudes.  These linear
transformations are derived using stars, and therefore are not
entirely appropriate for SNe.  This introduces an error in the
resulting magnitudes, one that is determined by the mismatch between
the actual instrumental passbands and the Landolt system.  Since the
passbands are shared across many observations, and because these
surveys cover a fairly limited redshift range, these uncertainties will be
correlated between different SN.  Second, and less
importantly, uncertainties in the modeling of atmospheric extinction
will be shared between most observations for a given survey.

However, in contrast to the earliest low-$z$ samples, the CfAIII
and CSP samples provide data in their natural systems.  Since what we
are really doing is comparing SN fluxes observed through low-$z$ natural
systems with SN fluxes in the MegaCam natural system, both of which
are much more accurately measured, using the natural system
frees us of the uncertainty related to the effective passbands
of the Landolt system, which is a significant advantage.

\citet{1993AJ....106.2392H} discuss the photometric transformations
used to calibrate the Cal\'an/Tololo sample, giving a typical
uncertainty of $\sim 0.02$ mag.  For the CfAI sample,
\citet{1999AJ....117..707R} quotes an rms of $0.02$ mags for the
typical residuals for the calibration stars around the mean.  No
specific zero-point uncertainty is given by \citet{2006AJ....131..527J}
for the CfAII sample, but uncertainties in the color terms are
estimated as $0.5 - 2\%$.  Together, these estimates suggest
that the zero points of the nearby data are quite well known.  However,
the treatments in these papers are generally focused on the
contribution of zero-point uncertainties to individual observations,
not the overall zero-point uncertainty.  To estimate this value, we
consider photometry of the same SNe from different sources.
\citet{1999AJ....117..707R} performs this comparison for two SNe,
while \citet{2006AJ....131..527J} carry out a similar analysis for
10 SNe and \citet{2009ApJ...700..331H} do so for 7.  However, the
latter is done in the Landolt system, and hence does not directly
apply to the natural system magnitudes we use in this analysis.  Some
of these SNe do not have extensive enough coverage to allow a detailed
comparison.  Considering these results, we estimate a global zero-point
uncertainty of about 0.015 mag in $BVR$.  We therefore adopt these
values for the three samples on the Landolt system.

\citet{2009ApJ...700..331H} quote zero-point systematic uncertainties
of 0.03 mag in $BVRIr^{\prime}i^{\prime}$, and 0.07 mag in $U$ for the
CfAIII sample.  However, this refers to the Landolt transformed
magnitudes, not the natural system magnitudes, which removes a
substantial part of the uncertainty.  The second major contribution
comes from a comparison of the pipeline reductions with those of an
earlier version used in \citet{2006AJ....131..527J}.  They note that
the scatter in the reduced magnitudes from object to object is
typically about 0.02 mag, and adopt this as their systematic
uncertainty.  However, the pipeline offset varies randomly from
object to object, and therefore is properly a statistical uncertainty
for our purposes.  The mean offsets between the two pipelines, however
(0.005 mag), are systematic uncertainties, since their effects do not
average down by combining distances to many SNe.  The appropriate
systematic uncertainty from this term is therefore 0.005 mag in each
filter.  In addition, we include the transformational uncertainties
for the magnitude of our fundamental flux standard.  While it is
likely that these are actually smaller at the color of this standard
than at color zero, we take the conservative approach of assuming that
the measurement uncertainties are smallest for zero color and
extrapolating.  Putting these contributions together, we arrive at
zero-point uncertainties of $0.011$, $0.007$, and $0.007$ mag for $BVR$
on the 4Shooter camera, and essentially identical values for
$BVr^{\prime}$ on Keplercam.

The zero-point uncertainties of the CSP sample have not been analyzed
in detail.  Comparison with photometry of SNe in common is complicated
because of the different photometric systems, but suggests the
uncertainties are on the order of 1\% or less.  The CSP quotes
zero-point uncertainties including the effects of bandpass
uncertainties, which we treat separately; removing these results in
0.8\% zero-point uncertainties.  There are 7 well observed SNe in
common between the CSP and CfAIII sample, so we can compare the
photometry.  Because they are in different systems, this comparison
must be mediated by an SN model.  We perform SiFTO fits to each SN and
use the resulting model SEDs to $K$-correct the photometry from one
system to another.  We are interested in mean offsets between the two
systems, so calculate the weighted mean difference and uncertainty,
including the estimated calibration correlations.  We find good
agreement in $Bg^{\prime}Vi^{\prime}$, and somewhat poorer in
$r^{\prime}$ (with differences of 0.8, 0.9, 1.5, 1.0, and 2.2$\sigma$,
respectively), after outlier rejection (which actually increases these
values), although the precision is only about 2\%.  The $i$
value should be treated with some caution, as at the low
redshift of the comparison objects it lies beyond the wavelength range
we normally consider in SiFTO and SALT2.  It is not clear how
seriously to take the disagreement in $r^{\prime}$, but fortunately
the $r^{\prime}$ zero point has virtually no impact on our results.

The effects are broken down into individual terms in
table~\ref{tbl:zp}.  The $B$ and $V$ filters can be seen to have the
largest effects, but overall the low-$z$ sample zero points are not a
major contributor to our final uncertainty budget.  Unsurprisingly,
including the SDSS SNe significantly reduces the importance of these
terms by providing an independent absolute magnitude anchor -- without
the SDSS sample, the combined statistical and systematic uncertainty
from the low-$z$ zero-point uncertainties is $\pm 0.074$ rather
than $\pm 0.064$.

\subsubsection{{\it HST} Calibration}
\label{subsubsec:hstcalib}
\nobreak 
In addition to the zero-point uncertainties, NICMOS suffers from
multiple types of non-linearity: an expected exposure length
non-linearity probably due to charge trapping, and a poorly understood
non-linearity related to the count rate of the observed object (plus
sky).  The NICMOS pipeline attempts to correct for both
non-linearities.  However, these corrections, particularly for the
latter effect, are not perfect, and therefore we include related terms
in our uncertainty budget.  The second nonlinearity is described in
\citet{2006hstc.conf..121D,2006nicm.rept....1D, 2008nicm.rept....3S}
and the scheme used to correct the SN data applied by
\citet{2007ApJ...659...98R} is described in
\citet{2006nicm.rept....3D}.

The correction model is parameterized by a power law such that count
rate $\propto$ flux$^{\alpha_N}$ with measured values (for NICMOS chip 2) of
$\alpha_N = 1.025 \pm 0.002$ for the F110W filter and $\alpha_N =
1.012 \pm 0.006$ for F160W.  Here the flux should include both the
object and background, which is important for faint sources such as
high-$z$ SN.  It is interesting that this effect seems to be wavelength
dependent.  Note that this correction is not based on a physical model, and so
may not be the most appropriate parameterization.  Furthermore, this
nonlinearity has only been tested down to count rates of a few per
second, while $z>1$ SN plus background observations are typically an
order of magnitude fainter.

The count rate non-linearity affects the observations of bright
standard stars differently than faint SNe, which means that uncertainty
in each $\alpha_N$ affects how the NICMOS calibration propagates to SN
observations, and hence introduces a systematic uncertainty in the 
light-curve parameters.  The NICMOS zero points are based on observations of
P330e, which is $\sim 11.5$ mag in F110W and F160W.  The proper way to
measure the impact of this effect would be to go back to the original
SN and standard star images and re-extract the photometry for
different values of $\alpha_N$.  However, we can obtain a decent
estimate simply by converting the magnitudes of the standard stars and
SNe to counts rates and comparing the flux ratio for different values
of $\alpha_N$.

Carrying out this program (and including the appropriate typical
background levels for the two passbands), the average uncertainty for
observations in the F110W and F160W filters induced by the uncertainties in
$\alpha_N$ are 0.022 and 0.060 mag, respectively, for {\it HST} SN
with data in those filters; the exact values depend on the flux level
of a particular photometry point, and so vary both from SN to SN and
with epoch for a particular SN.  Because the effects of the
uncertainties in the $\alpha_N$ are correlated across all of the SNe
observed with NICMOS, they have a quite substantial impact on the
weight of these SNe in the cosmological fits.  Furthermore, these
uncertainties may be optimistic, since they assume the power law model is
valid at such low flux levels.  Because of the high
background levels, these numbers are fairly constant as a function of
epoch, do not change much for different SNe in the {\it HST} sample,
nor are they strongly affected by the possible presence of a host
galaxy.

Changes in $\alpha_N$ make SNe either simultaneously fainter and
redder or bluer and brighter, so that correction for the
brighter-bluer relation partially cancels the effects of this
systematic term.  However, the residual effect is still substantial.
The effect on $w$ is small because the {\it HST} SNe have relatively little
weight in this fit, but this is not necessarily true of fits for other
cosmological parameters such as time-varying $w$. We can roughly
estimate the importance of this effect by calculating the change in
the inverse variance of the mean corrected magnitude of the {\it HST}
sample with and without this systematic -- this is essentially the
weight of the sample.  This shows that the uncertainty in the
nonlinearity correction reduces the weight of the {\it HST} sample by
17\%, approximately equal to removing three SNe; reducing the
uncertainties in $\alpha_N$ by a factor of two would be equivalent to
adding two more SNe. Fortunately, only the four highest redshift SNe are
strongly dependent on NICMOS F160W observations, where the nonlinearity
correction is the most uncertain.

In addition to the uncertainty in the NICMOS nonlinearity corrections,
we assign zero-point uncertainties of 2\% for the ACS bands longward of
7000 \AA\ following \citet{2007acs..rept....6B}, and 1\% below that.
For NICMOS, we additionally include an uncertainty of 0.009 and 0.0145
mag for F110W and F160W following the information on the NICMOS web
site.  The above terms nominally include the uncertainty in the
bandpasses, so we do not include separate terms for those.

\subsubsection{SDSS Zero Points}
\nobreak
\label{subsubsec:sdsszp}
The calibration adopted by the SDSS SN team is based on observations
of solar analogs with CALSPEC {\it HST} measured SEDs.  There are no
Landolt magnitudes available for these stars, so we take a slightly
less direct route by comparing their SEDs with that of \bd\ (which,
fortunately, is calibrated onto the same system).  For the uncertainty
in each zero point, we adopt values given in \S3.2 of K09, which are 0.4\%,
0.4\%, and 0.7\% in $gri$\footnote{These do not include the effects of
  the uncertainty in the SDSS bandpasses on the conversion from SDSS
  to AB magnitudes which we include in
  \S\ref{subsubsec:otherfilts}}. In addition to the individual filter
terms, because this calibration is not linked directly to the Landolt
system, the inter-calibration is subject to an additional ``gray''
uncertainty of 0.6\% arising from the uncertainty in the $V$
magnitude of \bd\ on the Landolt system, including transformational
uncertainties.  This additional term is the price paid for not
calibrating the SDSS (and CSP) observations to the same system as the
low-$z$ SNe, although the net calibration uncertainty is still small
compared to some other samples.

There are a small number of well-observed SNe in common between the
SDSS and low-$z$ samples, in particular two with the CSP and one from the
CfAIII sample.  Comparing the photometry as between the CSP
and CfAIII samples, we find good agreement in $gr$ but not in $i$,
with mean offsets of 1.2, 0.5, and 2.1$\sigma$ , respectively.  It is
even less clear how to interpret these numbers due to the very small
sample size.  Fortunately, there should be many more objects in
common between years 2-3 of the CSP and SDSS samples, so it will be
possible to carry out much more stringent comparisons in the future.

\subsubsection{System Bandpasses: SNLS}
\label{subsubsec:snlsfilts}
\nobreak
We next turn to uncertainties in the SNLS system bandpasses.  The four
MegaCam filters used by SNLS for the first 3 years of operation were
manufactured by SAGEM.  The filters scans show clear evidence of
variation with radial position in the array, which we account for in
our analysis (R09 \S7).  The largest systematic effect on our
measurements arises from shifts in the mean wavelengths of the
bandpasses, since this changes which portion of the SN SED is sampled at
different redshifts.  Alternative prescriptions, such as adding white
noise to the bandpasses \citep[as in][]{2007ApJ...666..694W},
have little effect on the cosmological parameters and therefore
underestimate the importance of accurate bandpass measurements.  The
mean bandpass can shift both from measurement uncertainties, and from the
effects of the dry environment on Mauna Kea on the interference
films\footnote{Fortunately, our filters do not sit in a vacuum, and
  hence should not experience the large shifts seen in the SDSS
  camera.}.  Therefore, we parameterize our bandpass uncertainties here
and in \S\ref{subsubsec:otherfilts} in terms of shifts in the mean
wavelengths.  Other effects, such as changing the width of the
bandpasses, have a much smaller effect.  There have been several filter
scans made of the MegaCam filters.  These are consistent, once the
radial and angular dependence of the filters is taken into account.

Two of the SNLS fields overlap with the footprint of SDSS, so we can
calculate accurate linear transformations between the two systems as
discussed in Appendix G of R09.  We can then compare these
transformations against the results of synthetic photometry to
estimate the size of any shifts in the mean bandpasses. We perform
synthetic photometry on a library of stellar SEDs and try to reproduce
the slope of the relation, but not the offset.  The latter is
important because by ignoring the offset these results are independent
of effects such as zero-point uncertainties.

There is some variation in the results depending on which spectral
library is used; we considered
\citet{1983ApJ...266..713O,1998PASP..110..863P,2005PASP..117..810S}.
The uncertainties in the transformation coefficients are of roughly
equal importance. For \gmeg\ we find a best fit shift of $3 \pm 4 \pm
4$ \AA , where the first uncertainty derives from the spread of
results from the different SED libraries, and the second from the
uncertainty in the measured color term between SDSS and MegaCam.  For
\rmeg\ we derive a shift of $1 \pm 1.5 \pm 7$ \AA , and for \imeg\ $12
\pm 6 \pm 5$ \AA .  The SED libraries do not extend far enough to the
red to allow us to test \zmeg .  While the different SED libraries
agree best in \rmeg , this bandpass is particularly sensitive to
uncertainties in the color term.  Taken together, the probability of
finding these shifts or larger due to measurement uncertainties in the case
of no shifts is 46\% .  We therefore interpret the results of the
above test as being consistent with the SAGEM filter scans having the
correct mean wavelengths. 

These limits are conservative for several reasons.  First, they do not
distinguish between uncertainties in the SDSS bandpasses and the SNLS
ones.  Furthermore, the SAGEM filter scans are thought to be much more
accurate than these uncertainties, and the other contributions to the
system bandpass all vary fairly slowly with wavelength.  We therefore
roughly average the above tests and adopt uncertainties of 7 \AA\ in
$g_M r_M i_M$.  For $z_M$, where we have not been able to carry out
the above test, we assume a bandpass uncertainty of 15 \AA\ by scaling
up the value from the other filters to include variations in the CCD
response, which set the red-side cutoff for this filter.  The
consequences are summarized in table~\ref{tbl:filts}.  We also
considered varying the width of the bandpasses, and the effects of
using the mean air mass when computing our bandpasses instead of the
air mass for a particular observation. We find little effect from
either, so do not include them in our uncertainty budget.  At an
air mass of 2 the mean air mass effects are less than 0.001 in the width
of the light-curve and 0.003 in color in the most extreme case.  Since
SNLS doesn't observe above air mass 1.6 in practice, this effect is
completely negligible.

\begin{table}
\centering
\caption{Bandpass Systematics \label{tbl:filts}}
\renewcommand{\arraystretch}{1.5}
\begin{tabular}{ll|ll}
\tableline\tableline
 Description & Uncertainty (\AA) & $w$ for \om=0.27 &
  Rel Area \\
\tableline
 Stat only           & \nodata & \allstatwfixomw & 1 \\
\tableline
 MegaCam Bandpasses  & \nodata & $-1.017 \pm 0.066$ & 1.20 \\
 $g_M$ only         & 7 & $-1.030 \pm 0.058$ & 1.01 \\
 $r_M$ only          & 7 & $-1.037 \pm 0.059$ & 1.01 \\
 $i_M$ only          & 7 & $-1.030 \pm 0.059$ & 1.02 \\
 $z_M$ only          & 15 & $-1.032 \pm 0.060$ & 1.04 \\
\hline
 low-$z$ Bandpasses      & \nodata & $-1.027 \pm 0.059$ & 1.04 \\
 CfAIII only           & 7 & \allstatwfixomw & 1.00 \\
 CSP only              & \nodata & \allstatwfixomw & 1.00 \\
 $B$ only              & 12 & $-1.029 \pm 0.059$ & 1.02 \\
 $V$ only              & 12 & $-1.028 \pm 0.059$ & 1.01 \\
 $R$ only              & 12 & \allstatwfixomw & 1.00 \\
\tableline
 SDSS Bandpasses  & \nodata & $-1.026 \pm 0.059$ & 1.02 \\
 $g$ only         & 7  & $-1.030 \pm 0.059$ & 1.00 \\
 $r$ only         & 16 & $-1.026 \pm 0.058$ & 1.00 \\
 $i$ only         & 25 & $-1.030 \pm 0.059$ & 1.01 \\
\hline
\tableline
\end{tabular}
\renewcommand{\arraystretch}{1}
\end{table}

\subsubsection{System Bandpasses: Other Samples}
\label{subsubsec:otherfilts}
\nobreak 
As described in Appendix~\ref{apndx:filterslowz}, we have applied
small shifts to the \citet{1990PASP..102.1181B} passbands to represent
the Landolt system.  There are uncertainties associated with these
shifts, as described in Appendix~A, and here we include them
in our uncertainty budget.  We adopt mean wavelength uncertainties of
12 \AA\ in the mean wavelength for $B$ and $V$ and 25 \AA\ for $R$.
In addition to the Landolt system, we also include terms representing
the uncertainties in the CfAIII natural system passbands.  As we did
for the SNLS filters, we adopt uncertainties of 7 \AA\ for these
passbands, which again are probably conservative.  The uncertainties
in the mean wavelength of the CSP filters are described in
\citet{2010AJ....139..519C}.  These are limited by the number of
spectrophotometric standards available, and are probably much larger
than the real uncertainties.  Due to the way in which the calibration
is constructed, the CSP bandpasses affect the zero points strongly, but
because this is currently such a small sample, the cosmological
effects are negligible.

For the SDSS bandpass uncertainties, we use the values given in
table~1 of K09. These systematic effects have very little effect in
fits where the SDSS SNe are included, although they do also affect the
SDSS to AB conversion and therefore the effective zero points.  The
effects of these shifts on the final uncertainty budget are given in
table~\ref{tbl:filts}.

\subsubsection{The Colors of the Flux Reference on the Landolt System}
\label{subsubsec:bdcolors}
\nobreak
As part of our analysis, we convert between magnitudes on the Landolt
system and fluxes.  This is necessary because we are comparing SNe
at a range of redshifts, and so the same regions of the rest frame
SN SEDs are sampled by different filters for different objects.
Fortunately, our measurement is not sensitive to the absolute flux scale,
so more precisely we convert the magnitudes to relative fluxes.
A major influence in our decision to calibrate MegaCam to the Landolt
system was the insensitivity to absolute calibration.  Were we to, for
example, calibrate our observations onto an AB system, we would have
included the uncertainty in the absolute flux scales of both systems
when comparing them.  This is the case for the SDSS sample.

Contrary to some statements found in the literature, the Landolt
system is not {\it defined} in terms of any particular magnitudes for
any particular star (i.e., Vega), nor is the
\citet{1953ApJ...117..313J} system on which it is based.  Instead,
both are defined based on particular observations of multiple stars --
Vega is one of the stars contributing to $U-V$ and $B-V$ for the
\citet{1953ApJ...117..313J} system.  We are free to choose any star
(or set of stars) with a known SED as long as we know its magnitudes
in the Landolt system and on the Landolt calibrated MegaCam system.
Ideally, we would be able to observe this star directly with
MegaCam. We have not yet been able to fully realize this goal, and
have instead settled on the star \bd\ as our fundamental flux
standard.  This star has a high-quality CALSPEC STIS SED
\citep{2004AJ....128.3053B} and Landolt magnitudes from
\citet{2007AJ....133..768L}.  Because it has colors similar to the
``average'' Landolt star we have high confidence that its magnitudes
have been accurately transferred to the \citet{1992AJ....104..340L}
catalog system , which most nearby observations are calibrated onto.
This is probably not true of the CALSPEC white-dwarfs also observed by
\citet{2007AJ....133..768L} -- due to the steepness of their SEDs,
their magnitudes are offset from the \citet{1992AJ....104..340L}
system by a non-negligible but not perfectly known amount, rendering
them unsuitable as fundamental flux standards for the Landolt system.
For a more extensive discussion of all of these issues, see \S8 and
S11 of R09.  We use the2010 February update to the \bd\ SED available on
the CALSPEC web
site\footnote{http://www.stsci.edu/hst/observatory/cdbs/calspec.html}.

The uncertainties in the magnitudes of \bd\ in both the Landolt system
and the SNLS measurements are a significant contribution to the final
uncertainty budget, as seen in table~\ref{tbl:bd}.  The largest
effects arise from the magnitudes of \bd\ in the SNLS filters, which
is partially due to the fact that \bd\ is a subdwarf, and hence may
transform slightly differently between photometric systems than the
average Landolt standard star.  This transformational uncertainty is
the most significant for the $z_M$ filter, which is reflected in this
term having the single largest effect on $w$.

Most previous SN analyses make use of Vega as their fundamental flux
standard.  This is a poor choice because the magnitudes of Vega on the
Landolt system are in fact poorly known, and it is too bright to
observe directly with almost any modern imager.  The uncertainties in
the Landolt magnitudes of \bd\ are the largest single identified
systematic uncertainty in our current analysis, and they would be
several times larger were we to use Vega instead.  These uncertainties have
been significantly underestimated in previous analyses that use Vega. If
we were able to replace the nearby SN sample with one calibrated onto
a system similar to the natural MegaCam one (e.g., the USNO or SDSS
system), these uncertainties would be much smaller.  Since
\bd\ effectively defines the USNO/SDSS system, this should be
practical in the future, and significantly reduce this contribution to
the final uncertainty budget.  

\begin{table}
\centering
\caption{Flux standard (\bd) uncertainties \label{tbl:bd}}
\renewcommand{\arraystretch}{1.5}
\begin{tabular}{ll|ll}
\tableline\tableline
 Description & Uncertainty (mag) & $w$ for \om=0.27 &
  Rel Area \\
\tableline
 Stat only           & \nodata & \allstatwfixomw & 1 \\
\tableline
 Colors      & \nodata & $-1.075 \pm 0.075$ & 1.31 \\
 $U-V$ only     & 0.0026 & $-1.031 \pm 0.058$ & 1.00 \\
 $B-V$ only     & 0.0015 & $-1.030 \pm 0.059$ & 1.01 \\
 $R-V$ only     & 0.0011 & $-1.030 \pm 0.058$ & 1.00 \\
 $I-V$ only     & 0.0013 & $-1.031 \pm 0.060$ & 1.05 \\
 $g_M-V$ only   & 0.0022 &  $-1.030 \pm 0.059$ & 1.01 \\
 $r_M-V$ only   & 0.0042 & $-1.006 \pm 0.062$ & 1.13 \\
 $i_M-V$ only   & 0.0022 & $-1.039 \pm 0.062$ & 1.06 \\
 $z_M-V$ only   & 0.0178 & $-1.094 \pm 0.071$ & 1.15 \\
\tableline
 SED          & \nodata & $-1.027 \pm 0.073$ & 1.23  \\
\end{tabular}
\renewcommand{\arraystretch}{1}
\tablecomments{The $UBVRI$ values have a correlated effect on 
 \snlsfilts\ via the linear transformation equations of R09, which
 are not included in the above numbers, but are included in the
 overall uncertainty budget.}
\end{table}

\subsubsection{The SED of \bd }
\nobreak
\label{subsubsec:bdsed}
In addition to the uncertainty in the magnitudes of \bd , we
also consider the uncertainties in the SED itself.  These
are essentially set by the quality of the reductions and
the repeatability of STIS in \citet{2004AJ....128.3053B}.
This is discussed in R09 (\S12.8 and table D.3), and
the cosmological effects are given in table~\ref{tbl:bd}.

\subsection{Comparisons of Different Light-Curve Fitters}
\nobreak
\label{subsec:lcfitters}
\S5 of G10 discusses comparisons between different light-curve fitting
packages, and shows that when SALT2 and SiFTO are trained using the
same incidental settings (such as the Landolt filter functions, colors
of \bd , etc.\ that are discussed elsewhere in this paper), the
resulting fits on the same data are very similar.  We consider this to
be a very encouraging result.  However, differences remain, which
should not be surprising given the very different approaches taken to
train the two models.  In essence, the SN data are not yet high-enough
quality to clearly prefer a particular modeling approach.  We
therefore include the differences in the derived light-curve
parameters between the two fitters for each SN as a systematic in our
analysis.  In addition, there is some uncertainty in how to
parameterize the model for the intrinsic uncertainty in the color
model.  We also include the difference between two parameterization
choices (the sigmoid and exppol models of \S4.4 in G10) in our
systematic uncertainty; the effects on the cosmological parameter
uncertainties are given in table~\ref{tbl:lcfit}.

\begin{table}
\centering
\caption{Light-Curve Fitter Systematics \label{tbl:lcfit}}
\renewcommand{\arraystretch}{1.5}
\begin{tabular}{l|ll}
\tableline\tableline
 Description & $w$ for \om=0.27 & Rel Area \\
\tableline
 Stat only   & \allstatwfixomw & 1 \\
\tableline
 Combined           & $-1.027 \pm 0.059$ & 1.02 \\
 SALT2 vs.\ SiFTO   & $-1.027 \pm 0.059$ & 1.01 \\
 Color Uncertainty Model & $-1.030 \pm 0.058$ & 1.00 \\
\end{tabular}
\renewcommand{\arraystretch}{1}
\end{table}

We do not carry out any comparisons with the other publicly available
light-curve fitting package, MLCS2k2 \citep{2007ApJ...659..122J}, as
discussed in \S5 of G10.  Unlike SALT2 and SiFTO, MLCS2k2 explicitly
attempts to separate intrinsic and extrinsic SN colors from
photometric data, and then assumes that the extrinsic colors arise
purely from dust, and that the remaining intrinsic color not related
to the shape of the light-curve does not affect SN luminosity.  SiFTO
and SALT2 do not make this distinction.  The merits of the two
approaches depend critically on how well this separation can be
performed, and how well SN intrinsic color can be predicted by the
light-curve shape.  The former depends on accurate models of the
distribution of extinction with redshift and how they combine with
selection effects \citep{2007ApJ...666..694W}.

If the two conditions are met, then MLCS, by incorporating additional
information beyond SN photometry, may be able to give tighter
statistical constraints on SN relative distances.  A test of how well
this procedure works is to check if the MLCS2k2 predictions of $A_V$
and $\Delta$ (the MLCS2k2 light-curve shape parameter) correlate with
residual from the Hubble diagram, since in principle the MLCS2k2 model
should already apply these corrections based on training with low-$z$
SNe.  Taking the MLCS2k2 fits from K09, we find that the Hubble
residuals exhibit $7 \sigma$ (statistical uncertainties only) evidence
for a linear relation with $A_V$ and $6 \sigma$ with $\Delta$.  This
suggests either a problem with the MCLS2k2 training sample, or with
the detailed extinction prior used.  Additionally, the issues with
observer-frame $U$-band data of low-$z$ SNe are a serious problem for
MLCS2k2 because it is currently not feasible to train it on high-$z$
data; therefore, we would effectively have to abandon most of our data
for SN at $z>0.7$ where the rest-frame $U$ is measured by
\rmeg\ and \imeg.  While we fully expect that this situation will be
addressed, especially in light of the improved low-$z$ data sets that
are becoming available, this means that comparing MLCS2k2 fits to
SiFTO and SALT2 at the current time is not productive.  Note that the
SALT2 and SiFTO models are capable of reproducing the MLCS2k2
assumptions if SN colors are dominated by normal dust extinction.  The
fact that the SiFTO and SALT2 derived color laws do not match the
Galactic color law, assumed in MLCS2k2, for any value of $R_V$
suggests that this assumption is incorrect.

\subsection{Peculiar Velocities, $z_{\mathrm{cut}}$, and the ``Hubble Bubble''}
\nobreak
\label{subsec:bubble}
\citet{2006PhRvD..73l3526H} discuss methods to calculate the
correlated uncertainty in SN luminosity distances due to peculiar
velocities given a survey geometry and radial selection function.
This method is applicable as long as the survey geometry does not
contain a large number of holes.  Unfortunately, for the current low-$z$
sample, the survey geometries consist almost entirely of holes, so
this formalism is not currently practical.

The flow model used in this analysis (\S\ref{subsec:pecvel}) is
parameterized by a quantity $\beta_I = \Omega_m^{0.55}/b_I$, where
$b_I$ is a ``biasing'' parameter for {\it IRAS}-selected galaxies.
$\beta_I$ should not be confused with the color-luminosity nuisance
parameter for SNe, $\beta$.  \citet{2005ApJ...635...11P} gives
$\beta_I = 0.49 \pm 0.04$ (statistical uncertainties only). We adopt
this value, but conservatively take the uncertainty in $\beta_I$ to be
five times larger, $\sigma_{\beta_I} = 0.2$ to reflect the spread in
values derived from other surveys and to account for simplifications
in the modeling, and therefore compare models with $\beta_I = 0.3,
0.5, 0.7$ to estimate the systematic effect on our sample.  The
results are given in table~\ref{tbl:overallsystematics}.  This is not
a major contributor to our uncertainty budget.

A related issue is the possibility of a monopole term in the local
expansion -- a so-called Hubble bubble.  Recently,
\citet{2007ApJ...659..122J} found evidence for such an effect using
light-curve fits to nearby SNe~Ia.  As discussed in
\citet{2007ApJ...664L..13C}, this is related to the interpretation of
SN colors. The result is that when the same data are analyzed in a
framework in which the relation between SN color and luminosity is
determined from SN data -- as is used here -- we see no evidence for a
Hubble bubble, and therefore we do not include this in our analysis.

\begin{figure}
\plotone{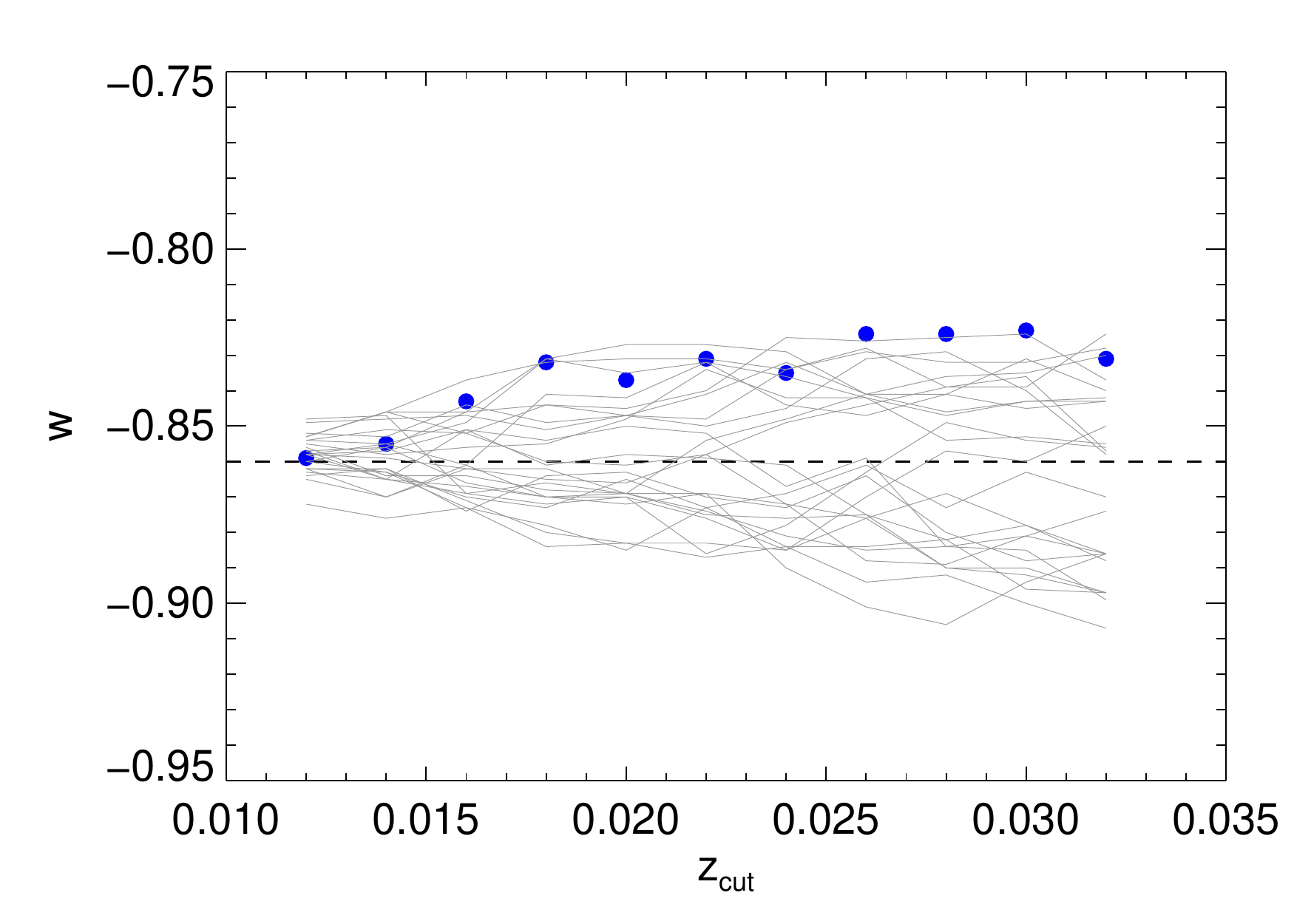}
\caption[Variation of $w$ with $z_{\mathrm{cut}}$]{Variation of $w$ with
  $z_{\mathrm{cut}}$, the minimum redshift low-$z$ SN included.  The dashed line
  is the result when $z_{\mathrm{cut}}=0.01$, the value used in our analysis.
  The points are highly correlated.  Taking this into account via
  Monte-Carlo simulation, we find that this trend is not statistically
  significant, and is consistent with shot noise. The gray
  lines are a sample of 30 independent Monte-Carlos simulations
  illustrating the shot noise as discussed in the text. \label{fig:zcutw}}
\end{figure}

Still, we can investigate the dependence of our results on the minimum
redshift.  K09 study this issue with a smaller nearby sample (\S9),
and include the variation as a systematic uncertainty.  We find a
similar trend, as shown for $w$ in figure~\ref{fig:zcutw}.  In order
to determine if this trend is significant, or is caused by shot noise
from removing SNe from the sample, we have carried out a set of
Monte-Carlo simulations where we note the number of nearby SN removed
by each redshift cut, and try randomly removing the same number from
the sample and refitting the cosmological parameters.  Neighboring
points are extremely correlated because $z_{\mathrm{cut}} = 0.02$ will also
remove all of the SNe that, for example, $z_{\mathrm{cut}}=0.016$ does.
Including these correlations in our simulation, we find no evidence
for a significant trend (a \chisq of 10.16 for 11 degrees of freedom).
Therefore, we find that the observed effect is entirely consistent
with shot-noise, and therefore is already included in our statistical
uncertainty budget.

\subsection{Malmquist Bias}
\label{subsec:Malmquistsys}
\nobreak 
As noted previously, we apply a redshift-dependent correction for
Malmquist bias to each SN sample.  Therefore, the uncertainty in this
correction should be included in our total uncertainty budget.  There
are a large number of individual terms, although only a few have
significant effects.  The SNLS Malmquist bias simulations and their
systematic uncertainties are described in \citet{2010AJ....140..518P}.
The systematic uncertainty on the Malmquist bias for SNLS is about
20\% of the correction, with some redshift dependence.  The dominant
terms are the uncertainty in the input \si\ and the value of
$\beta$ used in the calculation.

For SDSS, we follow the prescription of \S9 of K09 in varying the
spectroscopic selection model.  For the Cal\'an/Tololo, CfAI, and
CfAII samples, we assign 0.01 mag uncertainty in the mean Malmquist
bias, while for the CfAIII sample, we double the estimated
uncertainties for our model of the spectroscopic selection function
($\sigma_{\gamma}=0.62$, see \S\ref{subsubsec:nearbymalm}) in order to
compensate for the crudity of our model.  Within the redshift limits
we have set, the {\it HST} sample suffers from very small amounts of
Malmquist bias, so we do not include this as a systematic.

Some of the terms in the Malmquist bias corrections affect multiple
samples simultaneously. The bias estimates for the SNLS, SDSS, and
CfAIII samples depend on the assumed values of $\alpha, \beta$, and
\scriptm .  We assign uncertainties of 0.22, 0.22, and 0.05 mag to
these, which are twice the formal uncertainties
(table~\ref{tbl:cosmoresults}), and affect all three samples in a
correlated fashion.  Each simulation also depends on \si , and this is
varied separately for each sample by 0.01.  In total there are 22
individual terms in our Malmquist uncertainty; the combined effect is
given in table~\ref{tbl:overallsystematics}.  Malmquist correction is
one of the larger terms in our overall uncertainty budget, ranking
only behind calibration and the relation between host type and SN
absolute magnitudes.  About 80\% of the effect on the cosmological
parameters comes from the low-$z$ sample, which is not surprising because
the selection effects of the higher-redshift samples are much better
characterized.

\subsection{Contamination by Other Types of SNe in the SNLS Sample}
\nobreak 
\label{subsec:nonia}
All of the SNe used in our cosmological analysis are spectroscopically
typed, so contamination by non-SNe~Ia should be fairly minimal.
However, some is inevitable, particularly at the highest redshifts
where the signal to noise of the spectra can be quite low.  The
principle contaminants are expected to be SN~Ib/c, which are brighter
than other core-collapse SNe and whose spectra can be difficult to
distinguish from SN~Ia at certain phases with low signal-to-noise.
Unfortunately, we know relatively little about the demographics of
this population, which makes contamination effects difficult to
estimate.  Non-Ia contamination is also discussed in
\citet{2009A&A...507...85B}.

Our spectroscopic typing scheme is described in detail in
\citet{2005ApJ...634.1190H}.  In this scheme, SN candidates are typed
into various classes.  Those that concern this paper are SNIa, which
represent objects which are essentially certain SNe~Ia, and
SNIa$\star$, which probably are, but other types can't be conclusively
ruled out.  The false positive rate for SNIa is expected to be
negligible, but it is difficult to estimate the rate for SNIa$\star$.
Approximately 20\% of the SNLS sample presented here are SNIa$\star$
(50 out of 242), with the fraction increasing with redshift.

It is tempting to compare the corrected peak magnitudes of the
SNIa$\star$ sample to the SNIa sample, and to use this comparison to
estimate the systematic uncertainty on the cosmological parameters.
However, this test is severely flawed because luminosity affects our
ability to type spectra, so SNIa$\star$ will be fainter even in the
absence of contamination. Ideally, fainter SNe would be compensated
for with longer spectroscopic exposure times, but in practice this was
rarely realized at the 0.1-0.2 mag precision required due to weather,
the amount of time remaining in the queue, and the fact that exposure
times were estimated from a few pre-maximum points using preliminary,
real-time photometry.  Furthermore, the intrinsic luminosity of the SN is
correlated with the relative amount of host-galaxy contamination in
the spectra (the \%-increase of the SN+host relative to the host), which is
not as easy to compensate for, especially in poorer seeing conditions.
We do not have a good model for predicting the odds of a given SN
candidate getting an inconclusive spectroscopic type, and attempts to
develop one based on the observed demographics of the SNIa
versus SNIa$\star$ were somewhat circular, naturally explaining away any
observed difference in the apparent luminosity.  Note that this
implies that simply excluding all SNIa$\star$ from our analysis would
bias our measurement in a fashion we would not know how to correct
for very well.

Therefore, we adopt a simple model to estimate the contamination bias.
The critical factors in such a model are the number of contaminants
that can be confused with SN~Ia and the mean difference in magnitude
between the contaminants and SNe~Ia at that redshift.  There are a few
aspects of SNLS that help reduce the contamination.  First, we have
high-quality spectroscopic redshifts for all objects considered in
these fits.  Second, the majority of our SN have firm spectroscopic
types.  Combined, these two features mean that the outlier rejection
we apply (described in \S\ref{subsec:cuts}) is highly effective at
removing any contaminants with a different magnitude distribution than
SN~Ia -- and if the magnitude distributions were identical (after the
shape and color corrections are applied), there would be no bias even
if the contamination rate were high.  In addition, we reject SN that
are very poorly fit by the light-curve template, and preselect
candidates for spectroscopic follow up based on early-time light-curve
points \citep{2006AJ....131..960S}.  Finally, recall that SNIa$\star$
are probably SNe~Ia -- spectra which are truly uninformative are
not used in this analysis.  While the final three
items have considerable power at rejecting candidates, for our simple
model we will conservatively assume that they are completely
ineffective, significantly increasing the bias.

The basic ingredients for our model are the relative rate of SNe~Ia
and SNe~Ib/c and the SN~Ib/c luminosity distribution.  We assume that
the SN~Ib/c rate is proportional to the star-formation rate, which we
base on \citet{2006ApJ...651..142H}; this is a good approximation
because the progenitor systems are massive stars, and the delay time
between formation and explosion should be negligible for our purposes.
We take the SN~Ia rate from K.~Perrett et al.\ 2011, in preparation.
The most critical, but most uncertain, ingredient is the assumed
luminosity function of SN~Ib/c.  Recently, \citet{2010arXiv1006.4612L}
have studied the luminosity function of various SN types using the
galaxy targeted KAIT survey.  They find that SNe~Ib/c are, on average,
2.4 mag fainter than SNe~Ia, and that the SN~Ib/c rate is about 80\% of
the SN~Ia rate at low redshift.  If we take their luminosity function
at face value, the resulting bias is utterly negligible because they
essentially find no bright SN~Ib/c that could be confused with SN~Ia;
that is, all SNe from their sample would be rejected as outliers.  In
contrast, \citet{2002AJ....123..745R} suggest the existence a bright
component of the SN~Ib/c population which would be more easily
confused with SN~Ia.  It is difficult to be sure if this disagreement
is significant given the sample sizes, but we note that the KAIT
result is based on a galaxy targed survey which avoids low-luminosity
host galaxies, so if there is any relation between SN~Ib/c luminosity
and host type, as there is for SN~Ia, it could miss this portion of
the sample.  We will therefore conservatively assume that there is an
additional bright SN~Ib/c population missed by KAIT.  Combined with
the KAIT sample, the bright sample is $< 15\%$ of the total SN~Ib/c
population.  The magnitude distribution of core-collapse SN has also
been investigated photometrically by \citet{2009A&A...499..653B},
although this paper includes other types of core-collapse SNe which
are easier to distinguish from SNe~Ia.

The method is then to generate a simulated sample of a few
hundred thousand SN of both types, include observational uncertainties
representative of SNLS, apply the outlier rejection, assume
that all SN of any type which survive the outlier rejection
are included in the sample, and finally compute the difference
between the mean magnitude and the assumed cosmology in redshift bins
as the bias.  This is then multiplied by the actual fraction of
SNIa$\star$ in the corresponding real sample to estimate the
contamination bias (really, a conservative upper limit on the bias).

To describe this bright population, we use a Gaussian distribution
characterized by its mean offset from the SN~Ia population and its
width, $\Delta_{\mathrm{bc}}$ and $\sigma_{\mathrm{bc}}$.
\citet{2002AJ....123..745R} estimates $\Delta_{\mathrm{bc}}=0.7$ and
$\sigma_{\mathrm{bc}}=0.33$ mag, after color correction.  However,
these parameters are quite uncertain.  The amount of bias is maximized
for $\Delta_{\mathrm{bc}} = \sigma_{\mathrm{bc}}$ mag (for smaller
values of $\Delta_{\mathrm{bc}}$ the amount of contamination
increases, but the bias decreases, and for larger values the
contamination rapidly falls), increases with the fraction of bright
SN~Ib/c, and weakly increases as $\sigma_{\mathrm{bc}}$ decreases.  We
therefore tune these parameters to maximize the bias while remaining
vaguely plausible, adopting $\Delta_{\mathrm{bc}}=0.25$,
$\sigma_{\mathrm{bc}} = 0.25$, and also conservatively assume that the
bright component is 25\% of the total SN~Ib/c population.  The
results, including the effects of multiplying by the actual fraction
of SNIa$\star$ in each bin, are given in table~\ref{tbl:noniabias}.
The effects on the cosmology are smaller than we can measure.  The
bias increases with redshift, both because the larger observational
uncertainties make it harder for sigma clipping to reject contaminants, and
because the fraction of SNe with SNIa$\star$ increases.  Because our
outlier rejection is based on the cosmological fit, the fact that the
SNIa$\star$ fraction is large in the highest redshift bin does not
affect this process, since these SNe have little influence on the
cosmological models considered here.  However, if this data set is
used to investigate models where the Hubble relation changes sharply
above $z=0.9$, these results should be treated with some caution.

The predicted contamination is 0\% at low redshift, rising
to $\sim 10\%$ above $z=0.9$.  However, this model is deliberately
conservative, and does not match the observed properties of
the SNLS data very well. For example, it predicts that $> 60\%$ of the
candidates sent for spectroscopic typing would either be identified as
SN~Ib/c or SNIa$\star$, while we actually find $< 25\%$, and that is only
because we deliberately targeted some objects not expected
to be SN~Ia.  It also predicts that $>20\%$ of the candidates sent for
spectroscopy would be identified as bright SN~Ib/c, whereas there are
only a handful in reality.

While we find that for the SNLS3 sample the effects are small, this
may not be true for future surveys that will predominantly rely on
photometric rather than spectroscopic typing.  In this case, where
contamination may be large, the effects of sigma clipping are no
longer as simple or as beneficial, since the mean magnitude in a bin
may be severely affected when the types of most the SNe are
potentially uncertain.  This becomes even more complicated if only
photometric redshifts are available.  Such efforts will require a more
precise understanding of the properties and demographics of SNe~Ib/c
if they are to compete with spectroscopically typed surveys.

\begin{table}
\centering
\caption{Bias from non-Ia contamination \label{tbl:noniabias}}
\begin{tabular}{l|rrr}
\tableline\tableline
$z$ & Raw Bias (mag)&  SNIa$\star$\% & Effective Bias\tablenotemark{a} (mag)\\
\tableline
0.1  & 0.015 & 0\% & 0.0    \\
0.26 & 0.024 & 6\% & 0.001  \\
0.41 & 0.024 & 14\% & 0.003 \\
0.57 & 0.024 & 17\% & 0.004 \\
0.72 & 0.023 & 24\% & 0.006 \\
0.89 & 0.026 & 50\% & 0.013 \\
1.04 & 0.025 & NA & NA \\
\tableline
\end{tabular}
\tablenotetext{a}{\centering Includes the actual fraction of SNIa$\star$
 in each bin.  There are no SNLS SNe above $z=1.04$. }
\end{table}

\subsection{Milky-Way Extinction Correction}
\nobreak
\label{subsec:mwextcorr}
We correct all SNe for Milky-Way extinction.  In addition to the
random uncertainty in the $E(B-V)$ values, there is a correlated uncertainty
in the conversion from dust column density to extinction of 10\%, as
discussed in \citet{1998ApJ...500..525S}.  This affects the nearby and
distant SNe differently because distant SN searches generally target
regions of low galactic extinction and observe at longer rest-frame
wavelengths.  The overall effect is given in
table~\ref{tbl:overallsystematics}, and is one of the larger
contributions to our final uncertainty budget.

\subsection{Evolution}
\nobreak
\label{subsec:evolution}
The general lack of firm theoretical predictions makes it difficult to
quantify the effects of potential SN evolution.  Evolution in the
absolute magnitude of SNe~Ia with redshift is not constrainable
without a detailed physical model because it can in principle mimic
any cosmology.  Here we consider only models that we can put some
constraints on, in particular the possibility of evolution in the
stretch-luminosity and color-luminosity relations with redshift.
There is considerable evolution in the demographics of the SN
population between $z=0$ and $z=1$ \citep{2007astro.ph..1912H},
  but this does not bias the results as long as the demographic
  evolution is in parameters that are corrected for. If there are any
differences in the width- and color-luminosity corrections between
different sub-groups, we would expect there to be some evolution in
the mean values of $\alpha$ and $\beta$ with redshift.  In particular,
a change in the relative mixture of SN intrinsic colors and external
effects such as dust would affect $\beta$.

\begin{table}
\centering
\caption{Evolutionary Systematics \label{tbl:evolution}}
\renewcommand{\arraystretch}{1.5}
\begin{tabular}{ll|ll}
\tableline\tableline
 Description & dX/dz & $w$ for \om=0.27 & Rel Area \\
\tableline
 Stat only   & \nodata &  \allstatwfixomw & 1 \\
\tableline
 Combined           & \nodata & $-1.028^{+0.059}_{-0.058}$ & 1.02 \\
 $\alpha$ evolution & 0.07 & $-1.030 \pm 0.058$ & 1.00 \\
 $\beta$ evolution  & 1.0 & $-1.028^{+0.059}_{-0.058}$ & 1.02 \\
\end{tabular}
\renewcommand{\arraystretch}{1}
\end{table}

Recently K09 (\S10.2.3) have presented evidence for a strong decrease
in $\beta$ with redshift using SALT2 (although not the version used
here -- see Appendix A of G10 for details).  They find that $\beta$
decreases from $\sim 2.5$ at $z=0$ to 1 by $z=0.7$, but this evolution
is only seen in ESSENCE and first-year SNLS samples.  They also carry
out MLCS2k2 fits, but do not analyze them in the same fashion. The
relation between light-curve shape, color, and distance modulus for
MLCS2k2 has already been discussed in \S\ref{subsec:lcfitters}.  If we
extend this model to allow for evolution in the width- and
color-corrections with redshift, we find $d \alpha / d z = 0.25 \pm
0.07$ and $d \beta / dz = -1.86 \pm 0.43$, where the modified distance
modulus is $\mu^{\star} = \mu + \alpha \Delta - \beta A_V$.  Hence,
for the K09 sample, both SALT2 and MLCS2k2 show strong evidence for
evolving $\beta$ with redshift in the SDSS analysis, with MLCS2k2
showing more evolution.

\begin{figure}
\plotone{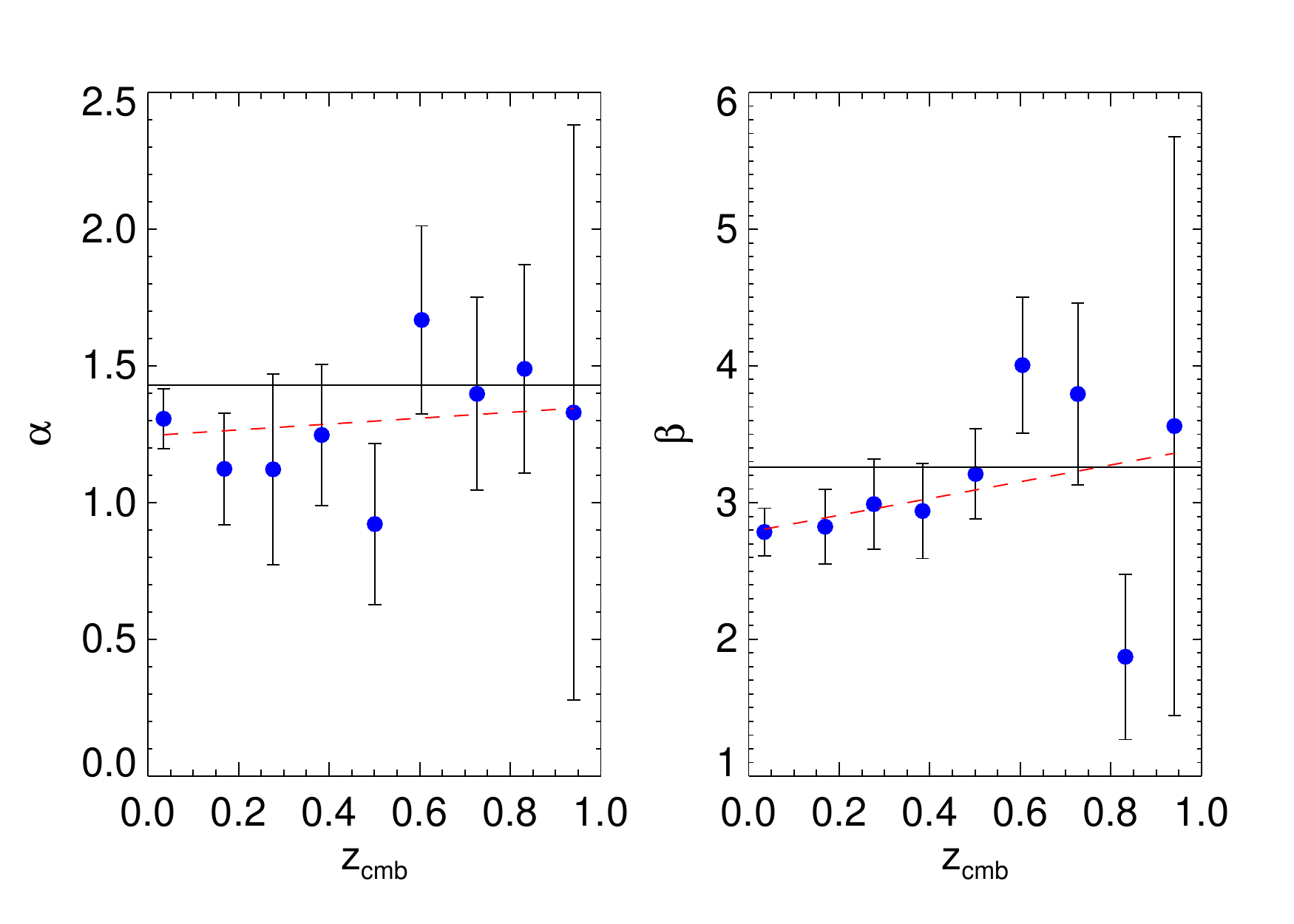}
\caption[Measured evolution in $\alpha$ and $\beta$ with redshift]{
  Evolution of $\alpha$ and $\beta$ with redshift in the SiFTO/SALT2
  combined light-curve fits.  The solid lines are the best-fit values
  assuming no evolution, including systematic effects, and simultaneously
  fitting the cosmological parameters, while the point in each
  redshift bin is evaluated with the same fixed cosmology.  The dashed
  line is the best linear fit to the points.
\label{fig:alphabeta}
}
\end{figure}

As shown in \S5.5 of G10, we find much weaker evidence for such a
trend in our data with SiFTO and an updated version of SALT2.  This is
probably related to the more sophisticated color-uncertainty modeling
compared with that used in K09.  Here we extend this analysis to
include external data sets and the combined SiFTO/SALT2 distances
(figure~\ref{fig:alphabeta}).  We find $d \alpha / d z = 0.021 \pm
0.07$ and $d \beta / d z = 0.588 \pm 0.40$, marginal evidence of
evolution.  Also note that the $\beta$ evolution has the opposite sign
than the SDSS results, increasing slightly with redshift.
Furthermore, SALT2 and SiFTO show small and opposite trends, with
SiFTO increasing while SALT2 decreases.

It is unlikely that the observed evolution in $\beta$ is real.  \S5.5
of G10 presents several lines of evidence to support this: the
opposite trends for SALT2 and SiFTO, the fact that artificially adding
selection effects at a different redshift moves the evolution to that
redshift, and the dependence on the assumed parameterization of
uncertainties in the light-curve intrinsic color model.  However,
while we can construct a plausible model that explains these trends,
we cannot completely rule out $\alpha$, $\beta$ evolution.  Therefore,
the conservative approach is to treat this effect as real and include
it in our systematic uncertainty budget.  Since we find no evidence
for $\alpha$ evolution, we adopt the uncertainty in the slope as our
systematic.  As in G10, for $\beta$ we conservatively adopt $d \beta /
dz = 1.0$. The effects of doing so are given in
table~\ref{tbl:evolution}.

Based on a model of inter-galactic dust, \citet{2010MNRAS.tmp.1064M}
predict a slow change in mean SN color with redshift without affecting
the absolute magnitudes that appears as a change in $\beta$ over a
long enough redshift interval. Unfortunately, over a small redshift
range, such as one of the redshift bins we use above, their model
manifests as a change in the offset in the color-luminosity relation
but does not change the slope ($\beta$), and therefore our test is not
sensitive to their model.  However, the $d \beta / dz$ value assumed
in our systematic is larger than that predicted by their model for our
data, so it is effectively included in our uncertainty budget.

\subsection{Environmental Dependence of SN Properties}
\nobreak
\label{subsec:environment}
As discussed in \S\ref{subsec:twoscriptm}, in our analysis we correct
for the recently discovered relationship between host-galaxy
properties and peak magnitude by splitting the sample on host-galaxy
mass and allowing \scriptm\ to be different for the two samples.
Because we do not have a physical model for this phenomenon, the cut
point is not well understood.  Furthermore, we are not certain if mass
is the best parameter, or if metallicity or star-formation rate are
more relevant.  Therefore, we consider systematic effects related to
this split.  \citet{2010MNRAS.406..782S} investigate the effects of
changing the split mass, and find that the effects on the
cosmological parameters are largest between $10^{9}$ and $10^{11}
M_{\odot}$.  We therefore estimate this systematic by comparing which
SNe change subsamples as we move the cut to these extremes from our
default cut at $10^{10} M_{\odot}$.

In addition, we consider changing variables to metallicity, and
similarly compute which SNe change subsamples.  Clearly it is
unreasonable to assume that both the appropriate cut differs from our
nominal mass cut and that mass is not the right variable, so we
equally weight each term (changing the cut and the two other
host-variables) so that the total weight is 1.  The final effects
are given in table~\ref{tbl:overallsystematics}, and are one of the
larger contributions to our final uncertainty budget (much less than
calibration, however).  If we do {\it not} apply this correction, as
in all previous analyses, but apply the difference purely as a
systematic uncertainty, the effect is approximately the same size as
all other uncertainties (statistical and systematic) combined.

\subsection{Gravitational Lensing}
\nobreak
\label{subsec:gravlensing}
Gravitational lensing is expected to cause increased dispersion in the
Hubble diagram of high redshift SNe.  While the mean amount of
magnification is unity at all redshifts, when there are a small number
of SNe in each redshift bin, the asymmetric nature of the lensing
probability (there is a long tail extending to high magnifications,
but a lower limit on the de-magnification), coupled with selection
effects, can produce biases in the peak luminosities.  For surveys
over very small areas, lensing will also induce correlations between
different SNe.  These issues are studied in
\citet{2005ApJ...631..678H}, who find that for surveys such as SNLS,
the number of SNe in each redshift range and the survey area are large
enough that these issues are minor.  Lensing does induce additional,
almost uncorrelated, scatter in the peak magnitudes, which we already
include in our statistical uncertainty budget using the model of
\citet{2010MNRAS.405..535J}, which gives $\sigma = 0.055 z$.
This term is largely degenerate with $\sigma_{\mathrm{int}}$, but we
keep track of it separately because it has a redshift dependence while
$\sigma_{\mathrm{int}}$ does not.  Substituting a different model for
the intrinsic scatter, such as \citet{2005ApJ...631..678H}
  ($\sigma = 0.088 z$) has almost no effect because
$\sigma_{\mathrm{int}}$ changes to compensate.  Therefore, besides
including the statistical term, we do not include any additional
lensing systematics in our uncertainty budget.

\section{COMPARISON OF UNCERTAINTY BUDGET WITH OTHER ANALYSES}
\label{sec:compare}
\nobreak 
It is somewhat difficult to compare our systematic uncertainties
directly with those of previous analyses because, in general, they do
not take into account that systematic uncertainties will usually shift
the measured cosmological parameters, which in turn affects the
marginalized uncertainties on each parameter.  For example, the
potential $\alpha, \beta$ evolution actually shrinks the marginalized
uncertainties slightly for fixed $\om$, although the area of the $\om,
w$ contour increases.  We shall first present two examples of the
classic form of systematics analysis discussed in
\S\ref{subsec:method}.  Note that we do not restrict ourselves to the
same data samples as the other papers in these comparisons. As noted
previously, all previous papers have not included the effects of
systematic uncertainties on the training of light-curve fitters.  The
principal effect of including these effects is to increase the
importance of any systematic uncertainty that affects SN colors.

The systematic contributions from the first year SNLS analysis are
given in table~6 of A06 in terms of the effects on each cosmological
parameter while holding all other parameters fixed at their best fit
value.  We can use the same technique to compare our results.  For
fixed $\Omega_m$, we find $\sigma_w = 0.067
\left(\mathrm{sys}\right)$, which is in reasonable agreement with the
0.07 value estimated by A06.  Examining individual terms in detail,
the fact that the values are similar in fact represents considerable
progress, as the A06 values were somewhat optimistic, particularly
regarding calibration. We include a large number of terms not
considered in the first year analysis, although most of these are
sub-dominant.  Our Malmquist bias uncertainty is much smaller, as
benefits the much more detailed simulations carried out for the
3-season sample.  Many of the individual uncertainty terms have gone
down, such as the SNLS zero-point uncertainty which has shrunk from
$\sigma_w = 0.05$ to 0.02.  However, A06 significantly underestimated
the uncertainties related to the flux standard, giving $\sigma_w =
0.03$ while we have 0.045.  This is particularly significant because
the A06 analysis was based on Vega, which has far less certain
magnitudes on the Landolt and MegaCam systems than the standard
adopted here, \bd .

\citet{2007ApJ...666..694W} discuss the systematics of the 3 year
ESSENCE sample.  \citet{2009ApJ...700.1097H} uses a very similar
analysis and obtain similar results.  They find a total systematic
uncertainty in $w$ for ESSENCE of 0.13, compared with 0.11 here.
However, their value is combined with the BAO constraint of
\cite{2005ApJ...633..560E}, which reduces the importance of systematic
effects that move the SN contours along their longest axis.  When this
combination is carried out, the equivalent value for SNLS3 is $\sim
0.06$, about a factor of 2 smaller -- see S11 for further
discussion.  This reduction is despite the fact that we again
include many additional systematic terms not considered by these
authors, such as the magnitudes of the flux standard, the low-$z$
zero points and filters, shifts in the mean passbands of our survey,
Malmquist uncertainties of the nearby sample, etc.  The major
difference arises from two terms: first, \citet{2007ApJ...666..694W}
includes a systematic related to the ``Hubble bubble'' discussed in
\S\ref{subsec:bubble} which we find no evidence for, and second, they
have significant systematic effects related to the prior on extrinsic
SN colors used in their MLC2k2 fits, which does not apply to SiFTO or
SALT2.  Without these two terms, their systematic uncertainty on $w$
(with the BAO measurement) is 0.054, similar to our value.

As discussed previously, MLCS2k2 trades sensitivity to an assumed
prior on the distribution of extinction for reduced statistical
uncertainties.  A major portion of the \citet{2007ApJ...666..694W}
systematic uncertainty budget arises from the uncertainty in this
prior, and the fact that survey selection effects cause it to evolve
strongly with redshift.  This tradeoff makes comparing our systematic
uncertainty budgets difficult.  SALT2 and SiFTO do not use such
priors, and Monte-Carlo studies of our selection biases indicate that
these have a very small effect on our color modeling.

We next consider the analysis of \citet{2008ApJ...686..749K}.  As
noted in \S\ref{subsec:method}, these authors use a more sophisticated
framework for including systematic uncertainties than the classic
quadrature method.  Their method is a good approximation to the one
used here when there is no structure in the covariances within a given
data set, and no significant redshift-dependent correlations between
datasets.  As shown in the right panel of figure~\ref{fig:corrmat},
the first is a reasonable assumption in some cases, but not in others,
particularly for surveys which cover a large redshift range such as
SNLS.  The second assumption neglects the highly redshift-dependent
correlations caused by calibration uncertainties related to the
fundamental flux standard, which are not treated.  One advantage of
\cite{2008ApJ...686..749K} is that, unlike this paper, it is a blind
analysis, which helps limit observer bias. Combining SN data with the
BAO measurement as above, they find a systematic uncertainty in $w$ of
0.085, significantly larger than our value of 0.067).  This difference
mostly arises from the extremely heterogeneous nature of their high-$z$
SN sample and the large uncertainties in relative zero point assigned
between them.

The analysis of \citet{2008ApJ...686..749K} was extended to a larger
sample and refined to allow for more subtle redshift dependence within
each sample in \citet{2010ApJ...716..712A}.  In both cases, the
authors fix $\alpha$ and $\beta$ at their best fit values when
computing systematic uncertainties, which will underestimate the final
uncertainties because of the correlation between the cosmological and
nuisance parameters, as discussed in \S\ref{subsec:wrongway}.  Their
calibration is based on Vega rather than \bd, adopting a magnitude
uncertainty of 0.01 mag in each of 6 bands.  As discussed in \S10 of
R09, the uncertainties in the magnitude of Vega on the Landolt system
are much larger, especially in $V-R$, so it seems likely that the
effects of calibration on their uncertainty budget are underestimated
by at least a factor of 2 to 3.  In general, the systematic
uncertainty as a function of redshift of this paper has a much simpler
structure than that presented here, although it is still a large
improvement over most treatments.

K09 is probably the most extensive treatment of systematic
uncertainties in the literature, although the authors use the
classical quadrature method.  They include many of the terms
considered here, although several are omitted (e.g., bandpass
uncertainties for SNLS, correlated peculiar velocities, the NICMOS
nonlinearity).  They present results for both MLCS2k2 and SALT2
(although not the improved version used here), but do not explicitly
include the comparison in their uncertainty budget.  Their largest
source of uncertainty is due to the rest-frame low-$z$ $U$ band, while
we exclude this data entirely based on the evidence presented in
\S\ref{subsec:nearbydata}.  They also find evidence for a large
variation of $\beta$ with redshift, but only include this term for
SALT2.  Our analysis shows that this is present in their data for both
fitters, and more strongly for MLCS2k2.  As shown in
\S\ref{subsec:evolution}, our analysis does not support evolution in
$\beta$.  Finally, they also include a large uncertainty for the
minimum $z_{\mathrm{cut}}$, which we have shown is consistent with shot noise
and therefore implicitly included in our statistical uncertainty.  If
we exclude these three terms, which we do not find evidence for, the
total systematic uncertainty budget for K09 is about 0.07 in $w$ when
combined with non-SN constraints, similar to the value for our sample.

\section{HOW THESE SYSTEMATICS WILL BE REDUCED}
\label{sec:improvement}
\nobreak 
The dominant systematic uncertainties in our analysis are due to
calibration, specifically the colors of \bd .  Other important terms
include the relationship between SN peak magnitudes and host galaxy
properties and corrections for Malmquist bias. A careful reading of
R09 will reveal that many of the important calibration systematic
effects are related to the necessity of cross-calibrating SNLS data to
the current low-$z$ SN sample on the Landolt system. We expect that a
number of significantly improved low- and intermediate-z samples
should become available in the next few years (a partial listing
includes the the SNfactory sample, the KAIT sample, and the other 2
years of SDSS data), and so it is worth examining exactly what
aspects of these samples will help us reduce these uncertainties.

The uncertainties related to the color of \bd\ include large terms
related to how well the Landolt magnitudes of this star can be
transferred to the MegaCam system; see \S10.2 of R09.
Because it is a slightly unusual star (a subdwarf), and because the
MegaCam bandpasses are very unlike the Landolt ones, these
uncertainties are large.  Furthermore, the effective bandpasses
of the Landolt system are simply not well understood, and likely never
will be, putting a fundamental limit on how well we can characterize
this uncertainty.  Were the low-$z$ sample replaced with one observed on
a better understood system more similar to the MegaCam one, such as
the USNO or SDSS systems, these transfer effects would be reduced
substantially reducing our calibration uncertainties.

Improved lower-z samples may allow us to understand various
degeneracies between intrinsic and extrinsic SN color, possibly by
studying SNe in a variety of environments or incorporating near-IR
observations.  Such samples may also lead to better luminosity
predictors beyond stretch and color, such as the spectral flux ratio
feature of \citet{2009A&A...500L..17B}.  Either possibility could lead
to qualitative improvements and not simply quantitative ones in SN
models.

The differences between light-curve models will also be reduced with
better training samples, or else it will become clear which model is a
good description of SN properties.  While including high-$z$ SN data in
the training has been very useful, especially in the near-UV, improved
low-$z$ samples can improve the situation even further.  A particular
weakness of the current high-$z$ data is the lack of multiple epochs of
spectroscopy at different phases of the light-curve, especially in the
near-UV.  This deficiency can be remedied with STIS or COS
spectroscopy of nearby SNe.  The low-$z$ sample zero points can obviously
be improved by better nearby samples.  The effects on the MegaCam
zero points are more subtle, but if SNLS is calibrated directly against
a similar system, they will be also be improved.

Since so many of these considerations relate directly to the
calibration of the nearby sample, we expect to obtain the best results
with low-$z$ SNe calibrated to an observing system that is closer to the
MegaCam one.  Therefore, we expect the best results in the near term
will be obtained when SNLS is properly combined with the SDSS SN
sample and the CSP samples, both of which are approximate
implementations of the USNO system.  In particular, due to the large
number of faint standard stars that have been observed with the SDSS
telescope \citep{2007AJ....134..973I}, the inter-calibration of SNLS
and SDSS should be extremely powerful.

We also hope that the CALSPEC calibration can be refined and extended
to more stars, particularly faint ones that can be observed directly
by survey telescopes (\bd\ is too bright for both MegaCam and SDSS),
and to stars of normal spectral types in the color range well sampled
by standard star catalogs, which will significantly improve the
reliability of the transformations between natural systems.  This
program should be carried out while the natural USNO, and MegaCam
systems still exist (although it is too late for the $i_M$ filter used
in this paper, which was destroyed in an unfortunate filter jukebox
malfunction, and for the SDSS imager, which has been decommissioned,
but there were many stars already observed by both); a major problem
with the Landolt system is that the original system no longer exists,
so there will always be limitations as to how accurately any given
flux standard can be tied to this system.  This finding has
implications for future absolute calibration programs.  From the
standpoint of SN observations, how well the calibrated flux standard
can be tied to the magnitude system in use is currently the limiting
factor rather than how well the SED itself is calibrated, so good
quality spectroscopic observations of fainter (and hence directly
observable) standards are vastly preferable to improved observations
of very bright standards such as Vega.

For $z > 0.7$ SNe, SNLS essentially observes in rest-frame $U$ and
$B$, and then relies on the relationship between $U-B$ and $B-V$ to
compare with lower-redshift SNe.  A survey which measures rest-frame
$B-V$ out to higher redshifts using the near-IR should be able to
reduce the importance of some of the calibration systematic effects
discussed here for high-$z$ SNe.  This would be particularly useful
for testing time-varying $w$ models.  However, such a survey then
becomes sensitive to the problems of inter-calibrating near-IR and
optical data, which are non-trivial.

\section{CONCLUSIONS}
\label{sec:conclusions}
\nobreak 
We have presented measurements of the cosmological parameters from the
first 3 years of the SNLS, combining them with external SN
data sets at higher and lower redshift and including a correction for
host-galaxy mass.  This is the largest, highest-quality moderate- to
high-redshift SN~Ia sample to date.  From SN data only, we find $w =
-0.91^{+0.16}_{-0.20}\left( \mathrm{stat} \right) ^{+0.07}_{-0.14}
\left( \mathrm{sys} \right)$, consistent with a
cosmological constant.  The data requires an accelerating universe at
$>99.999\%$ probability including all identified systematics.  The
combination with non-SN constraints is given in S11, and yields
considerable additional precision as well as allowing us to put
constraints on time variations in $w$.

We model systematic uncertainties using a covariance matrix, including
their effects on our empirical SN models.  This approach
should allow these uncertainties to be properly included in future
analyses\footnote{ The statistical and systematic covariance matrices
  used in this paper, as well as the light-curve parameters, are
  available at \url{https://tspace.library.utoronto.ca/snls}}.  For
SNe only, our current uncertainty budget is statistics dominated; when
external, non-SN constraints are included, which we simulate by fixing
\om , statistical and systematic uncertainties are comparable.  The
dominant identified systematic uncertainties are related to
calibration.  This limitation can best be addressed with
improved nearby SN data sets, particularly if they are calibrated onto
more modern, better understood photometric systems such as USNO/SDSS.
In this analysis, we have included external SNe samples calibrated onto
both the Landolt and USNO-like systems without taking full advantage
of the inter-calibration possibilities.  Future papers in
collaboration with the SDSS and CSP teams will address this point.
The next most important systematic term, the relation between host
mass and SN peak magnitudes can also be addressed with improved nearby
samples to ascertain whether the peak magnitude subsample effect
correlates best with metallicity or some other variable. Solid-angle,
rather than galaxy-targeted searches, will be the most useful for this
purpose because they will sample the underlying host galaxy population
in a less biased fashion.

In agreement with K09, we find that there are significant problems
with current low-$z$ observer-frame $U$-band observations, although we
are unsure where the issue lies; we have therefore excluded these
observations in our fits.  Interestingly, \up\ observations of low-$z$
SNe, which are calibrated to the USNO instead of the Landolt system,
but probe the same wavelength range, do not show any problematical
behavior.  We also note that the sign of the effect we observe is
opposite that predicted by the spectral studies of
\citet{2010arXiv1010.2211C,2010arXiv1010.2749F}.

In general, the systematic uncertainties of the SNLS3 sample are
smaller than those of previous analyses, even after correcting for the
fact that we do not find evidence for a number of pathologies found by
other authors, such as $\beta$ evolution of the Hubble bubble, and
therefore do not include them in our analysis.  The improvement in the
final uncertainty budget, despite the many additional effects we
consider, represents a significant advance, and makes clear the
benefits of large, homogenous SN samples.

\section{APPENDIX A: PASSBANDS FOR THE LOW-$Z$ SAMPLE}
\label{apndx:filterslowz}
\nobreak 
Determining the natural system response of the Landolt system, which
is the system of much of the low-$z$ data, is non-trivial. Fortunately,
this affects less than half of the nearby sample, and has no effect on
the SNLS, SDSS, or {\it HST} samples except indirectly through the
light-curve training\footnote{Although we report the peak $B$ band
  magnitude, because we work in SED space for both fitters, we are
  free to define this filter to be anything we want.  Therefore, any
  uncertainty on the Landolt filters only affects SNe whose
  observations are reported directly in that system.}.  There have
been many attempts to reconstruct the Landolt passbands, which are
nominally a realization of the Johnson-Morgan and Cousins systems.
These are natural systems, so it useful to attempt to reconstruct
them.  The standard work on this subject is
\citet{1990PASP..102.1181B}, which has been used in many previous SN
cosmological analyses.  Notable exceptions are A06, who used these
filters responses but shifted them by 41, 27, 21 and 25 \AA\ blueward
(for $BVRI$, respectively), and K09 (Appendix B) who use shifts of 15,
-12, -7, 45 \AA\ (blueward)\footnote{In fact, filter shifts were not
  used, but instead linear transforms were applied to bring Landolt
  system photometry onto the Bessell system.  These are equivalent
  because the implied shifts are small.  For larger shifts, using
  linear transforms is inadvisable.}.  The former was based on a
comparison of synthetic photometry of Landolt standard stars from
\citet{1992PASP..104..533H,1994PASP..106..566H} to their actual
magnitudes, and the latter on CALSPEC {\it STIS} standards using
magnitudes from \citet{2007AJ....133..768L}.

We have carried out a similar analysis to that of A06 for this
paper, but using a more recent spectrophotometric library of Landolt
stars \citep{2005PASP..117..810S}.  The CALSPEC library is attractive,
but unfortunately, as discussed in \S10.1 of R09, the magnitudes of
these stars from \citet{2007AJ....133..768L} are probably not on the
Landolt system to the required accuracy because the linear
transformations derived from normal stars were not appropriate to the
extremely blue CALSPEC white dwarfs. The \citet{2005PASP..117..810S}
compilation is essentially an extension of that of
\citet{1992PASP..104..533H,1994PASP..106..566H} to a much larger data
set by many of the same authors.

Using this sample, we find the best fits if we shift the Bessell
$BVRI$ by 9, 3, 21, and 14 \AA\ to the {\it red}.  Using bootstrap
with replacement, and folding in the uncertainty in the colors of the
fundamental flux standard, the uncertainties are about 12 \AA .
Except possibly in $R$, we find shifts which are consistent with
zero. Our results are in agreement with the analysis presented in
\citet{2005PASP..117..810S} using the same data, and those of
\citet{2006AJ....131.1184M}, who performed a similar test using a {\it
  HST}-based spectrophotometric library.  Repeating this test with the
\citet{1992PASP..104..533H,1994PASP..106..566H} libraries, we find
general agreement with the values used in A06.  Note that neither the
Hamuy nor the Stritzinger libraries have enough wavelength coverage to
enable this test for $U$.

A fraction of the Landolt standard stars also have magnitudes on the
USNO system \citep{2002AJ....123.2121S}.  Since the bandpasses of the
USNO system have been determined much more accurately, we can use the
spectrophotometry of these stars to test the
\citet{2005PASP..117..810S} library.  We find shifts of 4, -8, and 26
\AA\ for \gp \rp \ip, with uncertainties of about $\pm 16$ \AA
. Except for \ip , this library does an excellent job reproducing the
USNO magnitudes.  We can't apply this test to the
\citet{1992PASP..104..533H,1994PASP..106..566H} libraries because
there is no overlap with the USNO sample. Again, we find marginal
evidence for problems in the red, which may explain the shift found
for $R$, since the Landolt/Cousins $R$ filter has a much redder
response than \rp . This is particularly interesting because the
synthetic photometry tests agree well with the other libraries in our
tests of the MegaCam filters ( \S\ref{subsubsec:snlsfilts} ).
Fortunately, the $R$ filter plays a very minor role in our analysis.
We therefore adopt the best fit values as shifts applied to the
Bessell filters in our analysis, and treat the uncertainties in
\S\ref{subsubsec:otherfilts}.

In order to construct natural system bandpasses for the
\citet{2009ApJ...700..331H} sample, which combines Landolt-like and
USNO-like filters, we multiply the telescope/detector responses by an
assumed atmospheric transmission. We were unable to obtain an
atmospheric absorption curve specific to Mt.\ Graham, so substituted
that of Kitt Peak, which is at a similar altitude and shares a similar
climate.  This substitution is included in the filter bandpass
systematics discussed in \S\ref{subsubsec:otherfilts}.

\section{Appendix B: COSMOLOGICAL FITTING TECHNIQUES}
\label{apdx:margmin}
\nobreak 
We consider two different techniques for fitting the cosmological
parameters\footnote{Both codes are available from
  \url{http://casa.colorado.edu/\textasciitilde
    aaconley/Software.html}}.  In the first, we calculate the relative
probability of each value of the parameters over a grid and report the
expectation value of the resulting distribution for each parameter
(the marginalization approach).  In the second, we attempt to find the
value of the parameters that minimizes the \chisq (the minimization
approach).  In the first case, the uncertainties are derived by directly
computing the bounds that contain the desired fraction of the total
probability on each marginalized parameter.  In the second, we make
standard assumptions about the relation between the \chisq\ and the
parameters (namely, that the uncertainties are Gaussian and the model is
approximately linear in the parameters over the uncertainties) to estimate
the confidence limits.  The Markov-Chain Monte-Carlo results now used
in most CMB analyses are ideologically equivalent to the first
approach.

The results reported from these two approaches will generally not
agree because they do not have the same meaning mathematically.
This does not imply that either approach is incorrect; while
it would be comforting to be able to clearly choose one method
as more desirable, reality is not so kind.  Further discussion
of the differences between these two approaches can be found
in \citet{2005PhRvD..72f3501U}.

In the grid based approach, the $\chisq$ is converted into a relative
probability via $P \propto \exp \left(-1/2\, \chisq \right)$.  An
evenly spaced grid is used, which is equivalent to assuming a flat
prior on all parameters.  Properly speaking, since $\mathbf{C}$ 
  (the total covariance matrix of equation~\ref{eqn:chieqn}) is a
function of $\alpha, \beta$, we should have $P \propto \exp \left(
-1/2\, \chisq \right) / \sqrt{ \mathrm{det} \, \mathbf{C} }$, where
det is the determinant operator.  However, including this term results
in large negative biases on $\alpha$ and $\beta$ that are worse for
larger samples, whereas if it is omitted the biases are negligible for
data sets similar to ours.  See \citet{2007ApJ...665.1489K} for more
discussion of the $\mathrm{det} \, \mathbf{C}$ factor.

A weakness of the \chisq\ minimization approach is that the usual
method of estimating uncertainties (by searching for the boundary where the
\chisq\ increases by a certain amount) depends on the assumption that
the model is close to linear in the parameters over the size of the
uncertainties.  While this is true for the fits presented in this paper, it
is not necessarily accurate for more poorly constrained parameters, such
as the derivative of $w$, and can lead to significant underestimates of the
uncertainties.

\section{APPENDIX C: ANALYTIC MARGINALIZATION OVER $\mathcal{M}$}
\label{apndx:marg}
Because the uncertainties of individual SN do not depend on \scriptm , it is
possible to remove this parameter from our fits by analytically
marginalizing over it, following the technique described in
\citet{2001A&A...380....6G}.  Here we trivially extend this treatment
to the case of non-diagonal uncertainties and to the case of two values
of \scriptm\ split by some additional variable (such as host galaxy
stellar masses).

We marginalize by converting the \chisq\ to a relative probability,
integrating it over a prior $\pi$, and then converting back to
a \chisq . For the case of a single \scriptm :
\begin{displaymath}
 \chisq_{\scriptm\, marg} = -2 \log  \left[
   \int^{\infty}_{-\infty} \, d{\mathcal M} \,
   \exp \left( -\frac{1}{2} \chi^2 \right)
      \pi \left( {\mathcal M} \right)
 \right] .
\end{displaymath}
In our code we assume a flat prior on \scriptm\ (as we do implicitly
on all other parameters).   Following the discussion in \S\ref{subsec:method},
if we define the vector of residuals between the model magnitudes and
the observed magnitudes $\Delta \vec{\mathbf{m}}$, but this time 
omitting \scriptm\ from $m_{\mbox{mod}}$, then 
\begin{displaymath}
    \chi^2 = \left( \Delta \vec{\mathbf{m}} - \mathcal{M} \vec{\mathbf{1}} 
    \right)^T \cdot \mathbf{C}^{-1}
    \cdot \left( \Delta \vec{\mathbf{m}} - \mathcal{M} \vec{\mathbf{1}} \right)
\end{displaymath}
where $\mathbf{C}^{-1}$ is the inverse covariance matrix.  
Carrying out the integral,
\begin{equation}
 \chi^2_{\scriptm \, \textrm{marg}} = a + \log \frac{e}{2 \pi} - \frac{b^2}{e} .
 \label{eqn:chimarg}
\end{equation}
where $ a \equiv \Delta \vec{ \mathbf{m} }^{T} \cdot \mathbf{C}^{-1}
\cdot \Delta \vec{\mathbf{m}} $, $b \equiv \Delta \vec{ \mathbf{m}
}^{T} \cdot \mathbf{C}^{-1} \cdot \vec{\mathbf{1}} $, and $e \equiv
\vec{\mathbf{1}}^T \cdot \mathbf{C}^{-1} \cdot \vec{\mathbf{1}} $.
Note that the determinant of $\mathbf{C}$ does not appear in these
relations.  We do not analytically marginalize over $\alpha$ and
$\beta$ because the uncertainties of each SN depend on their values;
such a procedure was carried out in A06 by holding their values
fixed in $\mathbf{C}$, but this approach causes a significant
bias in the recovered values. Note that $a$, $b$, and $c$
all depend on parameters that we are fitting.

The formula is more complicated in the presence of two values of
\scriptm .  If we define two vectors $\vec{K_1}$ and $\vec{K_2}$ so
that $\vec{K_1}$ is one where the SN is in set 1 and 0 in set 2, and
vice-versa for $\vec{K_2}$, then $a$ remains the same, but the
expressions for $b$ and $e$ are modified to $b \equiv \Delta \vec{
  \mathbf{m} }^{T} \cdot \mathbf{C}^{-1} \cdot \vec{K_1} $ and 
$e \equiv \vec{K_1}^T \cdot \mathbf{C}^{-1} \cdot \vec{K_1} $.  Further
introducing $c \equiv \Delta \vec{ \mathbf{m}}^{T} \cdot
\mathbf{C}^{-1} \cdot \vec{K_2} $, $d \equiv \vec{K_1}^{T}
\cdot \mathbf{C}^{-1} \cdot \vec{K_2} $, $f \equiv 
\vec{K_2}^{T} \cdot \mathbf{C}^{-1} \cdot \vec{K_2} $, and $g \equiv e
f - d^2$ then
\begin{equation}
 \chi^2_{\scriptm \, \textrm{marg}} = a + \log \frac{e}{2 \pi} + 
 \log \frac{g}{2 \pi e} 
 - \frac{b^2 f}{g} - \frac{c^2 e}{g} + 2 \frac{b c d}{g}.
\end{equation}
Estimates for the two values of \scriptm\ are given by
$\scriptmo = (b f - c d)/g$ and $\scriptmt = (c e - b d)/g$.

Often there is a desire to evaluate the likelihood of some
cosmological parameters without fitting the nuisance
parameters.  One technique for doing this is to find the value of the
nuisance parameters that minimize the \chisq\ for a given set of
cosmological parameters and then evaluate the likelihood by
substituting those nuisance parameters.  So, for example, the
\scriptm\ that minimizes the \chisq\ is $\scriptm = b/e$.  The
resulting \chisq\ on substituting this value into
equation~\ref{eqn:chieqn} is equal to that of
equation~\ref{eqn:chimarg} to within an offset that depends on
$\alpha$ and $\beta$, but not \scriptm .  This trick is commonly used
\citep[e.g., in \texttt{CosmoMC},][] {2002PhRvD..66j3511L}, but
unfortunately significantly biases the cosmological results because
the \chisq\ for different values of $\alpha$ and $\beta$ can no longer
be compared.  Attempting to avoid fitting explicitly for $\alpha$ and
$\beta$ following this approach is also quite biased, because the
uncertainties depend on the nuisance parameters.  For the \chisq\
minimization fits, the results are not biased, but the uncertainty
estimates are incorrect, while for the marginalization fits both
the results and the uncertainties are biased.  Therefore it is
important to explicitly fit $\alpha$ and $\beta$ along with the
cosmological parameters.

\acknowledgements
The SNLS collaboration gratefully acknowledges the assistance of
Pierre Martin and the CFHT Queued Service Observations team.
Jean-Charles Cuillandre and Kanoa Withington were also indispensable
in making possible real-time data reduction at CFHT. This work is based on
observations obtained with MegaPrime/MegaCam, a joint project of CFHT
and CEA/DAPNIA, at the Canada-France-Hawaii Telescope (CFHT) which is
operated by the National Research Council (NRC) of Canada, the
Institut National des Sciences de l'Univers of the Centre National de
la Recherche Scientifique (CNRS) of France, and the University of
Hawaii. This work is based in part on data products produced at the
Canadian Astronomy Data Centre as part of the CFHT Legacy Survey, a
collaborative project of NRC and CNRS.  Canadian collaboration members
acknowledge support from NSERC and CIAR, French collaboration members
from CNRS/IN2P3, CNRS/INSU and CEA, and Portugese members
from Funda\c{c}\~ao para a Ci\^encia e Tecnologia.

We thank D.~J.\ Schlegel for making the updated SDSS calibration
available to us prior to publication, and A.~Riess, A.~Landolt, and
R.~Kessler for many useful discussions.  M.~Hicken provided useful
feedback about the CfAIII SN sample.  We thank R.~Bohlin for helping
us understand the uncertainties in the STIS SED of Vega, and
R.~de~Jong for clarifying the NICMOS non-linearity corrections.
Finally, thank the anonymous referee, whose extremely thoughtful and
thorough comments greatly improved this paper.

{\it Facilities:} \facility{CFHT (MegaCam)}, 
 \facility{Gemini:Gillett (GMOS-N)}, \facility{Gemini:South (GMOS-S)} 
 \facility{Keck:I (LRIS)}, \facility{VLT:Antu (FORS2)},
 \facility{VLT: Kueyen (FORS1)}


\end{document}